%% file: paper.tex
\documentclass[conference]{IEEEtran}

\usepackage{microtype}
\usepackage{graphicx}

\usepackage{booktabs} 

\usepackage{amsmath,amsfonts}

\usepackage{algorithmic}
\usepackage{graphicx}
\usepackage{textcomp}
\usepackage{xcolor}

\usepackage[ruled,linesnumbered]{algorithm2e}

\usepackage{mathrsfs}
\usepackage{amssymb}
\usepackage{amsmath}
\usepackage{amsthm}
\usepackage{epstopdf}
\usepackage{balance}
\usepackage{multirow}

\usepackage{epsfig,endnotes}

\usepackage{subfig}
\usepackage{subfloat}
\usepackage{grffile}
\usepackage{makecell}
\usepackage[font=bf]{caption}
\usepackage{color, url}
\usepackage{xspace} 
\usepackage{pdfx}

\usepackage{clipboard}
\newclipboard{myclipboard}
\openclipboard{myclipboard}

\newcommand{\myparatight}[1]{\smallskip\noindent{\bf {#1}:}~}
\newcommand{\argmax}{\operatornamewithlimits{argmax}}

\newcommand\CR[1]{\textcolor{black}{#1}}
\newcommand\CRR[1]{\textcolor{black}{#1}}

\newcommand{\lnorm}[1]{\ensuremath{\left\Vert#1\right\Vert}}

\usepackage{hyperref}

\allowdisplaybreaks

\begin{document}

\title{REaaS: Enabling Adversarially Robust Downstream Classifiers via Robust Encoder as a Service}

\author{%
  \IEEEauthorblockN{%
    Wenjie Qu$^{1}$\thanks{Wenjie Qu performed this research when he was an intern in Gong’s group.},
    Jinyuan Jia$^{2}$,
    Neil Zhenqiang Gong$^3$%
  }%
  \IEEEauthorblockA{$^1$ Huazhong University of Science and Technology, wen\_jie\_qu@outlook.com}%
  \IEEEauthorblockA{$^2$ University of Illinois Urbana-Champaign, jinyuan@illinois.edu}%
  \IEEEauthorblockA{$^3$ Duke University, neil.gong@duke.edu}%
}

\IEEEoverridecommandlockouts
\makeatletter\def\@IEEEpubidpullup{6.5\baselineskip}\makeatother
\IEEEpubid{\parbox{\columnwidth}{
    Network and Distributed System Security (NDSS) Symposium 2023\\
    28 February - 4 March 2023, San Diego, CA, USA\\
    ISBN 1-891562-83-5\\
    https://dx.doi.org/10.14722/ndss.2023.24444\\
    www.ndss-symposium.org
}
\hspace{\columnsep}\makebox[\columnwidth]{}}

\maketitle

\input{abstract}

\input{introduction}

\input{RelatedWork}

\input{Problem}

\input{method}

\input{evaluation}

\input{discussion}

\input{conclusion}

\balance
{
\bibliographystyle{IEEEtran}
\bibliography{refs}
}

\input{supplementary}

\end{document}

%% file: abstract.tex
\begin{abstract}

 \emph{Encoder as a service} is an emerging cloud service. Specifically, a service provider first pre-trains an encoder (i.e., a general-purpose feature extractor) via either supervised learning or self-supervised learning  and then deploys it as a cloud service API. A client queries the cloud service API to obtain feature vectors for its training/testing inputs when training/testing its classifier (called \emph{downstream classifier}). A downstream classifier is vulnerable to \emph{adversarial examples}, which are testing inputs with  carefully crafted perturbation that the downstream classifier misclassifies. Therefore, in safety and security critical applications, a client aims to build a robust downstream classifier and certify its robustness guarantees against adversarial examples. 
 
 What APIs should the cloud service provide, such that a client can use any certification method to certify the  robustness of its downstream classifier against adversarial examples while minimizing the number of queries to the APIs? How can a service provider pre-train an encoder such that clients can build more certifiably robust downstream classifiers? We aim to answer the two questions in this work. For the first question, we show that the cloud service only needs to provide two APIs, which we carefully design, to enable a client to certify the  robustness of its downstream classifier with a minimal number of queries to the APIs. 
 For the second question, we show that an encoder pre-trained using a spectral-norm regularization term enables clients to build more robust downstream classifiers. 
 
\end{abstract}

%% file: introduction.tex
\section{Introduction}

In an \emph{encoder as a service}, a service provider (e.g., OpenAI, Google, and Amazon) pre-trains a general-purpose feature extractor (called \emph{encoder}) and  deploys it as a cloud service; and a client  queries the cloud service APIs for the feature vectors of its training/testing inputs when training/testing a downstream classifier. For instance, the encoder could be pre-trained using supervised learning on a large amount of labeled data or self-supervised learning~\cite{he2019moco,chen2020simple} on a large amount of unlabeled data. A client could be a smartphone, IoT device, self-driving car, or edge device in the era of \emph{edge computing}.  Encoder as a service has been widely deployed by industry, e.g., OpenAI's GPT-3 API~\cite{GPT3-api} and Clarifai's  General Embedding API~\cite{clarifai_embedding}.   In the \emph{Standard Encoder as a Service (SEaaS)}, the service provides a single API (called \emph{Feature-API}) for clients and the encoder is pre-trained without taking the robustness of downstream classifiers into consideration. A client sends its training/testing inputs to the Feature-API, which returns their feature vectors to the client.

A downstream classifier is vulnerable to \emph{adversarial examples}~\cite{Szegedy14,carlini2017towards}. Suppose a testing input is correctly classified by the downstream classifier. An attacker adds a small carefully crafted perturbation to the testing input to induce misclassification. Such testing input with carefully crafted perturbation is called an adversarial example. Therefore, in security and safety critical applications such as user authentication and traffic sign recognition, a client desires to build a downstream classifier robust against adversarial examples. Many methods have been developed for an attacker to craft adversarial examples and the community keeps developing new, stronger ones. Therefore, instead of defending against a specific class of adversarial examples, a client aims to defend against all bounded adversarial perturbations via building a \emph{certifiably robust} downstream classifier. A classifier is certifiably robust if its predicted label  for a testing input is unaffected by  arbitrary perturbation added to the testing input once its size (measured by $\ell_2$-norm in this work) is less than a threshold, which is known as \emph{certified radius}. A larger certified radius indicates better certified robustness against adversarial examples.

In general, there are two categories of complementary methods to build a certifiably robust classifier and derive its certified radius for a  testing input, i.e., \emph{base classifier (BC) based certification}~\cite{wong2017provable,raghunathan2018certified,zhang2018crown,wang2018formal} and \emph{smoothed classifier (SC) based certification} (also known as \emph{randomized smoothing})~\cite{cao2017mitigating,lecuyer2019certified,cohen2019certified}. BC based certification aims to directly derive the certified radius of a given classifier (called \emph{base classifier})  for a testing input. BC based certification requires white-box access to the base classifier as it often requires propagating the perturbation from the input layer to the output layer of the base classifier layer by layer. SC based certification first builds a \emph{smoothed classifier} based on the base classifier via adding random noise (e.g.,  Gaussian noise) to a testing input and then derives the certified radius of the smoothed classifier for the testing input. To increase the testing inputs' certified radii, SC based certification often requires training the base classifier using training inputs with random noise. Moreover, to derive the predicted label and certified radius for a testing input,  SC based certification requires the base classifier to predict the labels of multiple noisy versions of the testing input.

\begin{figure*}[!t]
	 \centering
{\includegraphics[width=0.8\textwidth]{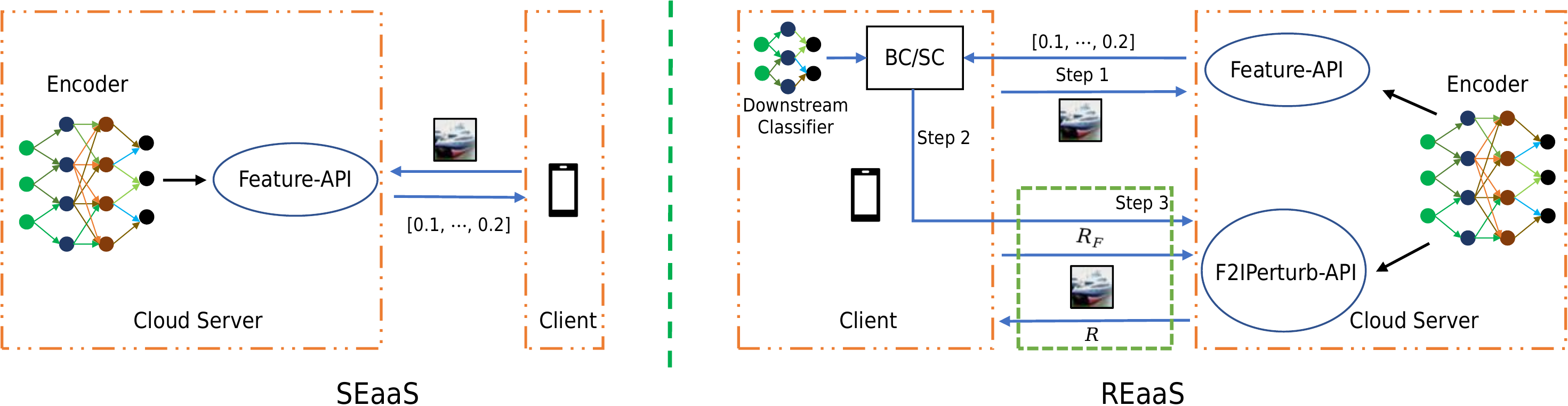}}
\vspace{-2mm}
\caption{\CR{SEaaS vs. REaaS.}}
\label{seaas_vs_reaas}
 \vspace{-8mm}
\end{figure*}

SEaaS faces three challenges when a client aims to build a certifiably robust downstream classifier and derive its certified radii for testing inputs. The first challenge is that a client cannot use BC based certification. In particular, the composition of the encoder and the client's downstream classifier is the base classifier that the client needs to certify in BC based certification. However, the client does not have white-box access to the encoder deployed on the cloud server, making BC based certification not applicable. 
The second challenge is that, although a client can use SC based certification by treating the composition of the encoder and its downstream classifier as a base classifier,  it incurs a large communication cost for the client and a large computation cost for the cloud server. Specifically, 
the client needs to query the Feature-API once for each noisy training input  in each training epoch of the downstream classifier because SC based certification trains the base classifier using noisy training inputs. Therefore, the client requires $e$ queries to the Feature-API \emph{per} training input, where  $e$ is the number of epochs used to train the downstream classifier. Moreover, to derive the predicted label and certified radius for a testing input, SC based certification requires the base classifier to predict the labels of $N$ noisy testing inputs. Therefore, the client requires $N$ queries to the Feature-API \emph{per} testing input. Note that $N$ is often a large number (e.g., 10,000)~\cite{cohen2019certified}. The large number of queries to the Feature-API imply 1) large communication cost, which is intolerable for resource-constrained clients such as smartphone and IoT devices, and 2) large computation cost for the cloud server. 
 The third challenge is that SC based certification achieves suboptimal certified radii. This is because the base classifier is the composition of the encoder and a client's downstream classifier, but a client cannot train/fine-tune the encoder as it is deployed on the cloud server.

\myparatight{Our work} We propose \emph{Robust Encoder as a Service (REaaS)} to address the three challenges of SEaaS. Figure~\ref{seaas_vs_reaas} compares SEaaS with REaaS. Our key idea is to provide another API called \emph{F2IPerturb-API}.\footnote{`F' stands for Feature and `I' stands for Input.} A downstream classifier essentially takes a feature vector  as  input and outputs a label. Our F2IPerturb-API enables a client to treat its downstream classifier \emph{alone} as a base classifier and certify the robustness of its base or smoothed downstream classifier in the \emph{feature space}. \CR{Specifically, a client performs three steps to derive the certified radius of a testing input in REaaS. First, the client obtains the feature vector of the testing input via querying the Feature-API. Second, the client views its downstream classifier alone as a base classifier and derives a \emph{feature-space certified radius} $R_F$ for the testing input using any BC/SC certification method. The client's base or smoothed downstream classifier predicts the same label for the testing input if the $\ell_2$-norm of the adversarial perturbation added to the testing input's feature vector is less than $R_F$. Third, the client sends the testing input and its feature-space certified radius $R_F$ to query the F2IPerturb-API, which returns the corresponding \emph{input-space certified radius} $R$ to the client. Our input-space certified radius $R$ guarantees the client's base or smoothed downstream classifier predicts the same label for the testing input if the $\ell_2$-norm  of the adversarial perturbation added to the testing input is less than $R$.}

The key challenge of implementing our F2IPerturb-API is how to find the largest input-space certified radius $R$ for a given testing input and its feature-space certified radius $R_F$. To address the challenge,  we formulate finding the largest $R$ as an optimization problem, where the objective function is to find the maximum $R$ and the constraint is that the feature-space perturbation is less than $R_F$. However, the optimization problem is challenging to solve due to the highly non-linear constraint. To address the challenge, we propose a binary search based solution. The key component of our solution is to check whether the constraint is satisfied for a specific $R$ in each iteration of binary search. Towards this goal, we derive an upper bound of the feature-space perturbation for a given $R$ and we treat the constraint satisfied if the upper bound is less than $R_F$. Our upper bound can be computed efficiently.

 F2IPerturb-API addresses the first two challenges of SEaaS. Specifically, 
  BC based certification is applicable in REaaS. Moreover,  SC based certification requires much less queries to the APIs in REaaS. Specifically, for any certification method, a client only requires one query to Feature-API per training input and two queries (one to Feature-API and one to F2IPerturb-API) per testing input in our REaaS.

To address the third challenge of SEaaS, we propose a new method to pre-train a robust encoder, so a client can derive larger certified radii even though it cannot train/fine-tune the encoder. Our method can be combined with standard supervised learning or self-supervised learning to enhance the robustness of a pre-trained encoder. 
An encoder is more robust if it produces more similar feature vectors for an input and its adversarially perturbed version. Our key idea is to derive an upper bound for the Euclidean distance between the feature vectors of an input and its adversarial version, where our upper bound is a product of a \emph{spectral-norm term} and the perturbation size. The spectral-norm term depends on the parameters of the encoder, but it does not depend on the input nor the adversarial perturbation. An encoder with a smaller spectral-norm term may produce more similar feature vectors for an input and its adversarial version. 
Thus, we use the spectral-norm term as a regularization term to regularize the pre-training of an encoder.

We perform a systematic evaluation on multiple datasets including CIFAR10, SVHN, STL10, and Tiny-ImageNet. 
Our evaluation results show that REaaS addresses the three challenges of SEaaS. First, REaaS makes BC based certification applicable. Second, REaaS incurs orders of magnitude less queries to the cloud service than SEaaS for SC based certification. 
For instance, REaaS  reduces the number of queries to the cloud service APIs respectively by $25\times$ and $5,000\times$ per training and testing input when a client trains its downstream classifier for $e=25$ epochs and uses $N=10,000$ for certification. Third, in the framework of  REaaS, our robust pre-training method achieves larger \emph{average certified radius (ACR)} for the testing inputs than existing  methods to pre-train encoders for both BC and SC based certification.  For instance, when the encoder is pre-trained on Tiny-ImageNet and the downstream classifier is trained on SVHN, the ACRs for MoCo (a standard non-robust self-supervised learning method)~\cite{he2019moco}, RoCL (an adversarial training based state-of-the-art robust self-supervised learning method)~\cite{kim2020adversarial}, and our method are respectively 0.011, 0.014, and 0.275 when a client uses SC based certification.

In summary, we make the following contributions:
\begin{itemize}
    \item \CR{We propose REaaS, which enables a client to build a certifiably robust downstream classifier and derive its certified radii using any certification method with a minimal number of queries to the cloud service. }
    \item \CR{We propose a method to implement F2IPerturb-API.}
    \item \CR{We propose a spectral-norm term to regularize the pre-training of a robust encoder.}
    \item We extensively evaluate REaaS and compare it with SEaaS on multiple datasets. 
\end{itemize}

%% file: RelatedWork.tex
\section{Related Work}

\subsection{Adversarial Examples}
We discuss adversarial examples~\cite{Szegedy14,goodfellow2014explaining} in the context of encoder as a service. We denote by $f$ a pre-trained encoder and $g$ a downstream classifier. Given a testing input $\mathbf{x}$, the encoder outputs a feature vector for it, while the downstream classifier takes the feature vector as input and outputs a label. For simplicity, we denote by $f(\mathbf{x})$ the feature vector and  $g\circ f(\mathbf{x})$ the predicted label for $\mathbf{x}$, where $\circ$ represents the composition of the encoder and downstream classifier. In an adversarial example, an attacker adds a carefully crafted small perturbation $\delta$ to $\mathbf{x}$ such that its predicted label changes,  i.e., $g\circ f(\mathbf{x}+\delta)\neq g\circ f(\mathbf{x})$. The carefully perturbed input $\mathbf{x}+\delta$ is called an adversarial example. Many methods (e.g.,~\cite{Szegedy14,carlini2017towards,madry2017towards}) have been proposed to find an adversarial perturbation $\delta$ for a given input $\mathbf{x}$. In our work, we focus on certified defenses, which aim to defend against any bounded adversarial perturbations no matter how they are found. Therefore, we omit the details on how an attacker can find an adversarial perturbation. 

\subsection{Certifying Robustness of a Classifier} 
\label{certificationmethods}

\myparatight{Definition of certified radius} A classifier is certifiably robust against adversarial examples if its predicted label for an input is unaffected by any perturbation once its size is bounded~\cite{wong2017provable,lecuyer2019certified,cohen2019certified}. Formally, a classifier $h$ is certifiably robust if we have the following guarantee for an input $\mathbf{x}$:
\begin{align}
     h(\mathbf{x}+\delta)=h(\mathbf{x}), \forall  \lnorm{\delta}_2 < R,
\end{align}
where $R$ is known as \emph{certified radius}. Note that certified radius $R$ may be different for different inputs $\mathbf{x}$, but we omit the explicit dependency on $\mathbf{x}$ in the notation for simplicity.

A certification method against adversarial examples aims to build a certifiably robust classifier and derive its certified radius $R$ for any input $\mathbf{x}$. There are two general categories of certification methods, i.e.,  \emph{base classifier (BC) based certification}~\cite{wong2017provable,raghunathan2018certified,zhang2018crown,wang2018formal} and \emph{smoothed classifier (SC) based certification}~\cite{cao2017mitigating,lecuyer2019certified,cohen2019certified}. Both categories of methods may be adopted in different scenarios depending on certification needs. On one hand,   
BC based certification often produces \emph{deterministic} guarantees (i.e., the derived certified radius is absolutely correct), while SC based certification often provides \emph{probabilistic} guarantees (i.e., the derived certified radius may be incorrect with a small \emph{error probability}). On the other hand, SC based certification often derives a larger certified radius than BC based certification due to its probabilistic guarantees.

\myparatight{Base classifier (BC) based certification} BC based certification aims to directly derive the certified radius $R$ of a given classifier (called \emph{base classifier}) for an input $\mathbf{x}$. These methods often propagate perturbation from the input $\mathbf{x}$ to the output of the base classifier layer by layer in order to derive the certified radius. Therefore, they require white-box access to the base classifier. 
\CR{Suppose $F$ is a base classifier that maps an input $\mathbf{x}$ to one of $c$ classes $\{1,2,\cdots, c\}$. We use $H(\mathbf{x})$ to denote the base classifier's last-layer output vector for $\mathbf{x}$, where $H_l(\mathbf{x})$ represents the $l$th entry of $H(\mathbf{x})$ and $l=1,2,\cdots,c$. $F(\mathbf{x})$ denotes the predicted label for $\mathbf{x}$, i.e.,  $F(\mathbf{x})=\argmax_{l=1,2,\cdots,c}H_{l}(\mathbf{x})$. Next, we overview how to derive the certified radius $R$ using CROWN~\cite{zhang2018crown}, a state-of-the-art BC based certification method.  CROWN shows that each entry $H_{l}(\mathbf{x})$ can be bounded by two  linear functions $H^{L}_{l}(\mathbf{x})$ and $H^{U}_{l}(\mathbf{x})$. Suppose the base classifier predicts label $y$ for $\mathbf{x}$ when there is no adversarial perturbation, i.e., $F(\mathbf{x})=y$. CROWN finds the largest $r$ such that the lower bound of the $y$th entry (i.e., $\min_{\lnorm{\delta}_2 < r} H^{L}_y(\mathbf{x} + \delta)$) is larger than the upper bounds of all other entries (i.e., $\max_{l\neq y} \max_{\lnorm{\delta}_2 < r} H^{U}_{l}(\mathbf{x}+\delta)$) and views it as the certified radius $R$ for the input $\mathbf{x}$. The complete details of CROWN can be found in Appendix~\ref{crownexample}.  } 
In the context of encoder as a service, the composition of the encoder and downstream classifier (i.e., $g\circ f$) is a base classifier \CR{F}, whose certified radius a client aims to derive. However, in SEaaS, a client does not have white-box access to the encoder $g$ since it is deployed on the cloud server. As a result,  a client cannot use BC based certification to derive the certified radius of  $g\circ f$.

\myparatight{Smoothed classifier (SC) based certification} SC based certification first builds a \emph{smoothed classifier} based on the base classifier and then derives the certified radius $R$ of the smoothed classifier.   In SEaaS,  a client builds a smoothed classifier $h$ based on the base classifier $g\circ f$ via adding random Gaussian noise $\mathcal{N}(0, \sigma^2 \mathbf{I})$ to an input $\mathbf{x}$, where $\sigma$ is the standard deviation of the Gaussian noise. Specifically, given a testing input $\mathbf{x}$, the client constructs $N$ noisy inputs  $\mathbf{x}+\mathbf{n}_1, \mathbf{x}+\mathbf{n}_2,\cdots, \mathbf{x}+\mathbf{n}_N$, where $\mathbf{n}_i$ ($i=1,2,\cdots,N$) is sampled from $\mathcal{N}(0, \sigma^2 \mathbf{I})$. The client uses the base classifier $g\circ f$ to predict the label of each noisy input. Moreover, the client computes the \emph{label frequency} $N_l$  of each label $l$ among the noisy inputs, i.e.,  $N_l = \sum_{j=1}^{N}\mathbb{I}(g\circ f(\mathbf{x}+\mathbf{n}_j)=l)$, where $\mathbb{I}$ is an indicator function. The smoothed classifier predicts the label with the largest label frequency for the original testing input $\mathbf{x}$. Moreover, the client can derive the certified radius $R$ of the smoothed classifier for $\mathbf{x}$ based on the label frequencies. Due to the random sampling, the derived certified radius may be incorrect with an error probability $\alpha$, which can be set by the client.  In Appendix~\ref{randomizedsmoothingexample}, we take Cohen et al.~\cite{cohen2019certified} as an example to discuss more technical details on SC based certification.

To improve certified radius, the base classifier $g\circ f$ is often trained using noisy training inputs~\cite{cohen2019certified}. In encoder as a service (both SEaaS and REaaS), a client does not have white-box access to the encoder $g$ and thus can only train its downstream classifier $f$ using noisy training inputs. Specifically, for SEaaS, in each epoch of training the downstream classifier, a client adds random Gaussian noise from $\mathcal{N}(0, \sigma^2 \mathbf{I})$  to each training input, queries the Feature-API to obtain the feature vector of each noisy training input, and uses the feature vectors to update the downstream classifier via stochastic gradient descent.

SC based certification faces two challenges in SEaaS. First, a client needs to query the cloud service many times, leading to a large communication cost for the client and a large computation cost for the cloud server. Specifically, for each testing input $\mathbf{x}$, a client needs to query the Feature-API $N$ times to obtain the feature vectors of the $N$ noisy inputs in order to compute the label frequencies. Moreover, the client queries the Feature-API $e$ times \emph{per} training input, where $e$ is the number of epochs used to train the downstream classifier. We note that \cite{salman2020denoised} proposed to prepend a denoiser to a base classifier instead of training it with  noisy training inputs, which can reduce the number of queries from $e$ to 1  per training input  when applied to SEaaS. 
However, it is hard for a client with a small amount of data to train such a denoiser.  
Second, the derived certified radius is suboptimal because a client cannot fine-tune the encoder,  which is not pre-trained to support certified robustness.

\subsection{Pre-training an Encoder} 

\subsubsection{Pre-training Non-robust Encoders} We discuss both standard supervised learning and self-supervised learning methods to pre-train encoders, which do not take robustness against adversarial examples into consideration. 

\myparatight{Supervised learning} The idea of using supervised learning to pre-train an encoder is to first train a deep neural network classifier using labeled training data and then use the layers excluding the output layer as an encoder. Specifically, supervised learning defines a loss function $l(i)$ (e.g., cross-entropy loss) for each labeled training example $(\mathbf{x}_i, y_i)$, where $y_i$ is the ground truth label of $\mathbf{x}_i$. Then, supervised learning iteratively trains a deep neural network classifier by minimizing the sum of the losses over the labeled training examples. Such paradigm of training a deep neural network classifier via supervised learning and using the layers excluding the output layer as an encoder is also known as \emph{transfer learning}~\cite{pan2009survey}.

\myparatight{Self-supervised learning} Unlike supervised learning, self-supervised learning~\cite{he2019moco,chen2020simple} aims to pre-train an encoder using unlabeled data, which has attracted growing attention in the past several years in the AI community.   A basic component of self-supervised learning is \emph{data augmentation}. Specifically, given an image, the data augmentation component applies a series of (random) data augmentation operations (e.g., random cropping, color jitter, and flipping) sequentially to produce an \emph{augmented image}. 
Roughly speaking, the main idea of self-supervised learning is to pre-train an encoder such that it produces similar feature vectors for two augmented images produced from the same image, but dissimilar  feature vectors  for two augmented images produced from different images. Next, we take MoCo~\cite{he2019moco}, a state-of-the-art self-supervised learning algorithm, as an example to elaborate more details. 

MoCo uses an auxiliary encoder (called \emph{momentum encoder}) that has the same architecture as the  encoder and a queue (denoted as $\Gamma$). In particular, the queue is used to cache the output of the momentum encoder for the augmented images and is dynamically updated. For simplicity, we respectively use $f$ and $f_e$ to denote the encoder and the momentum encoder, and use $\theta$ and $\theta_e$ to denote their encoder parameters. Suppose we have a mini-batch of unlabeled images which are denoted as $\mathbf{x}_i, i=1,2,\cdots, m$. We apply the data augmentation component to each image in the mini-batch twice. We use $\mathbf{x}_i^1 $ and $ \mathbf{x}_i^2$ to denote the two augmented images produced for the image $\mathbf{x}_i$, respectively. Given $\mathbf{x}_i^1$, $\mathbf{x}_i^2$, and the queue $\Gamma$, MoCo defines a loss function for $\mathbf{x}_i$ as follows:
\begin{align}
\label{moco_contrastive_loss_def}
    \ell(i) = - \log (\frac{\exp(Sim(f(\mathbf{x}_i^1),f_e(\mathbf{x}_i^2))/\tau)}{\sum_{\mathbf{z}\in \Gamma \cup \{f_e(\mathbf{x}_i^2) \}} \exp(Sim(f(\mathbf{x}_i^1),\mathbf{z})/\tau)}),
\end{align}
where $\tau$ is a temperature parameter and $Sim(\cdot, \cdot)$ measures the similarity of two feature vectors (e.g., cosine similarity). MoCo uses gradient descent to minimize $\frac{1}{m}\cdot \sum_{i=1}^{m}\ell(i)$ to update the parameters $\theta$ in the  encoder $f$. The queue $\Gamma$ is dynamically updated in each step, where $\{f_e(\mathbf{x}_1^2),f_e(\mathbf{x}_2^2),\cdots, f_e(\mathbf{x}_m^2)\}$ are enqueued and the $m$ ``oldest'' vectors are dequeued.

\subsubsection{Pre-training Empirically Robust Encoders} 
Adversarial training~\cite{goodfellow2014explaining,madry2017towards} is a standard method to train empirically robust classifiers in supervised learning. The key idea is to generate adversarial examples based on training examples during training, and use the adversarial examples to augment the training data.  The layers excluding the output layer of the classifier are then used as an encoder.  
Several studies~\cite{chen2020adversarial,kim2020adversarial,jiang2020robust} generalized adversarial training to pre-train a robust  encoder in self-supervised learning. Roughly speaking, the idea is to first generate adversarial examples that incur large loss and then use them to pre-train an encoder.  For instance, Kim et al.~\cite{kim2020adversarial} proposed \emph{Robust Contrastive Learning (RoCL)} to pre-train robust encoders. 
Specifically, given a training image, RoCL uses  Projected Gradient Descent (PGD)~\cite{madry2017towards} to generate an adversarial perturbation for an augmented version of the training image such that the adversarially perturbed augmented version incurs a large loss. Then, RoCL pre-trains an encoder such that it outputs similar feature vectors for the adversarially perturbed augmented version and other augmented versions of the training image.

The goal of these studies is to pre-train \emph{empirically} rather than \emph{certifiably} robust encoders. As a result, the pre-trained encoders achieve suboptimal certified radii for a client as shown by our experimental results. In this work, we propose a new  method, which can be combined with  either supervised learning or self-supervised learning to pre-train a robust encoder that can achieve larger certified radii against adversarial examples for a client.  

%% file: Problem.tex
\section{Problem Formulation}

\myparatight{Threat model} We consider an attacker can use adversarial examples to induce misclassification for a client. Specifically, given a testing input, an attacker can add a carefully crafted perturbation to it such that the client's downstream classifier predicts a different label. To defend against adversarial examples, the client aims to build a certifiably robust classifier, which provably predicts the same label for a testing input no matter what perturbation is added to it once the $\ell_2$-norm of the perturbation is less than a threshold (called \emph{certified radius}). Note that we do not constrain on what method an attacker can use to find the perturbation since we aim to defend against all bounded perturbations.

\myparatight{Problem definition} We aim to design an encoder as a service. In particular, when designing an encoder as a service, we essentially aim to answer two key questions: 1) what APIs should the cloud service provide for a client? and 2) how to pre-train the encoder? 

\myparatight{Design goals} We aim to design an encoder as a service to achieve three goals,  \emph{generality}, \emph{efficiency}, and \emph{robustness}, which we elaborate in the following: 

\begin{itemize}
    \item {\bf Generality.} As we discussed in Section~\ref{certificationmethods},  BC and SC based certification methods are complementary and may be adopted by different clients due to their different needs.  
      We say an encoder as a service achieves the generality goal if a client can use any certification method   to build a certifiably robust  classifier and derive its certified radius for any given testing input. We note that  SEaaS can only support SC based certification.

    \item {\bf Efficiency.} We use the number of queries sent to the cloud service to measure the communication cost between a client and the cloud server. Moreover, we use computation time  to measure the computation cost for a client and the cloud server. We aim to design an encoder as a service to achieve a small communication cost and computation cost.

    \item {\bf Robustness.} A certified radius measures the certified robustness of a classifier for a testing input. We use the average certified radius of testing inputs to measure the certified robustness of a classifier. We aim to design an encoder as a service that enables  a client to build a downstream classifier with a large average certified radius in both BC and SC based certification. 
    
\end{itemize}

We note that SEaaS does not achieve any of the three goals. Specifically, SEaaS cannot support BC based certification; SEaaS incurs a large communication cost between a client and the cloud server as well as a large computation cost for the cloud server in SC based certification; and SEaaS achieves suboptimal average certified radius for SC based certification because certified robustness is not taken into consideration when pre-training the encoder.

%% file: method.tex
\section{Our REaaS}

\subsection{Overview}

To achieve the generality and efficiency goals,  our key idea is to enable a client to treat its own downstream classifier as a base classifier and certify the robustness of its base downstream classifier or smoothed downstream classifier in the \emph{feature space}. Towards this goal, other than the \emph{Feature-API} provided in SEaaS, our REaaS provides another API (called \emph{F2IPerturb-API}).  In particular,  
 since a downstream classifier takes a feature vector as input, we propose a client first derives a feature-space certified radius $R_F$ of its base downstream classifier or smoothed downstream classifier for a testing input. Then, the client  transforms the feature-space certified radius $R_F$ to the input-space certified radius $R$ by querying the F2IPerturb-API. To achieve the robustness goal, we further propose a new method to pre-train a robust encoder, which uses a spectral-norm term to regularize the pre-training of an encoder. Our pre-trained encoder aims to produce similar feature vectors for an input and its adversarially perturbed version.

\subsection{Feature-API and F2IPerturb-API}

\subsubsection{Feature-API}
\label{sec:feature-api}
We first introduce the input and output of Feature-API, and then its implementation. 

\myparatight{Input and output for a client} In {Feature-API}, the input from a client is an image $\mathbf{x}$ and the output returned to the client is the input's feature vector $\mathbf{v}$. Formally, Feature-API is represented as $\mathbf{v}=\text{\emph{Feature-API}}(\mathbf{x})$.

\myparatight{Implementation on the server} Given an input $\mathbf{x}$, the cloud server uses a pre-trained encoder $f$ to compute its feature vector $\mathbf{v}$. In particular, we have $\mathbf{v} = f(\mathbf{x})$. 

We note that SEaaS only has this Feature-API.

\subsubsection{F2IPerturb-API} Like Feature-API, we first introduce the input and output of F2IPerturb-API, and then its implementation on the cloud server.

\myparatight{Input and output for a client} F2IPerturb-API transforms a feature-space certified radius  to an input-space certified radius.   The input from a client contains an input image $\mathbf{x}$ and a feature-space certified radius $R_F$ for $\mathbf{x}$. In particular, the client's downstream classifier predicts the same label for $\mathbf{x}$ once the $\ell_2$-norm of the perturbation to $\mathbf{x}$'s feature vector  $\mathbf{v}$ is bounded by $R_F$. The client can use any BC or SC based method to derive $R_F$ for $\mathbf{x}$ by treating its downstream classifier alone as a base classifier and $\mathbf{x}$'s feature vector $\mathbf{v}$ as an ``input'' to the base classifier. 

The output of F2IPerturb-API is an input-space certified radius $R$ such that when the $\ell_2$-norm of the adversarial perturbation added to the input image  $\mathbf{x}$ is smaller than $R$, the $\ell_2$-norm of the perturbation introduced to the feature vector $\mathbf{v}$ is smaller than $R_F$. Formally, given an input image $\mathbf{x}$ and a feature-space certified radius $R_F$,  F2IPerturb-API is represented as follows: $R = \text{\emph{F2IPerturb-API}}(\mathbf{x},R_F)$.

\myparatight{Implementation on the server} A larger $R$ enables the client to derive a larger certified radius. The key challenge of implementing the F2IPerturb-API is how to derive the largest $R$ for a given input $\mathbf{x}$ and $R_F$.   To address the challenge, we formulate the input-space certified radius $R$ as the solution to the following optimization problem:
\begin{align}
\label{api_3_opt_1}
   & R = \max_{r} r \\
    \textit{ s.t. }& \max_{\lnorm{\delta}_2 < r}
    \label{api_3_opt_2} \lnorm{f(\mathbf{x}+\delta)-f(\mathbf{x})}_2 < R_F,
\end{align}
where $f$ is an encoder and $\delta$ is an adversarial perturbation. However, the optimization problem is challenging to solve because the constraint is highly non-linear when the encoder is a complex neural network. To address the challenge, we propose a binary search based method to solve $R$ in the optimization problem. In particular, we search in the range $[\rho_{k}^L,\rho_{k}^U]$ in the $k$th round of binary search, where we set $\rho_{1}^L$ to be $0$ and $\rho_{1}^U$ to be a large value (e.g., $10$ in our experiments) in the first round.  Moreover, we denote $\rho_k= \frac{\rho_{k}^L+\rho_{k}^U}{2}$ for simplicity. In the $k$th round, we check whether $r = \rho_k$ satisfies the constraint in Equation~(\ref{api_3_opt_2}).  
If the constraint is satisfied, then we can search the range $[\rho_k,\rho_{k}^U]$ in the $(k+1)$th round, i.e., $\rho_{k+1}^L = \rho_k$ and $\rho_{k+1}^U=\rho_{k}^U$. Otherwise, we search the range $[\rho_{k}^L,\rho_k]$ in the $(k+1)$th round, i.e., $\rho_{k+1}^L=\rho_{k}^L$ and $\rho_{k+1}^U = \rho_k$. We stop the binary search when $\rho_{k}^U-\rho_{k}^L \leq \beta$ and treat $\rho_{k}^L$ as $R$, where $\beta$ is a parameter characterizing the binary-search precision.

\begin{algorithm}[tb]
   \caption{\emph{F2IPerturb-API}}
   \label{alg:api_2}
\begin{algorithmic}
   \STATE {\bfseries Input from client:} image $\mathbf{x}$ and feature-space certified radius $R_F$
   \STATE {\bfseries Output for client:} image-space certified radius $R$ \\
   \STATE $\rho^L, \rho^U \gets 0, \text{large value (e.g., } 10)$ \\
    \WHILE{$\rho^U - \rho^L > \beta$} 
    \STATE $\rho_k = \frac{\rho^L+\rho^U}{2}$ \\
    \FOR{$i=1,2,\cdots, d$} 
    \STATE $f_i^L, f_i^U \gets \textsc{Crown}(i,\mathbf{x},f)$ \\
   \STATE \CRR{$L_i= \min_{\lnorm{\delta}_2 \leq \rho_k}f_i^L(\mathbf{x}+\delta)-f_i(\mathbf{x})$} \\
   \STATE \CRR{$U_i= \max_{\lnorm{\delta}_2 \leq \rho_k}f_i^U(\mathbf{x}+\delta)-f_i(\mathbf{x})$} \\
    \ENDFOR
    \STATE $R_F^\prime = \sqrt{\sum_{i=1}^d\max(L_i^2, U_i^2)}$ \\
    \IF{$R_F^\prime < R_F$}
    \STATE $\rho^L = \rho_k$ \\
    \ELSE
    \STATE $\rho^U = \rho_k$ \\
    \ENDIF
    \ENDWHILE
   \STATE \textbf{return} $\rho^L$
\end{algorithmic}
\end{algorithm}

Our binary search based solution faces a key challenge, i.e.,  how to check whether $r = \rho_k$ satisfies the constraint in Equation~(\ref{api_3_opt_2}). Our key idea to address the challenge is to derive an upper bound for the left hand side of the constraint (i.e., $\max_{\lnorm{\delta}_2 < \rho_k} \lnorm{f(\mathbf{x}+\delta)-f(\mathbf{x})}_2$) and decide that the constraint is satisfied if the upper bound is smaller than $R_F$, where the upper bound can be efficiently computed for any $\rho_k$.  Suppose the  encoder $f$ maps an input $\mathbf{x}$ to a $d$-dimensional feature vector $f(\mathbf{x})$, where $f_i(\mathbf{x})$ represents the $i$th entry of $f(\mathbf{x})$. 
An  encoder $f$ is essentially a deep neural network. Therefore, according to CROWN~\cite{zhang2018crown}, we have the following lower bound and upper bound for $f_i(\mathbf{x} + \delta)$ when $\lnorm{\delta}_2 < \rho_k$:
\begin{align}
\label{inequality_crown_method}
  \min_{\lnorm{\delta}_2 < \rho_k} f^{L}_i(\mathbf{x} + \delta) \leq f_i(\mathbf{x}+\delta) \leq \max_{\lnorm{\delta}_2 < \rho_k} f^{U}_{i}(\mathbf{x}+\delta),
\end{align} 
where  $f^{L}_i$ and $f^{U}_i$ are two linear functions and $i=1,2,\cdots, d$. \CR{In Appendix~\ref{tightness_of_crown_equation}, we show that Equation~\ref{inequality_crown_method} is tight when $f$ consists of one linear layer.}  \CRR{As $\min_{\lnorm{\delta}_2 \leq \rho_k} f^{L}_i(\mathbf{x} + \delta) \leq \min_{\lnorm{\delta}_2 < \rho_k} f^{L}_i(\mathbf{x} + \delta)$ and $\max_{\lnorm{\delta}_2 < \rho_k} f^{U}_{i}(\mathbf{x}+\delta) \leq \max_{\lnorm{\delta}_2 \leq \rho_k} f^{U}_{i}(\mathbf{x}+\delta)$, we have the following when $\lnorm{\delta}_2 < \rho_k$:}
\begin{align}
   \CRR{ \min_{\lnorm{\delta}_2 \leq \rho_k} f^{L}_i(\mathbf{x} + \delta) \leq f_i(\mathbf{x}+\delta) \leq \max_{\lnorm{\delta}_2 \leq \rho_k} f^{U}_{i}(\mathbf{x}+\delta),} 
\end{align}
\CRR{Therefore}, we have the following inequalities for $\forall \lnorm{\delta}_2 < \rho_k$:
\begin{align}
&f_i(\mathbf{x}+\delta) - f_i(\mathbf{x}) \geq L_i, \\
& f_i(\mathbf{x}+\delta) - f_i(\mathbf{x}) \leq U_i, 
\end{align}
where \CRR{$L_i=\min_{\lnorm{\delta}_2 \leq \rho_k} f^{L}_i(\mathbf{x} + \delta) - f_i(\mathbf{x})$ and $U_i=\max_{\lnorm{\delta}_2 \leq \rho_k} f^{U}_{i}(\mathbf{x}+\delta) - f_i(\mathbf{x})$. }
Based on the above two inequalities, we have the following:
\begin{align}
    \max_{\lnorm{\delta}_2 < \rho_k} (f_i(\mathbf{x}+\delta) - f_i(\mathbf{x}))^2 \leq  \max(L_i^2,  U_i^2).
\end{align}
Therefore, we can derive an upper bound for $\max_{\lnorm{\delta}_2 <\rho_k} \lnorm{f(\mathbf{x}+\delta)-f(\mathbf{x})}_2$ as follows:
\begin{align}
   & \max_{\lnorm{\delta}_2 < \rho_k} \lnorm{f(\mathbf{x}+\delta)-f(\mathbf{x})}_2 \\
   \leq & \sqrt{ \sum_{i=1}^d \max_{\lnorm{\delta}_2< \rho_k} (f_i(\mathbf{x}+\delta) - f_i(\mathbf{x}))^2}    \\
   \label{upperboundfs}
    \leq & \sqrt{\sum_{i=1}^d \max( L_i^2,  U_i^2)} \\
    \triangleq & R_F'.
\end{align}
If the upper bound $R_F'$ is smaller than $R_F$, then we have  $r = \rho_k$ satisfies the constraint in Equation~(\ref{api_3_opt_2}). We note that $r = \rho_k$ may also satisfy the constraint even if the upper bound $R_F'$  is no smaller than $R_F$. However, such cases do not influence the correctness of our binary search. 
Note that \CRR{$\min_{\lnorm{\delta}_2 \leq \rho_k} f^{L}_i(\mathbf{x} + \delta)$ and $\max_{\lnorm{\delta}_2 \leq \rho_k} f^{U}_{i}(\mathbf{x}+\delta)$ have closed-form solutions for $i=1,2,\cdots,d$~\cite{zhang2018crown}.} Therefore, $L_i$, $U_i$, and the upper bound $R_F'$ can be computed efficiently. In other words, we can efficiently check whether $r = \rho_k$ satisfies the constraint in Equation~(\ref{api_3_opt_2}) for any $\rho_k$.  Algorithm~\ref{alg:api_2} shows our {F2IPerturb-API}, where the function \textsc{Crown} obtains the lower bound and upper bound linear functions for each $f_i(\mathbf{x})$. 

Our binary search based solution correctly finds a lower bound of the optimal $R$ of the optimization  problem in Equation~(\ref{api_3_opt_1}) because the constraint in Equation~(\ref{api_3_opt_2})  is guaranteed to be satisfied in each round of our binary search. 

\myparatight{Image rescaling} We note that, in our above discussion on the two APIs, a client's input image size  is the same as the input size of the cloud server's  encoder. 
When the size of a client's input image is different, the cloud server can rescale it to be the input size of its  encoder using the standard bilinear interpolation. The bilinear interpolation can be viewed as a linear transformation. In particular, suppose $\mathbf{x}_b$ and $\mathbf{x}_a$ respectively represent the image before and after rescaling. Then, we have $\mathbf{x}_a = \mathbf{W} \cdot \mathbf{x}_b$, where $\mathbf{W}$ is the matrix used to represent the linear transformation. The cloud server can implement this linear transformation (i.e., rescaling)  by adding a linear layer whose weight matrix is $\mathbf{W}$ before the encoder. Moreover, the cloud server can view the linear layer + the encoder as a ``new''  encoder to implement the two APIs.

\subsection{Pre-training Robust Encoder}
Our REaaS is applicable to any encoder. However, a more robust encoder enables a client to derive a larger certified radius for its testing input. Therefore, we further propose a method to pre-train robust encoders. An encoder $f$ is more robust if it produces more similar feature vectors for an input and its adversarially perturbed version, i.e., if $f(\mathbf{x}+\delta)$ and $f(\mathbf{x})$ are more similar. In particular, based on our implementation of the F2IPerturb-API, if $\lnorm{f(\mathbf{x}+\delta)-f(\mathbf{x})}_2$ is smaller for any adversarial perturbation $\delta$, then F2IPerturb-API would return a larger input-space certified radius to a client for a given feature-space certified radius. 
Therefore, our key idea is to reduce $\lnorm{f(\mathbf{x}+\delta)-f(\mathbf{x})}_2$ when pre-training an encoder $f$. Next, we derive an upper bound of  $\lnorm{f(\mathbf{x}+\delta)-f(\mathbf{x})}_2$, based on which we design a regularization term to regularize the pre-training of an encoder. 

A neural network (e.g., an encoder) can often be decomposed  into the composition of a series of linear transformations~\cite{Szegedy14}. 
 In particular, we can do so if each layer of the neural network (e.g., linear layer, convolutional layer, and batch normalization layer) can be expressed as a linear transformation. 
\CRR{ We denote an encoder as the composition of $n$ linear transformations, i.e., $f(\cdot) = T^{n}\circ T^{n-1}\circ \cdots \circ T^1 (\cdot)$}. 
\cite{Szegedy14} showed that the difference between the outputs of any neural network $f$ ($f$ is an encoder in our case) for an input and its adversarially perturbed version can be bounded as follows:
\CRR{\begin{align}
    \lnorm{f(\mathbf{x}+\delta)-f(\mathbf{x})}_2 \leq \prod_{j=1}^{n}\lnorm{T^j}_s \cdot \lnorm{\delta}_2,
\end{align}}
where $\mathbf{x}$ is an input, $\delta$ is an adversarial perturbation, and $\lnorm{\cdot}_s$ represents spectral norm. \CRR{The product of the spectral norms of the $n$ linear transformations (i.e., $\prod_{j=1}^{n}\lnorm{T^j}_s$) }is independent with input $\mathbf{x}$ and adversarial perturbation $\delta$. \CRR{Therefore, our idea is 
to add $\prod_{j=1}^{n}\lnorm{T^j}_s$ as a regularization term (called \emph{spectral-norm regularization}) when pre-training an encoder.} Minimizing such regularization term may enforce the encoder to produce more similar feature vectors for an input and its adversarially perturbed version, i.e., $\lnorm{f(\mathbf{x}+\delta)-f(\mathbf{x})}_2$ may be smaller. In particular, we minimize the following loss function for each mini-batch of inputs when pre-training an encoder:   
\CRR{\begin{align}
    \frac{1}{m}\cdot \sum_{i=1}^{m}\ell(i) + \lambda \cdot \prod_{j=1}^{n}\lnorm{T^j}_s,
\end{align}}
where $\ell(i)$ is a loss for a training input in pre-training, $m$ is batch size, and $\lambda$ is a hyper-parameter used to balance the two terms. For instance, when using supervised learning to train a classifier, whose layers excluding the output layer are used as an encoder, the loss  $\ell(i)$ is often the cross-entropy loss; when using self-supervised learning algorithm MoCo~\cite{he2019moco}  to pre-train an encoder, $\ell(i)$ is defined in Equation~(\ref{moco_contrastive_loss_def}). We adopt the \emph{power method}~\cite{mises1929praktische} to estimate the spectral norms of the linear transformations when pre-training an encoder.

\subsection{Certifying Robustness for a Client}
\label{3_api_bcbc}
In REaaS, a client can treat its own downstream classifier as a base classifier. 
We discuss how a client can use our two APIs to train a base downstream classifier and derive the certified radius of the base downstream classifier in BC based certification or the smoothed downstream classifier in SC based certification for a testing input.

\myparatight{BC based certification} When training a base downstream classifier, a client queries the Feature-API to obtain a feature vector for each training input. Then, given the feature vectors and the corresponding training labels, the client  can use any training method (e.g., standard supervised learning) to train a base downstream classifier. Given a testing input, the client queries the Feature-API to obtain its feature vector and uses the base downstream classifier to predict its label.  
Moreover, the client can use any BC based certification method to derive a feature-space certified radius for the testing input by treating its feature vector as an ``input'' to the base downstream classifier. 
Then, the client queries the F2IPerturb-API to transform the feature-space certified radius to an input-space certified radius. 

\myparatight{SC based certification} Similar to BC based certification, a client queries the Feature-API to obtain a feature vector for each training input when training a base downstream classifier. However, unlike BC based certification, the client adds noise to the training feature vectors in SC based certification. 
In particular, the client adds random noise  (e.g., Gaussian noise) to each feature vector in each mini-batch of training feature vectors in each training epoch. Note that the client does not need to query the Feature-API again for the noisy feature vector. Given a testing input, the client queries the Feature-API to obtain its feature vector and uses the smoothed downstream classifier to predict its label and derive its feature-space certified radius. In particular, the client constructs $N$ noisy feature vectors by adding random noise to the feature vector and uses it's base downstream classifier to predict their labels. Based on the predicted labels, the client can derive the predicted label and feature-space certified radius for the original feature vector. Then, the client queries the F2IPerturb-API to transform the feature-space certified radius to an input-space certified radius.

\begin{table}[!t]\renewcommand{\arraystretch}{1.5} 
\fontsize{7}{6}\selectfont
\setlength{\tabcolsep}{0.5mm}
\centering
\caption{Comparing the communication and computation cost per training/testing input in SEaaS and REaaS. $e$ is the number of epochs used to train a base downstream classifier. $N$ is the number of noisy inputs per testing input in SC. $T_f$ (or $T_g$) and $M_f$ (or $M_g$) are respectively the number of layers and the maximum number of neurons in a layer in an encoder (or a downstream classifier). $K_f$ (or $K_g$) is the number of parameters in an encoder (or a downstream classifier).}
\subfloat[\fontsize{9}{8}\selectfont Communication cost]
{
\begin{tabular}{|c|c|c|c|c|}
\hline
\multirow{3}{*}{Service} & \multicolumn{4}{c|}{\#Queries} \\\cline{2-5}
& \multicolumn{2}{c|}{Per training input} & \multicolumn{2}{c|}{Per testing input}  \\ \cline{2-5}
                   & BC & SC  & BC & SC   \\ \hline
SEaaS                 & N/A   & $e$ &  N/A  & $N$ \\ \hline
REaaS                 & \multicolumn{2}{c|}{1}      & \multicolumn{2}{c|}{2} \\ \hline
\end{tabular}
\label{table:theoreical_communication}
}

\subfloat[\fontsize{9}{8}\selectfont Computation cost]
{
\begin{tabular}{|c|c|c|c|c|c|}
\hline
\multirow{3}{*}{Service}& \multirow{3}{*}{Entity}& \multicolumn{4}{c|}{Computational complexity} \\\cline{3-6}
& & \multicolumn{2}{c|}{Per training input}& \multicolumn{2}{c|}{Per testing input}  \\ \cline{3-6}
             &      & BC & SC  & BC & SC   \\ \hline
\multirow{2}{*}{SEaaS}   &  \makecell{Cloud \\server }           &  \multirow{2}{*}{N/A}  & $O(e\cdot K_f)$ &  \multirow{2}{*}{N/A}  & $O(N\cdot K_f)$ \\ \cline{2-2}\cline{4-4}\cline{6-6}
&Client& &$O(e\cdot K_g)$&&$O(N\cdot K_g)$ \\ \hline
\multirow{2}{*}{REaaS}        &    \makecell{Cloud \\server }         &  $O(K_f)$   &$O(K_f)$& $O(K_f+T_f^2\cdot M_f^3)$  & $O(K_f+T_f^2 \cdot M_f^3)$ \\ \cline{2-6}
&Client&$O(e\cdot K_g)$ &$O(e\cdot K_g)$&$O(K_g+T_g^2\cdot M_g^3)$& $O(N\cdot K_g)$ \\ \hline
\end{tabular}
\label{table:theoreical_computation}
}
\vspace{-6mm}
\end{table}

\subsection{Theoretical Communication and Computation Cost Analysis}
\vspace{-2mm}
\myparatight{Communication cost} The number of queries to the APIs characterizes the communication cost for a client and the cloud server. In both BC and SC based certification, a client only queries the {Feature-API} once for each training input in REaaS. Therefore, the number of queries per training input is $1$ in REaaS. In both BC and SC based certification, a client only queries the {Feature-API} and {F2IPerturb-API} once to derive the predicted label and certified radius for a testing input. Therefore, the number of queries per testing input is $2$ in REaaS. 
Note that the client only sends an image $\mathbf{x}$ to the cloud server when querying the Feature-API, while it also sends the feature-space certified radius $R_f$ to the cloud server when querying the {F2IPerturb-API}. However, $R_f$ is a real number whose communication cost is negligible, compared to that of the image $\mathbf{x}$. Thus, we consider querying Feature-API and querying F2IPerturb-API have the same communication cost in our analysis for simplicity. 
Table~\ref{table:theoreical_communication} compares the number of queries per training/testing input in SEaaS and REaaS. Compared with SEaaS, REaaS makes BC based certification applicable and incurs a much smaller communication cost in SC based certification.

\myparatight{Computation cost} Table~\ref{table:theoreical_computation} compares the computational complexity of REaaS and SEaaS for the cloud server and a client. In both REaaS and SEaaS, the computation cost for the cloud server to process a query to the Feature-API is linear to the number of encoder parameters, i.e., $O(K_f)$, where $K_f$ is the number of parameters in the encoder. In REaaS, we use binary search to process a query to the F2IPerturb-API. Given the initial search range $[\rho_1^L, \rho_1^U]$ and binary-search precision $\beta$, the number of rounds of binary search is $\lceil \log_2(\frac{\rho_1^U-\rho_1^L}{\beta}) \rceil$. In practice, we can set $\rho_1^L$, $\rho_1^U$, and $\beta$ to be constants, e.g., $\rho_1^L=0$, $\rho_1^U=10$, and $\beta=10^{-50}$, and thus $\lceil \log_2(\frac{\rho_1^U-\rho_1^L}{\beta}) \rceil$ can be viewed as a constant. From~\cite{zhang2018crown},  the computational complexity is $O(T_f^2\cdot M_f^3)$ in each round of binary search, where $T_f$ and $M_f$ are respectively the number of layers and the maximum number of neurons in a layer in an encoder. Thus, the computational complexity for the cloud server to process a query to the F2IPerturb-API is $O(T_f^2\cdot M_f^3)$.

On the client side, the computational complexity of gradient descent is $O(K_g)$ for each training input per epoch when training a base downstream classifier in both BC and SC based certification, where $K_g$ is the number of parameters in the base downstream classifier. Therefore, the computational complexity of training a base downstream classifier is $O(e \cdot K_g)$ per training input, where $e$ is the number of training epochs. The computational complexity for a client to  derive the feature-space certified radius of a testing input is $O(T_g^2\cdot M_g^3)$ in BC based certification~\cite{zhang2018crown}, where $T_g$ and $M_g$ are respectively the number of layers and the maximum number of neurons in a layer in the base downstream classifier. Moreover, the computational complexity of using the base downstream classifier to predict a label for a (noisy) feature vector is $O(K_g)$.

As SEaaS does not support BC based certification, we focus on comparing the computation cost of SEaaS and REaaS for SC based certification. First, we observe that the computation cost per training/testing input is the same for a client in SEaaS and REaaS. Second, REaaS incurs a smaller computation cost per training input for the cloud server than SEaaS, because REaaS incurs much fewer queries than SEaaS. Third, REaaS often incurs a smaller computation cost per testing input for the cloud server than SEaaS, because $N$ is often large to achieve a large certified radius as shown in our experiments.

%% file: evaluation.tex
\section{Evaluation}
\subsection{Experimental Setup}
\label{section_experimental_setup}
\vspace{-2mm}
\myparatight{Datasets} We use CIFAR10~\cite{krizhevsky2009learning}, SVHN~\cite{netzer2011reading}, STL10~\cite{coates2011analysis}, and Tiny-ImageNet~\cite{tinyimagenet}  in our experiments. CIFAR10 has 50,000 training and 10,000 testing images from ten classes. SVHN   contains 73,257 training  and 26,032 testing images from ten classes. STL10   contains 5,000 training  and 8,000 testing images from ten classes, as well as 100,000 unlabeled images. Tiny-ImageNet contains 100,000 training  and 10,000 testing images from 200 classes.

We rescale each image in all datasets to $32 \times 32$ by the standard bi-linear interpolation. Therefore, the input image size in a downstream dataset is the same as the input size of the pre-trained encoder. However, we will also explicitly explore the scenarios in which the input image size of a downstream dataset is different from the input size of the pre-trained encoder.

\myparatight{Pre-training encoders} We use STL-10 and Tiny-ImageNet as pre-training datasets to pre-train encoders. We adopt these two datasets because they contain more images than CIFAR10 and SVHN. In particular, we use the unlabeled data of STL10 to pre-train an encoder when STL10 is used as a pre-training dataset. When Tiny-ImageNet is used as a pre-training dataset, we use its training dataset to pre-train an encoder.
Unless otherwise mentioned, we adopt MoCo~\cite{he2019moco} as the pre-training algorithm in SEaaS, while we adopt MoCo with our spectral-norm regularization as the pre-training algorithm in REaaS, since they only need unlabeled data. Moreover, we adopt the public implementation of MoCo~\cite{moco_implementation} in our experiments. When calculating the spectral norm of the encoder during pre-training,  we run 10 iterations of power iteration in each mini-batch. 
The architecture of the encoder can be found in Table \ref{architecture} in Appendix. We pre-train an encoder for 500 epochs with a learning rate 0.06 and a batch size  512.

\myparatight{Training downstream classifiers} As we have four datasets, we use the other three datasets as downstream datasets when a dataset is used as a pre-training dataset. Moreover, when a dataset is used as a downstream dataset, we adopt its training dataset as the downstream training dataset and testing dataset as the downstream testing dataset. We use the downstream training dataset to train a base downstream classifier. In particular, in BC based certification, we use standard supervised learning to train a base downstream classifier on the feature vectors of the training inputs. We note that some works~\cite{balunovic2019adversarial,zhang2019towards} proposed new methods to train a base classifier to improve its certified robustness in BC based certification. These methods are also applicable in our REaaS, but we do not evaluate them since our focus is to show the applicability of BC based certification in REaaS instead of its optimal certified robustness. For SC based certification, we train a base downstream classifier via adding Gaussian noise $\mathcal{N}(0, \sigma^2\mathbf{I})$ to the training inputs in SEaaS and the feature vectors of the training inputs in REaaS. 
 
We use a fully connected neural network with two hidden layers as a base downstream classifier. We respectively adopt ReLU and Softmax as the activation functions in the two hidden layers and the output layer.  The number of neurons in both hidden layers is 256. We train a base downstream classifier for 25 epochs using cross-entropy loss, a learning rate of 0.06, and a batch size of 512. 

\myparatight{Certification methods} For BC based certification, we adopt CROWN~\cite{zhang2018crown} to derive the certified radius of a base downstream classifier for a testing input in REaaS. 
We adopt the public implementation of CROWN~\cite{crown_implementation}. For SC based certification, we adopt Gaussian noise based randomized smoothing~\cite{cohen2019certified} to build a smoothed classifier and derive its certified radius for a testing input. In SEaaS, a client treats the composition of the encoder and its downstream classifier as a base classifier, while a client treats its downstream classifier alone as a base classifier in REaaS. 
We use the public code~\cite{randomsmooth_implementation} for Gaussian noise based randomized smoothing. Appendix~\ref{crownexample} and~\ref{randomizedsmoothingexample} show the technical details of CROWN and Gaussian noise based randomized smoothing, respectively.

\myparatight{Evaluation metrics}
Recall that REaaS aims to achieve three design goals. We can evaluate the generality goal by showing that REaaS supports both BC and SC based certification. For the efficiency goal, we use \emph{\#Queries per training (or testing) input} to measure the communication cost between a client and the cloud server. Moreover, we use \emph{running time per testing input} on the cloud server to measure its computation cost. We do not consider running time on a client as it is the same in SEaaS and REaaS. Note that \#Queries per training input also characterizes the computation cost per training input for the cloud server as it is linear to the number of queries. For the robustness goal, we use \textit{average certified radius (ACR)} of the correctly classified testing examples to measure the certified robustness of a base or smoothed classifier. 

Note that there often exists a trade-off between robustness and accuracy for a classifier. Therefore, we further consider accuracy under adversarial perturbation as an evaluation metric. In particular, we consider the widely adopted \emph{certified accuracy @ a perturbation size}, which is the fraction of testing inputs in a downstream testing dataset whose labels are correctly predicted  and whose certified radii are no smaller than the given perturbation size. Certified accuracy @ a perturbation size is the least testing accuracy that a classifier can achieve no matter what adversarial perturbation is added to each testing input once its $\ell_2$-norm is at most the given perturbation size. The certified accuracy @ a perturbation size decreases as the perturbation size increases. 
ACR is the area under the certified accuracy vs. perturbation size curve (details are shown in  Appendix~\ref{relationship_of_ACR}).   
Therefore, ACR can also be viewed as a metric to measure the robustness-accuracy trade-off of a classifier, where a larger ACR indicates a better trade-off.

\myparatight{Parameter settings}  F2IPerturb-API has the following three parameters: $\rho_1^L$ and $\rho_1^U$ which specify the range of $R$ in the first round of binary search, and $\beta$ which characterizes the binary-search precision. We set $\rho_1^L$ to be $0$ and set $\rho_1^U$ to be $10$. Note that they do not impact experimental results once $\rho_1^L$ is set to $0$ and $\rho_1^U$ is set to a large value (e.g., 10). We set the default value of $\beta$ as $0.001$. We note that $\beta$ has a negligible impact on certified accuracy and ACR. In particular, the absolute difference between the certified accuracy (or ACR) when $\beta=0.001$ and that when $\beta$ is an arbitrarily small value (e.g., $10^{-50}$) is smaller than $0.001$.

Randomized smoothing has the following three parameters:  the number of Gaussian noise $N$,     standard deviation $\sigma$ of the Gaussian noise, and error probability $\alpha$. 
Following prior work~\cite{cohen2019certified}, unless otherwise mentioned, we set  $N=100,000$, $\sigma=0.5$, and $\alpha=0.001$.
We set the default value of the hyperparameter $\lambda$ in our pre-training method as $0.00075$.  We normalize pixel values to $[0,1]$.

\begin{table}[!t]\renewcommand{\arraystretch}{1.5}
\centering
\fontsize{9}{8}\selectfont
\setlength{\tabcolsep}{1mm}
\caption{ ACR and \#Queries in SEaaS and REaaS.}

\subfloat[\fontsize{9}{8}\selectfont Pre-training dataset is Tiny-ImageNet]
{\begin{tabular}{|c|c|c|c|c|c|}
\hline
\multirow{3}{*}{\makecell{Service}}&\multirow{3}{*}{\makecell{Certification \\method}} & \multirow{3}{*}{ \makecell{Downstre-\\am dataset}}  & \multirow{3}{*}{ ACR}   &  \multicolumn{2}{c|}{\#Queries} \\ \cline{5-6}
&&&& \makecell{Per train-\\ing input} & \makecell{Per test-\\ing input} \\ \hline
\multirow{6}{*}{SEaaS} & \multirow{3}{*}{BC} & CIFAR10     & \multicolumn{3}{c|}{\multirow{3}{*}{N/A}}    \\ \cline{3-3}
 & & SVHN     &  \multicolumn{3}{l|}{}    \\ \cline{3-3}
 &&STL10&\multicolumn{3}{l|}{}  \\ \cline{2-6}
 &\multirow{3}{*}{SC} & CIFAR10     & 0.157 & \multirow{3}{*}{25}    &\multirow{3}{*}{$1 \times 10^5 $}   \\ \cline{3-4}
 & & SVHN     & 0.226 &      & \\ \cline{3-4}
&& STL10&0.134&& \\ \hline \hline
\multirow{6}{*}{REaaS} & \multirow{3}{*}{BC} & CIFAR10   &0.138  & \multirow{6}{*}{1}&   \multirow{6}{*}{2} \\ \cline{3-4}
 & & SVHN     & 0.258  &&   \\ \cline{3-4}
 &&STL10& 0.090&& \\ \cline{2-4}
 &\multirow{3}{*}{SC} & CIFAR10     & 0.171 &    &   \\ \cline{3-4}
 & & SVHN     & 0.275 &      & \\ \cline{3-4}
&& STL10&0.143&& \\ \hline 
\end{tabular}}

\subfloat[\fontsize{9}{8}\selectfont Pre-training dataset is STL10]{\begin{tabular}{|c|c|c|c|c|c|}
\hline
\multirow{3}{*}{\makecell{Service}}&\multirow{3}{*}{\makecell{Certification \\method}} & \multirow{3}{*}{ \makecell{Downstre-\\am dataset}}  & \multirow{3}{*}{ ACR}   &  \multicolumn{2}{c|}{\#Queries} \\ \cline{5-6}
&&&& \makecell{Per train-\\ing input} & \makecell{Per test-\\ing input} \\ \hline
\multirow{7}{*}{SEaaS} & \multirow{4}{*}{BC} & CIFAR10     & \multicolumn{3}{c|}{\multirow{4}{*}{NA}}    \\ \cline{3-3}
 & & SVHN     &  \multicolumn{3}{l|}{}    \\ \cline{3-3}
 &&\makecell{Tiny-\\ImageNet}&\multicolumn{3}{l|}{}  \\ \cline{2-6}
 &\multirow{4}{*}{SC} & CIFAR10     & 0.155 & \multirow{4}{*}{25}    &\multirow{4}{*}{$1 \times 10^5 $}   \\ \cline{3-4}
 & & SVHN     & 0.244 &      & \\ \cline{3-4}
&& \makecell{Tiny-\\ImageNet}&0.016&& \\ \hline \hline
\multirow{7}{*}{REaaS} & \multirow{4}{*}{BC} & CIFAR10   &0.139  & \multirow{7}{*}{1}&   \multirow{7}{*}{2} \\ \cline{3-4}
 & & SVHN     & 0.272  &&   \\ \cline{3-4}
 &&\makecell{Tiny-\\ImageNet}& 0.027&& \\ \cline{2-4}
 &\multirow{4}{*}{SC} & CIFAR10     & 0.173 &    &   \\ \cline{3-4}
 & & SVHN     & 0.278 &      & \\ \cline{3-4}
&& \makecell{Tiny-\\ImageNet}&0.033&& \\ \hline 
\end{tabular}}

\label{table:overallcomparison}
\vspace{-6mm}
\end{table}

\begin{table}[!t]\renewcommand{\arraystretch}{1.5}
\fontsize{9}{8}\selectfont
\setlength{\tabcolsep}{1mm}
\center
\caption{\CR{Comparing the running time per testing input for the cloud server in SC for SEaaS and REaaS. The pre-training dataset is Tiny-ImageNet.} }
\begin{tabular}{|c|c|c|}
\hline
Service                & \makecell{Downstream \\dataset} & Running time (s) per testing input \\ \hline
\multirow{3}{*}{SEaaS} & CIFAR10 &     73.77                               \\ \cline{2-3} 
                       & SVHN    &      72.65                              \\ \cline{2-3} 
                       & STL10   &     73.48                               \\ \hline\hline
\multirow{3}{*}{REaaS} & CIFAR10 &          1.05                          \\ \cline{2-3} 
                       & SVHN    &        1.06                            \\ \cline{2-3} 
                       & STL10   &       1.04                             \\ \hline
\end{tabular}
\label{compare_computation_time_of_service_provider}
\end{table}

\begin{table}[!t]\renewcommand{\arraystretch}{1.5}
\fontsize{9}{8}\selectfont
\setlength{\tabcolsep}{1mm}
\centering
\caption{\CR{Training without noise vs. training with noise for SC in SEaaS. The pre-training dataset is Tiny-ImageNet.}}
\label{table_impact_N_ACR_SEaaS_training}
\begin{tabular}{|c|cc|}
\hline
\multirow{2}{*}{\makecell{Downstream \\dataset}} & \multicolumn{2}{c|}{ACR}                                          \\ \cline{2-3} 
                         & \multicolumn{1}{c|}{Training with noise} & Training without noise \\ \hline
CIFAR10                  & \multicolumn{1}{c|}{0.157}               & 0.106                  \\ \hline
SVHN                     & \multicolumn{1}{c|}{0.226}                    &   0.155                    \\ \hline
STL10                    & \multicolumn{1}{c|}{0.134}                    &   0.088                     \\ \hline
\end{tabular}
\end{table}

\begin{table}[!t]\renewcommand{\arraystretch}{1.5}
\fontsize{9}{8}\selectfont
\setlength{\tabcolsep}{1mm}
\centering
\caption{\CR{Impact of $N$ on ACR for SC in SEaaS. The pre-training dataset is Tiny-ImageNet.}}
\begin{tabular}{|c|cccc|}
\hline
\multirow{2}{*}{\makecell{Downstream \\dataset}} & \multicolumn{4}{c|}{N}                                                                          \\ \cline{2-5} 
                         & \multicolumn{1}{c|}{100}   & \multicolumn{1}{c|}{1,000} & \multicolumn{1}{c|}{10,000} & 100,000 \\ \hline
CIFAR10                  & \multicolumn{1}{c|}{0.091} & \multicolumn{1}{c|}{0.132} & \multicolumn{1}{c|}{0.148}  & 0.157   \\ \hline
SVHN                     & \multicolumn{1}{c|}{0.130}      & \multicolumn{1}{c|}{0.186}      & \multicolumn{1}{c|}{0.211}       & 0.226   \\ \hline
STL10                    & \multicolumn{1}{c|}{0.079}      & \multicolumn{1}{c|}{0.111}      & \multicolumn{1}{c|}{0.127}       & 0.134   \\ \hline
\end{tabular}
\label{table_impact_N_ACR_SEaaS}
\vspace{-4mm}
\end{table}

\subsection{Experimental Results}
\label{sec:experimental_results}

\CR{We first show that REaaS achieves our three design goals, but SEaaS does not. Then,  we  show the impact of relevant factors on REaaS. In particular, we consider 1) different ways to pre-train an encoder, 2) image scaling, and 3) different hyperparameters of certification methods such as $N$, $\sigma$, and $\alpha$ for randomized smoothing. Note that we fix all other parameters to their default values when studying the impact of one parameter on REaaS.}

\myparatight{REaaS achieves the generality, efficiency, and robustness goals} 
In  SC based certification, a client respectively adds Gaussian noise to images and  their feature vectors to train a base downstream classifier  in SEaaS and REaaS. Thus, the certified robustness  of the smoothed classifiers are not comparable even if we use the same standard deviation $\sigma$ of Gaussian noise in SEaaS and REaaS. Therefore, we try multiple values of $\sigma$ and report the largest ACR for each service. Moreover, we select $\sigma$ values such that the largest ACR is not reached at the smallest or largest value of $\sigma$, to ensure the largest ACR is found for each service. In particular, we try $\sigma=0.125, 0.25, 0.5, 0.75, 1$ for both SEaaS and REaaS. 
\CR{We note that $\sigma$ controls a tradeoff between certified accuracy without attacks (i.e., perturbation size is 0) and robustness. Specifically, a smaller $\sigma$ can achieve a larger certified accuracy without attacks but also make the curve drop more quickly (i.e., less robust). ACR measures such trade-off, and thus we adopt the $\sigma$ that achieves the largest ACR for each method when comparing the certified accuracy of SC based certification in SEaaS and REaaS.}

\begin{table}[!t]\renewcommand{\arraystretch}{1.3}
\fontsize{9}{8}\selectfont
\setlength{\tabcolsep}{1mm}
\center
\caption{Comparing the ACRs in REaaS for different downstream datasets when the encoders are pre-trained by different self-supervised learning methods. The pre-training dataset is Tiny-ImageNet.}
\subfloat[CIFAR10]{\begin{tabular}{|c|c|c|}
\hline
\makecell{Certification Method} & \makecell{Pre-training Method}  & ACR   \\ \hline
\multirow{3}{*}{BC} &Non-robust MoCo & 0.012        \\ \cline{2-3}
 &RoCL &   0.016      \\ \cline{2-3}
 & Ours & 0.138            \\ \hline\hline

\multirow{3}{*}{SC} & Non-robust MoCo & 0.020       \\ \cline{2-3}
 &RoCL &  0.024     \\ \cline{2-3}
 & Ours & 0.171      \\ \hline
\end{tabular}}

\subfloat[SVHN]{\begin{tabular}{|c|c|c|}
\hline
\makecell{Certification Method} & \makecell{Pre-training Method}  & ACR  \\ \hline
\multirow{3}{*}{BC} & Non-robust MoCo & 0.009      \\ \cline{2-3}
 &RoCL & 0.015     \\ \cline{2-3}
 & Ours &   0.258   \\ \hline\hline

\multirow{3}{*}{SC} & Non-robust MoCo & 0.011     \\ \cline{2-3}
 &RoCL  &   0.014      \\ \cline{2-3}
 & Ours & 0.275     \\ \hline
\end{tabular}}

\subfloat[STL10]{\begin{tabular}{|c|c|c|}
\hline
\makecell{Certification Method} & \makecell{Pre-training Method}  & ACR  \\ \hline
\multirow{3}{*}{BC} & Non-robust MoCo &0.011           \\ \cline{2-3}
 &RoCL & 0.014    \\ \cline{2-3}
 & Ours &0.090    \\ \hline\hline

\multirow{3}{*}{SC} & Non-robust MoCo & 0.015      \\ \cline{2-3}
 & RoCL & 0.020   \\ \cline{2-3}
 & Ours & 0.143    \\ \hline
\end{tabular}}
\label{table:differentpretraining}
\vspace{-6mm}
\end{table}

\begin{table}[!t]\renewcommand{\arraystretch}{1.3} 
\fontsize{9}{8}\selectfont
\setlength{\tabcolsep}{1mm}
\center
\caption{Comparing the ACRs in REaaS for different downstream datasets when the encoders are pre-trained by different  self-supervised learning methods.  The pre-training dataset is STL10.}
\subfloat[CIFAR10]{\begin{tabular}{|c|c|c|}
\hline
\makecell{Certification Method} & \makecell{Pre-training Method}  & ACR   \\ \hline
\multirow{3}{*}{BC} &Non-robust MoCo &  0.010       \\ \cline{2-3}
 &RoCL &   0.012      \\ \cline{2-3}
 & Ours &  0.139           \\ \hline\hline

\multirow{3}{*}{SC} & Non-robust MoCo &  0.014 \\ \cline{2-3}
 &RoCL &   0.017    \\ \cline{2-3}
 & Ours &    0.173   \\ \hline
\end{tabular}}

\subfloat[SVHN]{\begin{tabular}{|c|c|c|}
\hline
\makecell{Certification Method} & \makecell{Pre-training Method}  & ACR  \\ \hline
\multirow{3}{*}{BC} & Non-robust MoCo &   0.006    \\ \cline{2-3}
 &RoCL &  0.009    \\ \cline{2-3}
 & Ours &    0.272 \\ \hline\hline

\multirow{3}{*}{SC} & Non-robust MoCo &   0.007   \\ \cline{2-3}
 &RoCL  &   0.012     \\ \cline{2-3}
 & Ours &  0.278    \\ \hline
\end{tabular}}

\subfloat[Tiny-ImageNet]{\begin{tabular}{|c|c|c|}
\hline
\makecell{Certification Method} & \makecell{Pre-training Method}  & ACR  \\ \hline
\multirow{3}{*}{BC} & Non-robust MoCo & 0.003          \\ \cline{2-3}
 &RoCL &  0.004   \\ \cline{2-3}
 & Ours &  0.027  \\ \hline\hline

\multirow{3}{*}{SC} & Non-robust MoCo &  0.003     \\ \cline{2-3}
 & RoCL &  0.004  \\ \cline{2-3}
 & Ours &    0.033 \\ \hline
\end{tabular}}
\label{table:differentpretraining_stl10}
\vspace{-6mm}
\end{table}

Table~\ref{table:overallcomparison} compares ACR and \#Queries per training/testing input in SEaaS and REaaS, while Table~\ref{compare_computation_time_of_service_provider} compares the running time per testing input for the server in SC for SEaaS and REaaS. We have the following observations. First, REaaS supports both BC and SC. Therefore,  REaaS achieves the generality goal. In contrast, SEaaS only supports SC.
Second, REaaS achieves the efficiency goal as it is much more efficient than SEaaS. Specifically, 
\#Queries per training/testing input in REaaS is orders of magnitude smaller than that in SEaaS for SC. We note that a client using SEaaS could choose to train a base downstream classifier without adding noise to its training inputs to reduce the \#Queries per training input to 1 or use a small $N$ to reduce the \#Queries per testing input. However, the smoothed classifier achieves (much) smaller ACRs in such cases as shown in Table~\ref{table_impact_N_ACR_SEaaS_training} and~\ref{table_impact_N_ACR_SEaaS}. Base on  Table~\ref{compare_computation_time_of_service_provider}, REaaS also incurs a much lower computation cost for the server than SEaaS.

Third, REaaS achieves the robustness goal as it achieves larger ACRs than SEaaS for SC. The reason is that, in SEaaS, the base classifier is the composition of an encoder and a base downstream classifier, but the client can only train the base downstream classifier with noise. In contrast, a client can build a smoothed classifier upon a base downstream classifier alone which can be trained with noise and the encoder is pre-trained in a robust way in REaaS. \CR{Figure~\ref{fig:overallcomparison1} in Appendix further compares the certified accuracy vs. perturbation size of SC in SEaaS and REaaS.}  
We find that REaaS can achieve a better trade-off between accuracy without attacks and robustness than SEaaS. Specifically, REaaS  achieves larger certified accuracy than SEaaS when the perturbation size is small. Moreover, the gap between the certified accuracy of SEaaS and REaaS is much larger when the perturbation size is small than that when the perturbation size is large.

\begin{figure}[!t]
\center
\vspace{-4mm}
\subfloat[BC]{\includegraphics[width=0.24\textwidth]{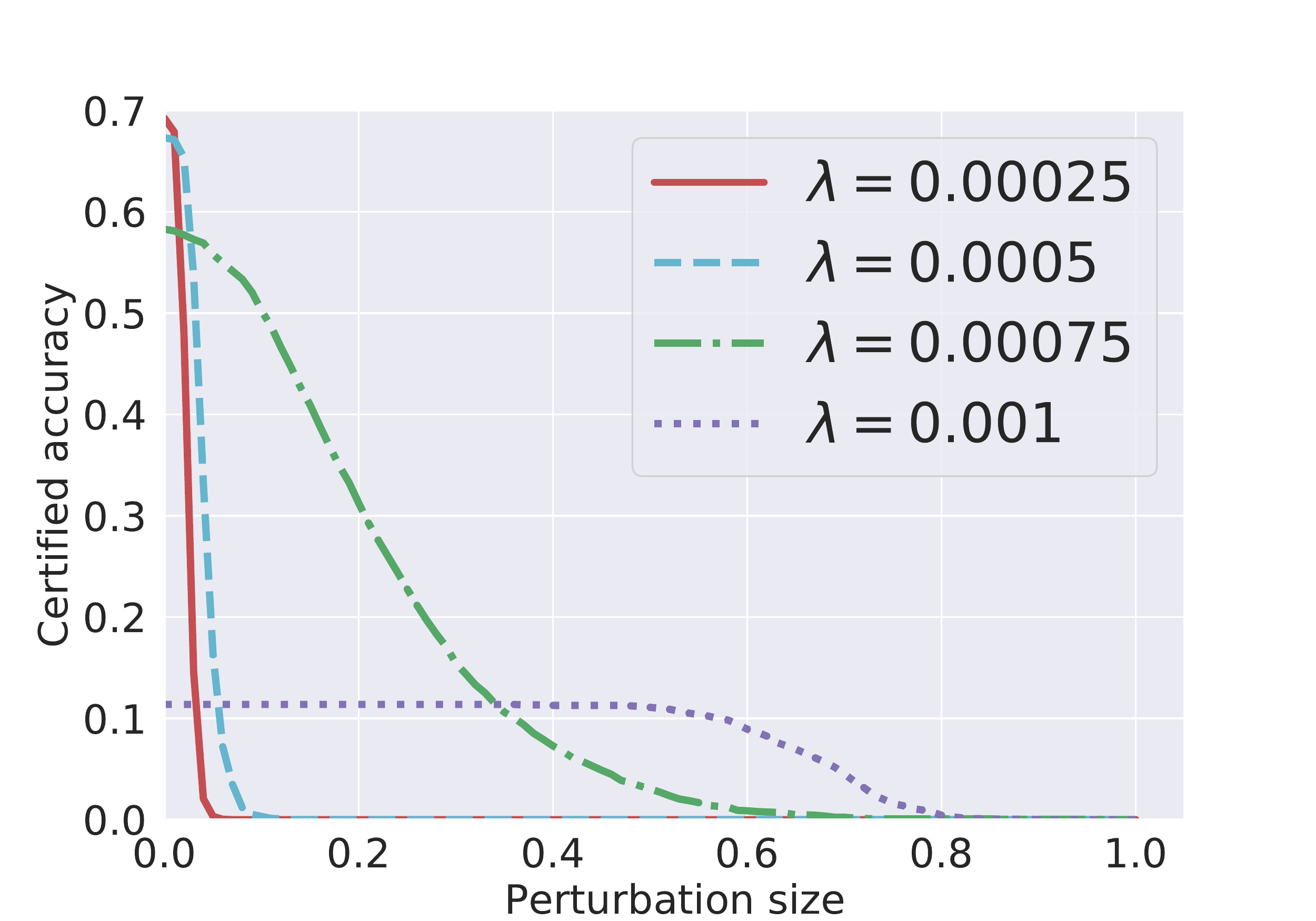}}
\subfloat[SC]{\includegraphics[width=0.24\textwidth]{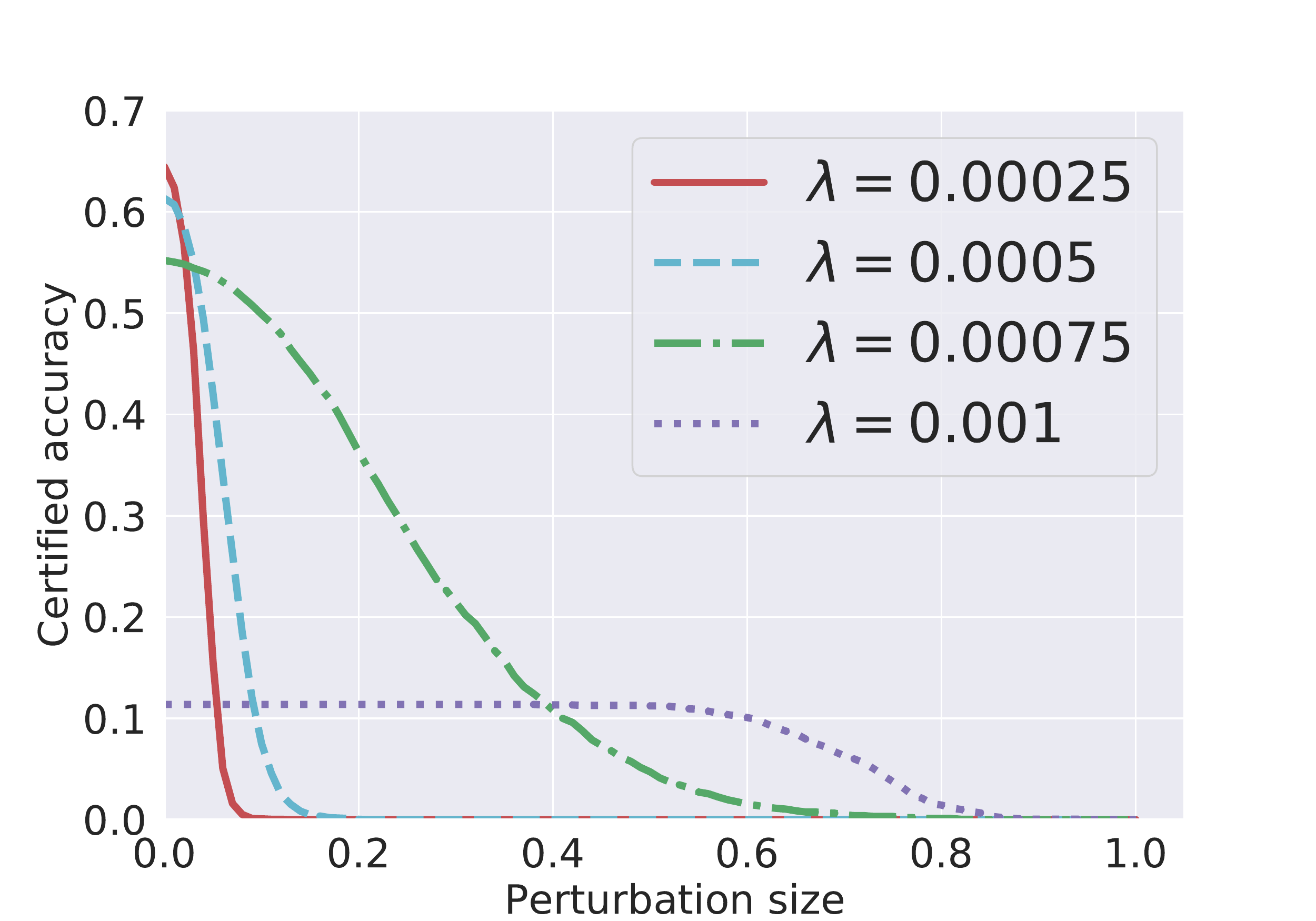}}
\caption{Impact of $\lambda$ on certified accuracy vs. perturbation size for BC  and SC  in REaaS. The pre-training dataset is Tiny-ImageNet and the downstream dataset is CIFAR10.}
\label{fig:hyperparameterCA}
 \vspace{-6mm}
\end{figure}

\myparatight{Impact of methods to pre-train encoders} We can use different methods to pre-train an encoder in REaaS.  Table~\ref{table:differentpretraining} and~\ref{table:differentpretraining_stl10} show ACRs in REaaS when different self-supervised learning methods are used to pre-train encoders. In particular, we consider non-robust MoCo~\cite{he2019moco},  RoCL~\cite{kim2020adversarial}, and our robust pre-training method (i.e., MoCo with our spectral-norm regularization).  Table~\ref{table:differentpretrainingSL} shows ACRs of REaaS when different supervised learning methods are used to pre-train encoders.  In particular, we consider a standard, non-robust supervised learning method, adversarial training~\cite{madry2017towards} (we use the default parameter settings in the authors' public implementation), and our robust pre-training method (i.e., standard supervised learning with our spectral-norm regularization). We only show results when the pre-training dataset is Tiny-ImageNet for supervised pre-training methods, as STL10 dataset only has a small number of labeled training images which are insufficient to pre-train high-quality encoders using supervised learning. We try $\sigma=0.125, 0.25, 0.5, 0.75, 1$ and report the largest ACR for each pre-training method. As the results show, our robust pre-training method achieves substantially larger ACRs than existing methods for both supervised learning and self-supervised learning. Our method is better than RoCL and adversarial training because they aim to train empirically robust rather than certifiably robust encoders, and is better than MoCo and standard supervised learning because the encoders pre-trained by them are non-robust.

\begin{table}[!t]\renewcommand{\arraystretch}{1.5} 
\fontsize{9}{8}\selectfont
\setlength{\tabcolsep}{1mm}
\center
\caption{Comparing the ACRs in REaaS for different downstream datasets when the encoders are pre-trained by different  supervised learning (SL) methods. The pre-training dataset is Tiny-ImageNet.}  

\subfloat[CIFAR10]{\begin{tabular}{|c|c|c|}
\hline
\makecell{Certification Method} & \makecell{Pre-training Method}  & ACR   \\ \hline
\multirow{3}{*}{BC} &Non-robust SL & 0.019       \\ \cline{2-3}
 & Adversarial Training &  0.035        \\ \cline{2-3}
 & Ours & 0.174 \\ \hline \hline

\multirow{3}{*}{SC} & Non-robust SL & 0.022       \\ \cline{2-3}
 &  Adversarial Training   &  0.041 \\ \cline{2-3}
 & Ours & 0.172 \\ \hline
\end{tabular}}

\subfloat[SVHN]{\begin{tabular}{|c|c|c|}
\hline
\makecell{Certification Method} & \makecell{Pre-training Method}  & ACR  \\ \hline
\multirow{3}{*}{BC} & Non-robust SL & 0.008      \\ \cline{2-3}
 & Adversarial Training & 0.018 \\ \cline{2-3}
 & Ours &   0.268 \\ \hline\hline

\multirow{3}{*}{SC} & Non-robust SL & 0.011     \\ \cline{2-3}
 & Adversarial Training &   0.021  \\ \cline{2-3}
 & Ours & 0.292\\ \hline 
\end{tabular}}

\subfloat[STL10]{\begin{tabular}{|c|c|c|}
\hline
\makecell{Certification Method} & \makecell{Pre-training Method}  & ACR  \\ \hline
\multirow{3}{*}{BC} & Non-robust SL & 0.012           \\ \cline{2-3}
 & Adversarial Training& 0.026  \\ \cline{2-3}
 &  Ours & 0.100 \\ \hline\hline

\multirow{3}{*}{SC} & Non-robust SL & 0.018      \\ \cline{2-3}
 & Adversarial Training& 0.034 \\ \cline{2-3}
 &  Ours & 0.114  \\ \hline
\end{tabular}}
\label{table:differentpretrainingSL}
\vspace{-5mm}
\end{table}

\begin{figure}[!t]
\center
\vspace{-4mm}
\subfloat[BC]{\includegraphics[width=0.24\textwidth]{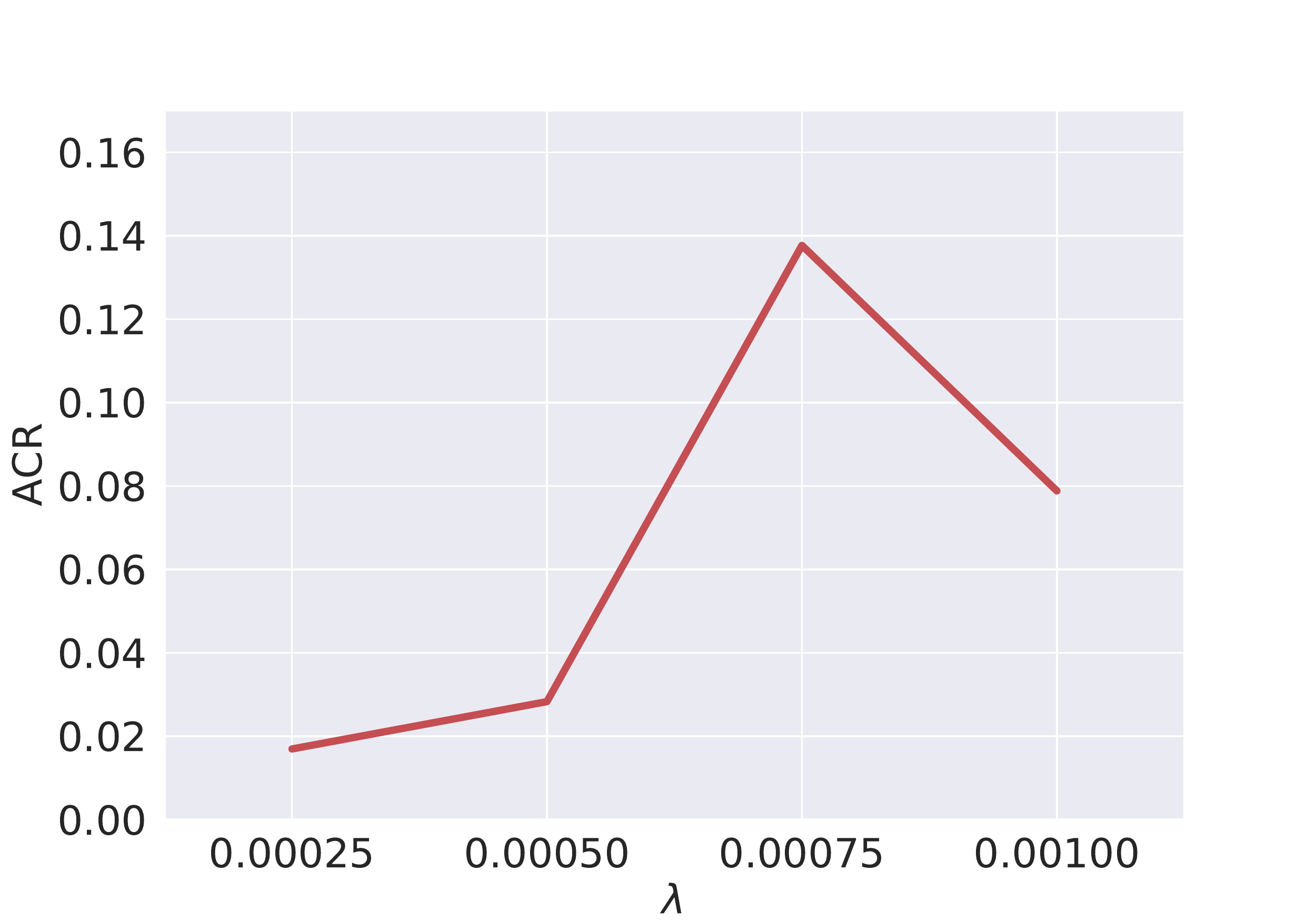}}
\subfloat[SC]{\includegraphics[width=0.24\textwidth]{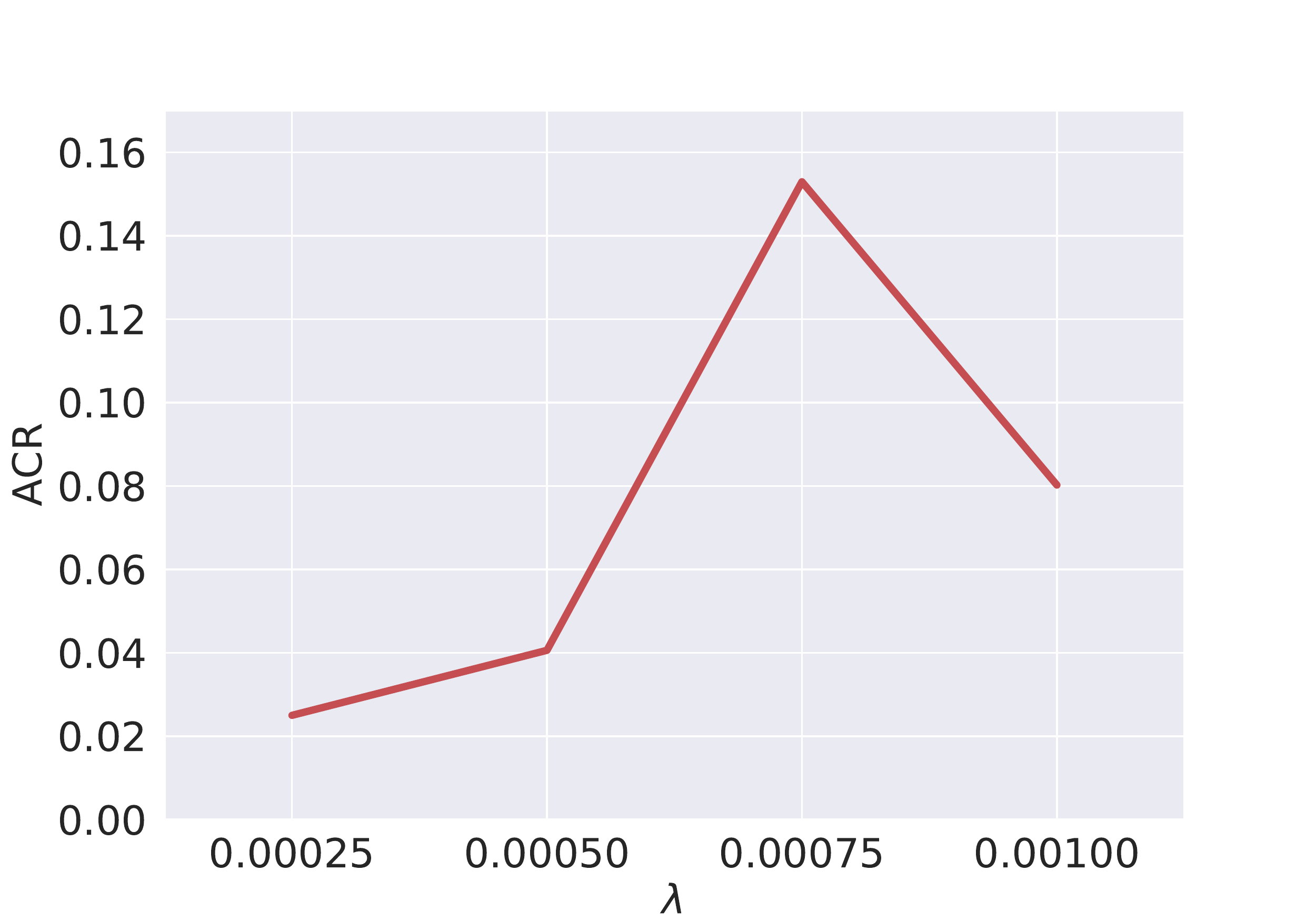}}
\caption{Impact of $\lambda$ on ACR for BC  and SC in REaaS. The pre-training dataset is Tiny-ImageNet and the downstream dataset is CIFAR10.}
\label{fig:hyperparameter}
\vspace{-6mm}
\end{figure}

\begin{figure*}[!t]
\centering
\subfloat{\includegraphics[width=0.33\textwidth]{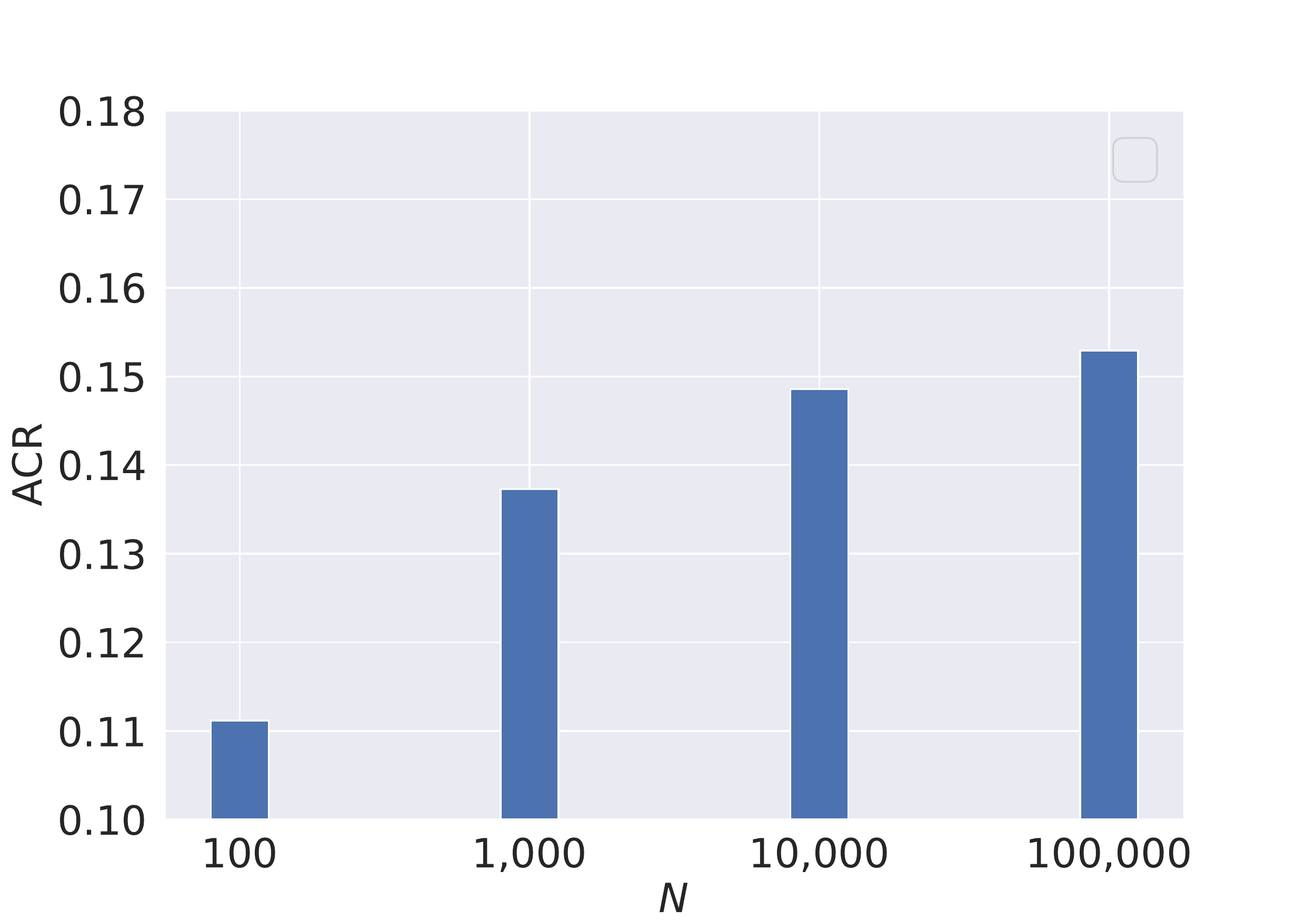}}
\subfloat{\includegraphics[width=0.33\textwidth]{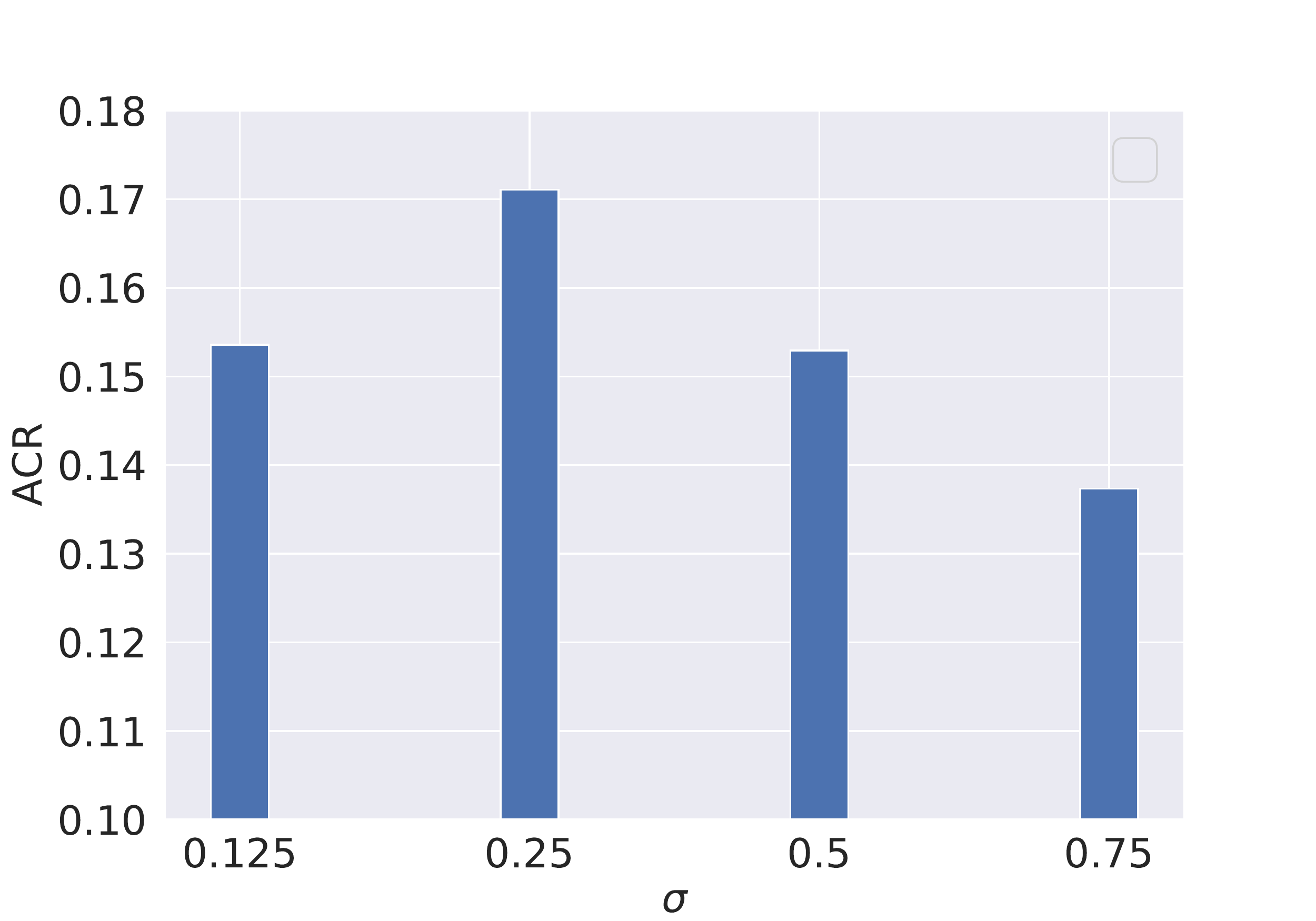}}
\subfloat{\includegraphics[width=0.33\textwidth]{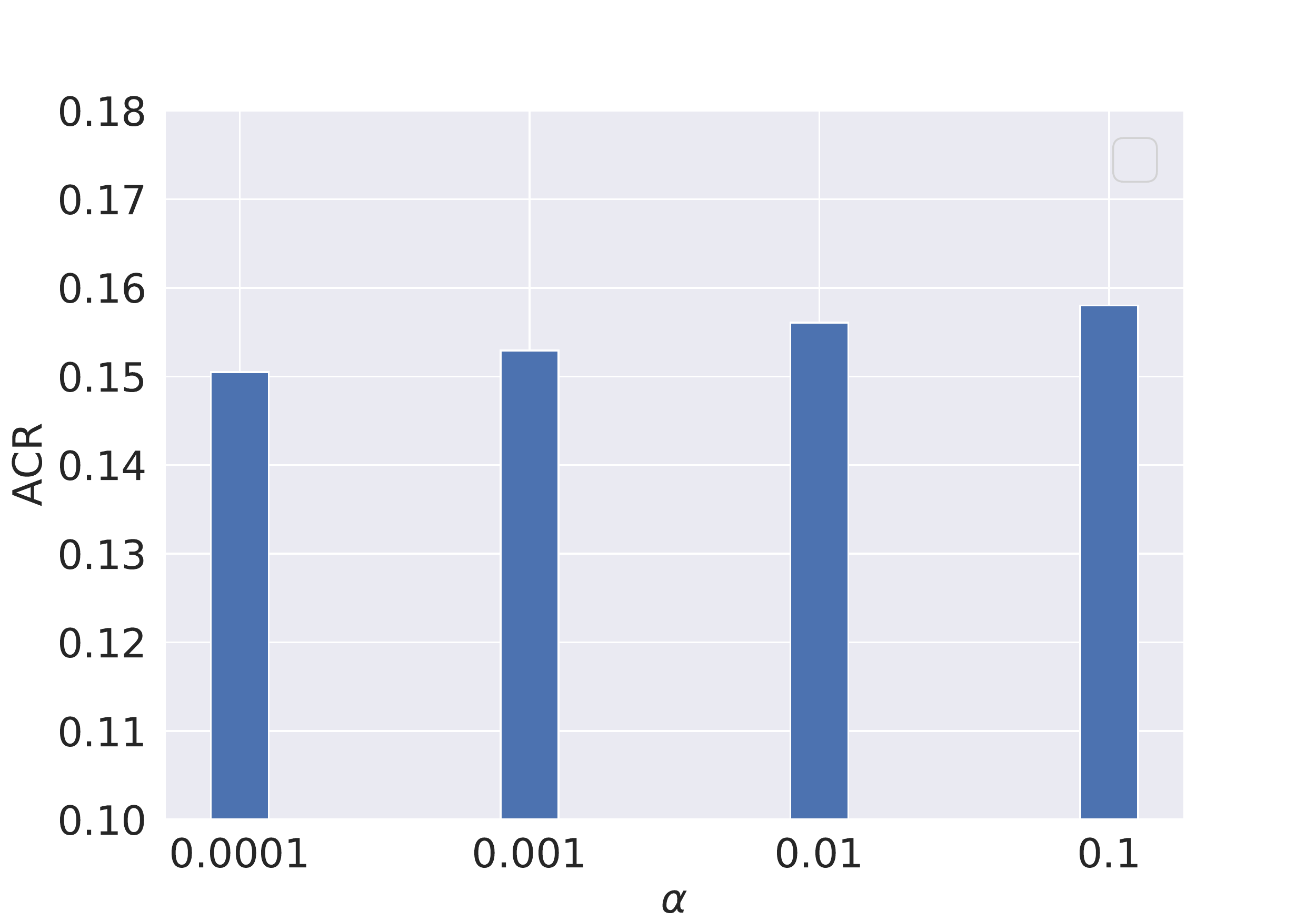}}
\caption{Impact of $N$, $\sigma$, and $\alpha$ on ACR of SC in REaaS. The pre-training dataset is Tiny-ImageNet and the downstream dataset is CIFAR10.}
\label{fig:SCparameterACR}
\vspace{-7mm}
\end{figure*}

\myparatight{Impact of hyperparameter $\lambda$} Figure~\ref{fig:hyperparameterCA} shows the impact of $\lambda$ on certified accuracy in REaaS. We find that $\lambda$ measures a trade-off between accuracy without attacks (i.e., perturbation size is 0) and robustness. In particular, when $\lambda$ is smaller, the accuracy without attacks is larger, but the certified accuracy decreases more quickly as the perturbation size increases.  Figure~\ref{fig:hyperparameter} shows the impact of $\lambda$ on ACR. Our results show that, for  both BC and SC, ACR first increases as $\lambda$ increases and then decreases after $\lambda$ is larger than a certain value. The reason is that a larger or smaller $\lambda$ leads to a worse trade-off between accuracy without attacks and robustness as shown in Figure~\ref{fig:hyperparameterCA}.

\begin{figure}[!t]
\centering
\subfloat[BC]{\includegraphics[width=0.24\textwidth]{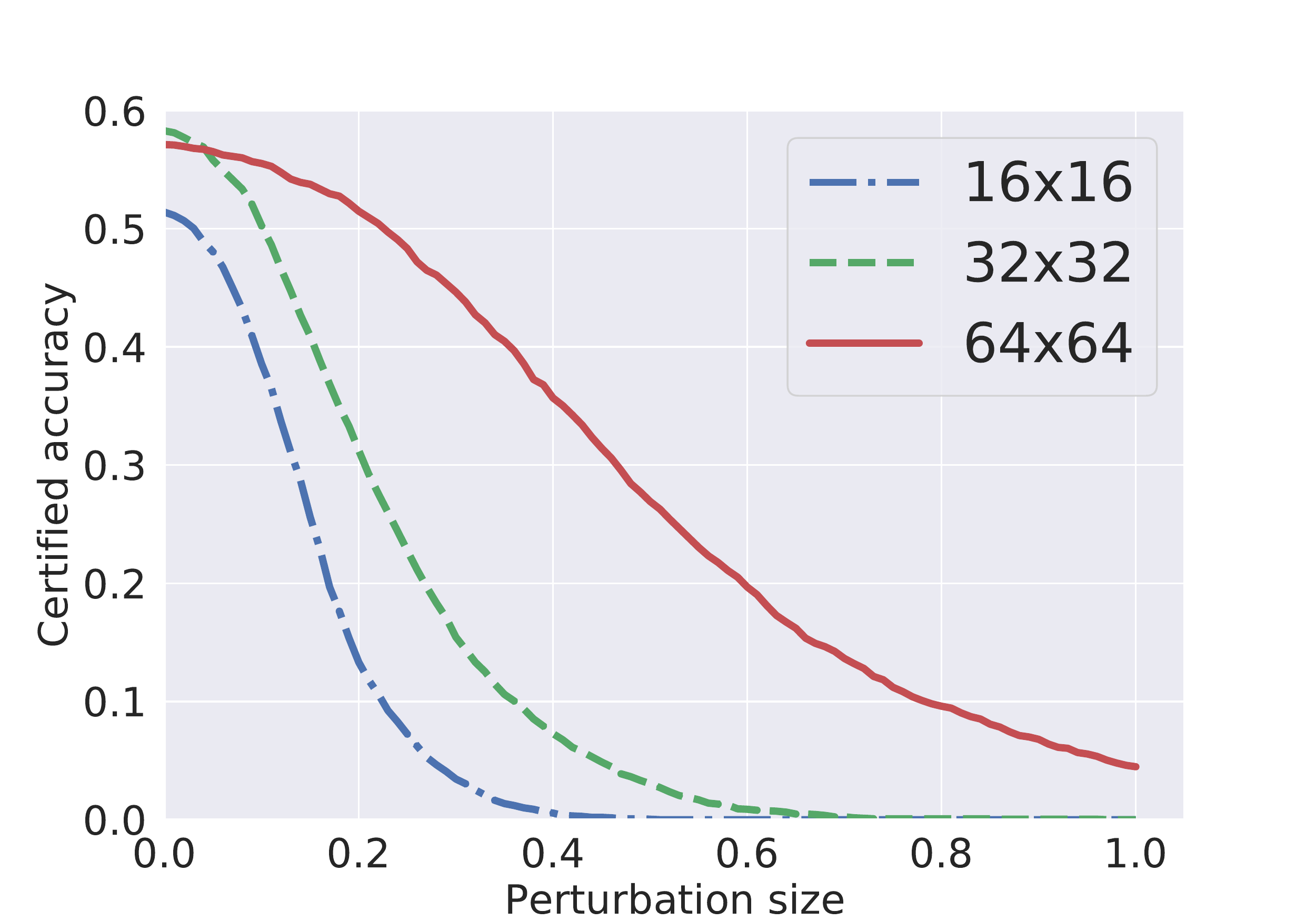}}
\subfloat[SC]{\includegraphics[width=0.24\textwidth]{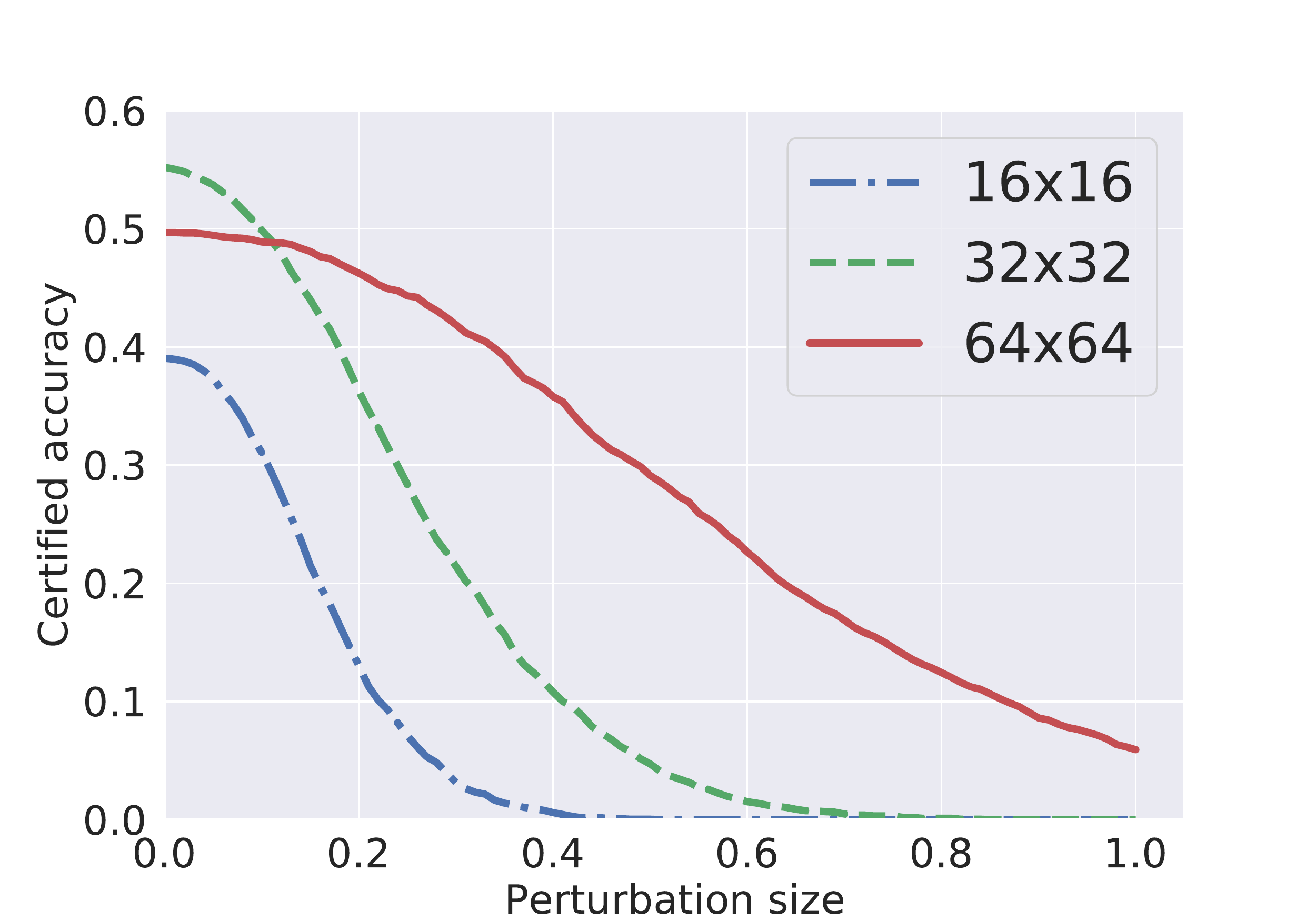}}
\caption{Impact of downstream input size on certified accuracy vs. perturbation size for BC  and SC  in REaaS. The pre-training dataset is Tiny-ImageNet and the downstream dataset is  (or created from)  CIFAR10.  The  input size  of the  pre-trained  encoder is 32x32.}
\label{fig:impact_input_size_ca}
\end{figure}

\myparatight{Impact of image rescaling} To study the impact of image rescaling, we create downstream datasets with different input image sizes via resizing images in CIFAR10. 
Table~\ref{table:scaling} shows the results on ACR and Figure~\ref{fig:impact_input_size_ca} shows the results on certified accuracy. We find that, when the size of the images in a downstream dataset is larger (or smaller) than the input size of the encoder, the downstream input-space ACR is larger (or smaller) for both BC and SC. The reason is that down-scaling (or up-scaling) the downstream input images to be the same size as the input size of the encoder reduces (or enlarges) the perturbation in the downstream image space.

\begin{table}[!t]\renewcommand{\arraystretch}{1.5}
\fontsize{9}{8}\selectfont
\setlength{\tabcolsep}{1mm}
\center
\caption{Impact of image rescaling on ACR in REaaS. The pre-training dataset is Tiny-ImageNet and the downstream dataset is (or created from) CIFAR10. The input size of the encoder is 32x32.}
\begin{tabular}{|c|c|c|}
\hline
\makecell{Certification Method} & \makecell{Size of Images in\\ Downstream Dataset}  & ACR  \\ \hline
\multirow{3}{*}{BC} & 16x16 &  0.082                \\ \cline{2-3}
 & 32x32 &   0.138                \\ \cline{2-3}
 & 64x64 &      0.303            \\ \hline\hline

\multirow{3}{*}{SC} & 16x16 &   0.068                \\ \cline{2-3}
 & 32x32 &   0.153                \\ \cline{2-3}
 & 64x64 &   0.305                \\ \hline

\end{tabular}
\label{table:scaling}
\end{table}

\myparatight{Impact of $N$, $\sigma$, and $\alpha$ for SC} Figure~\ref{fig:SCparameterACR} and~\ref{fig:SCparameterCA} (in Appendix) shows the impact of $N$, $\sigma$, and $\alpha$ on ACR and certified accuracy of SC in REaaS. We have the following observations. First, both   ACR and certified accuracy increase as $N$ or $\alpha$ increases. The reason is that the estimated certified radii are larger when $N$ or $\alpha$ is larger. Second, we find that $\sigma$ achieves a trade-off between accuracy without attacks (i.e., perturbation size is 0) and robustness. In particular, a smaller $\sigma$ can achieve a larger accuracy without attacks, but the curve drops faster as the perturbation size increases. Third, ACR first increases and then decreases as $\sigma$ increases. The reason is that a smoothed classifier  is less accurate without attacks when  $\sigma$ is larger  and is less robust when $\sigma$ is smaller.

\myparatight{REaaS vs. white-box access to the encoder} In REaaS, a client has black-box access to the encoder. We compare REaaS with the scenario where a client has white-box access to the encoder, e.g., the cloud server shares its encoder with a client. Specifically, with white-box access to the encoder, a client can  use either BC or SC by treating  the composition of the encoder and its downstream classifier as a base classifier. For BC, the client  can use CROWN~\cite{crown_implementation} to derive the certified radius of its base classifier for a testing input. For SC, the client can train/fine-tune the base classifier (both the encoder and downstream classifier) using training inputs with noise. The white-box scenario represents the upper-bound robustness a client can achieve. Therefore, comparing with the robustness in the white-box scenario enables us to understand how close our REaaS with the two APIs is to such upper bound. Table~\ref{table:compare_with_whitebox} compares the ACRs of REaaS and such white-box scenario. We find that REaaS can achieve comparable ACRs with the white-box scenario.

\begin{table}[!t]\renewcommand{\arraystretch}{1.5}
\fontsize{9}{8}\selectfont
\setlength{\tabcolsep}{1mm}
\center
\caption{Comparing the ACRs of REaaS  and the white-box scenario for different downstream datasets. The pre-training dataset is Tiny-ImageNet.}
\subfloat[CIFAR10]{\begin{tabular}{|c|c|c|}
\hline
\makecell{Certification Method} & \makecell{Service}  & ACR   \\ \hline
\multirow{2}{*}{BC} &White-box & 0.157        \\ \cline{2-3}
 & REaaS & 0.138            \\ \hline\hline

\multirow{2}{*}{SC} & White-box & 0.188       \\ \cline{2-3}
 & REaaS & 0.171      \\ \hline
\end{tabular}}

\subfloat[SVHN]{\begin{tabular}{|c|c|c|}
\hline
\makecell{Certification Method} & \makecell{Service}  & ACR  \\ \hline
\multirow{2}{*}{BC} & White-box & 0.286      \\ \cline{2-3}
 & REaaS &   0.258   \\ \hline\hline

\multirow{2}{*}{SC} & White-box & 0.302     \\ \cline{2-3}
 & REaaS & 0.275     \\ \hline
\end{tabular}}

\subfloat[STL10]{\begin{tabular}{|c|c|c|}
\hline
\makecell{Certification Method} & \makecell{Service}  & ACR  \\ \hline
\multirow{2}{*}{BC} & White-box &0.102          \\ \cline{2-3}
 & REaaS &0.090    \\ \hline\hline

\multirow{2}{*}{SC} & White-box & 0.151     \\ \cline{2-3}
 & REaaS & 0.143    \\ \hline
\end{tabular}}
\label{table:compare_with_whitebox}
\vspace{-6mm}
\end{table}

%% file: discussion.tex
\section{Discussion}
\label{section-discussion}

\begin{figure}[!t]
\centering
\vspace{-2mm}
\subfloat[$\ell_1$-norm]{\includegraphics[width=0.24\textwidth]{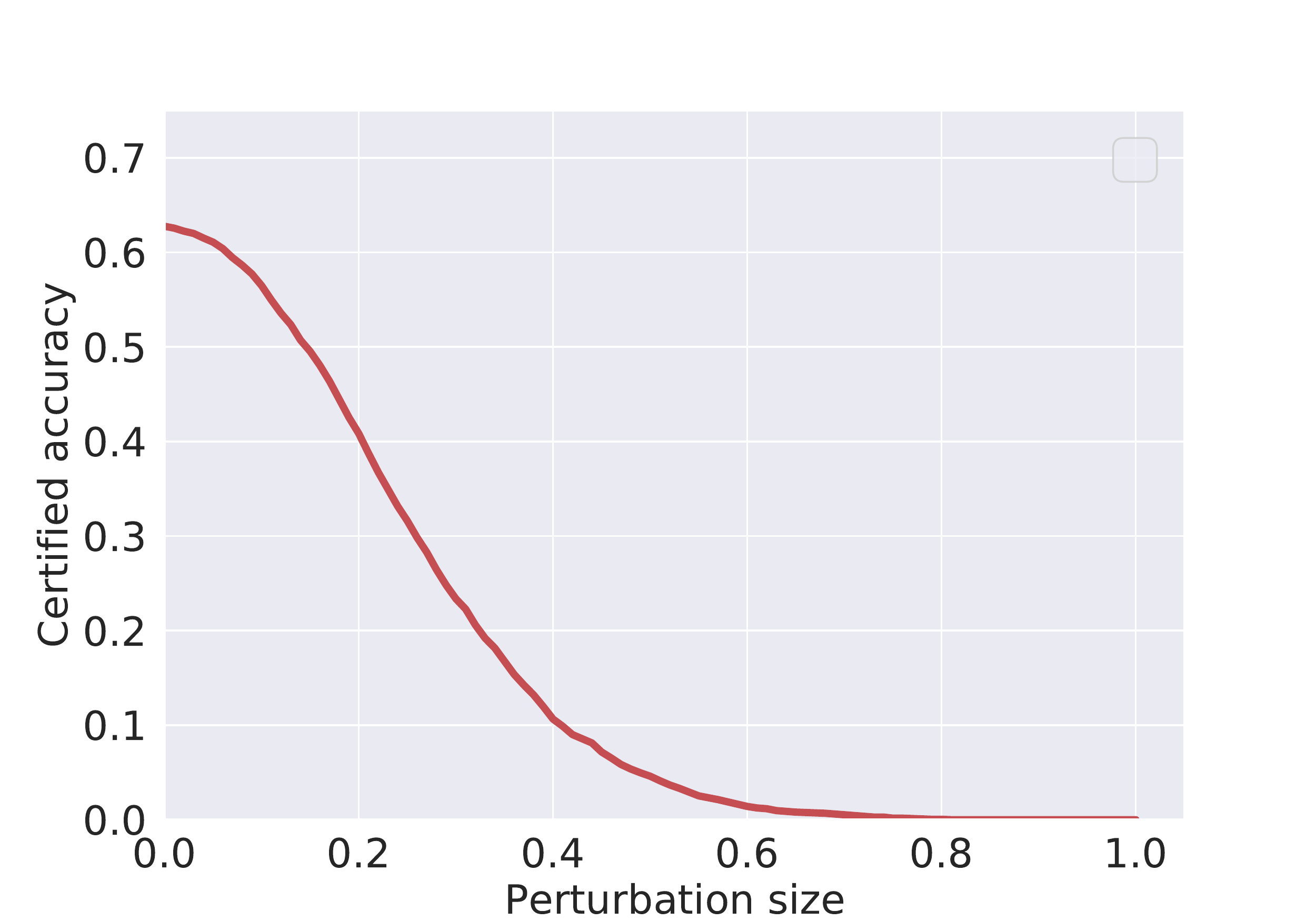}}
\subfloat[$\ell_\infty$-norm]{\includegraphics[width=0.24\textwidth]{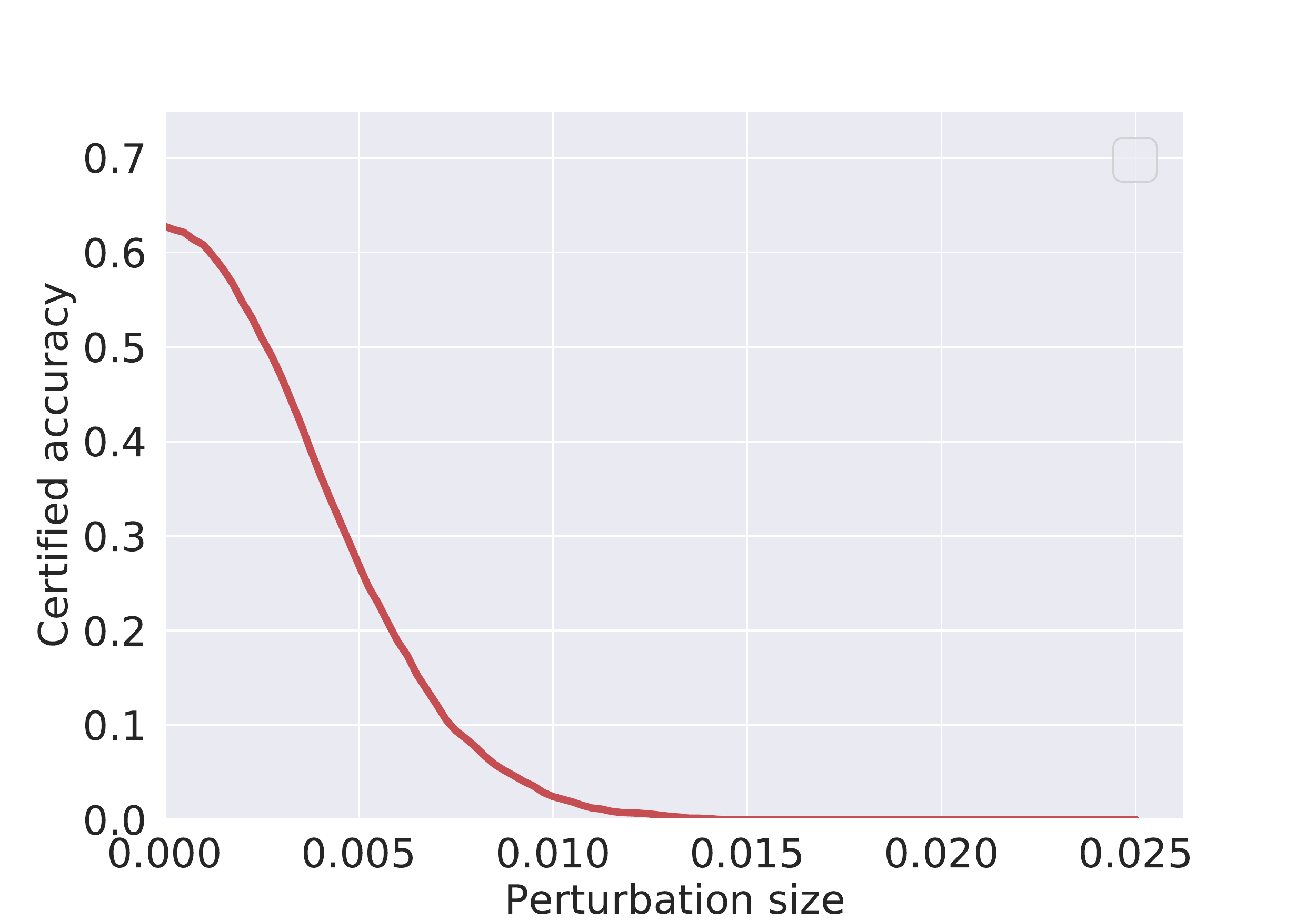}}
\caption{Certified accuracy vs. perturbation size of SC in REaaS  under  $\ell_1$-norm  and $\ell_\infty$-norm adversarial perturbations. The pre-training dataset is Tiny-ImageNet and the downstream dataset is CIFAR10.}
\label{fig:L2toLinf}
\vspace{-5mm}
\end{figure}

\myparatight{Extension to $\ell_p$-norm adversarial perturbations} We focus on certified robustness against $\ell_2$-norm adversarial perturbation in this work. The certified robustness  can be extended to other $\ell_p$-norms, e.g., via leveraging the relationship between $\ell_2$-norm and other $\ell_p$-norms. 
For instance, suppose the certified radius is $R$ for an image in $\ell_2$-norm; the certified radius in $\ell_1$-norm and $\ell_\infty$-norm can respectively be computed as  $R$ and $\frac{R}{\sqrt{dim}}$, where $dim$ is the product of the number of pixels and the number of channels in the image. Figure~\ref{fig:L2toLinf}  shows the certified accuracy of SC in REaaS for $\ell_1$-norm and $\ell_\infty$-norm adversarial perturbations, where the $\ell_1$-norm and $\ell_\infty$-norm certified radii are obtained from $\ell_2$-norm certified radius with $N=100,000$, $\sigma=0.5$, and $\alpha=0.001$.

\myparatight{\CR{Extending REaaS to natural language processing (NLP) domain}}\CR{An attacker can make a text classifier predict an incorrect label for a text by substituting a small number of words as their synonyms~\cite{alzantot-etal-2018-generating,ren-etal-2019-generating,huang2019achieving}. Our REaaS can also be applied to enable adversarially robust downstream text classifiers against those attacks by slightly adapting our F2IPerturb-API (please refer to Appendix~\ref{reaas_nlp_domain} for details). 
Given a text and a feature-space certified radius, our adapted F2IPerturb-API returns an input-space certified radius, which is the maximum number of words that can be substituted such that the downstream classifier's predicted label for the text is unchanged. Table~\ref{table:nlp} shows our experimental results (please refer to Appendix~\ref{reaas_nlp_domain} for details of the experimental setup). Our results show that our REaaS is also applicable to NLP domain. }

\myparatight{\CR{Encoder stealing}}\CR{Our REaaS introduces a new F2IPerturb-API. A natural question is whether the new F2IPerturb-API makes the encoder more vulnerable to stealing attacks. We argue that the answer is probably no. The reason is that our new API returns a certified radius for a query image, which can also be obtained by an attacker via calling the existing Feature-API many times. However, an attacker may obtain such certified radius with less queries using our new API. We explore whether certified radii can be exploited to assist encoder stealing. In particular, we extend StolenEncoder~\cite{liu2022stolenencoder}, which uses Feature-API to steal encoder, to steal encoders using both Feature-API and F2IPerturb-API (see Appendix~\ref{stolenencoder} for the details of StolenEncoder and its extended version as well as the experimental setup).  Table~\ref{table:stealencoder} shows our experimental results, where $\gamma$ is a hyperparameter. Note that the total number of queries to the APIs made by the extended StolenEncoder is twice of StolenEncoder in our comparison. Our results show that the downstream classifiers built upon stolen encoders obtained by StolenEncoder and its extended version achieve comparable accuracy, which implies that certified radii may not be able to assist encoder stealing. }

\begin{table}[!t]\renewcommand{\arraystretch}{1.5}
\centering
\fontsize{9}{8}\selectfont
\setlength{\tabcolsep}{1mm}
\caption{\CR{ACR and \#Queries of REaaS in NLP domain, where BC is used. The pre-training dataset is SST-2~\cite{socher-etal-2013-recursive} and the downstream dataset is IMDB~\cite{maas2011learning}.}}

{\begin{tabular}{|c|c|c|}
\hline
 \multirow{2}{*}{ACR}   &  \multicolumn{2}{c|}{\#Queries} \\ \cline{2-3}
& \makecell{Per training input} & \makecell{Per testing input} \\ \hline
2.517  & 1&   2 \\ \hline
\end{tabular}}

\label{table:nlp}
\end{table}

\begin{table}[!t]\renewcommand{\arraystretch}{1.5}
\centering
\fontsize{9}{8}\selectfont
\setlength{\tabcolsep}{1mm}
\caption{ \CR{Comparing StolenEncoder and its extended version using our F2IPerturb-API. The pre-training dataset is Tiny-ImageNet and the downstream dataset is CIFAR10.}}
\begin{tabular}{|c|c|c|c|c|c|}
\hline
 & StolenEncoder & \multicolumn{4}{c|}{Extended StolenEncoder} \\ \hline
$\gamma$ &0& $10^{-4}$ & $10^{-3}$ & $10^{-2}$ & $10^{-1}$ \\ \hline
Stolen Accuracy (\%) & 62.3 & 62.6 & 63.4 & 61.5 & 59.4 \\ \hline
\end{tabular}
\label{table:stealencoder}
\end{table}

\myparatight{Privacy-preserving encoder as a service}In both SEaaS and REaaS, a client sends his/her raw images to the cloud server. Therefore, an untrusted service provider may compromise the privacy of the client, especially when the downstream datasets contain sensitive images such as facial and medical images. We believe  it is an interesting future work to develop privacy-preserving encoder as a service. For instance, we can leverage (local) differential privacy~\cite{duchi2013local,erlingsson2014rappor,wang2017locally}, secure hardware~\cite{costan2016intel}, and cryptography~\cite{bost2015machine,gilad2016cryptonets} based methods.

\myparatight{\CR{Other attacks to pre-trained encoders}}\CR{In this work, we focus on adversarial examples~\cite{Szegedy14,carlini2017towards}. Some recent studies~\cite{jia2022badencoder,carlini2021poisoning,liu2022poisonedencoder} show that pre-trained encoders are also vulnerable to poisoning and backdoor attacks, which are orthogonal to our work. We believe it is an interesting future work to extend our framework to defend against those attacks. }

%% file: conclusion.tex
\section{Conclusion and Future Work}
In this work, we show that, via providing two APIs, a cloud server 1) makes it possible for a client to certify robustness of its downstream classifier against adversarial perturbations using any certification method and 2) makes it orders of magnitude more communication efficient and more computation efficient  to certify robustness using smoothed classifier based certification. Moreover, when the cloud server pre-trains the encoder via considering our spectral-norm regularization term, it achieves better certified robustness for the clients' downstream classifiers. Interesting future work includes 
\CR{extending REaaS to poisoning and backdoor attacks} 
as well as designing both robust and privacy-preserving encoder as a service.

\section*{Acknowledgements}

We thank the anonymous reviewers for the constructive comments. This work was supported by NSF under grant No.
2131859, 2112562, and 1937786, as well as ARO
under grant No. W911NF2110182.

%% file: supplementary.tex
\appendices

\section{ACR is the Area under the Certified Accuracy vs. Perturbation Size Curve}
\label{relationship_of_ACR}
Suppose the largest certified radius of a testing input is $R_{max}$ and the certified radius of the testing inputs follows a probability distribution in the interval [0, $R_{max}$], whose probability density function is $p(R)$. The certified accuracy (CA) at a perturbation size $R$ can be formally defined as follows:
\begin{align}
 CA(R)=\int_R^{R_{max}}p(r) dr.  
\end{align}
By definition,  the area under the certified accuracy vs. perturbation size curve is calculated as follows:

\begin{align}
 Area&=\int_0^{R_{max}}CA(R) dR\\
 &=  CA(R) \cdot R|_0^{R_{max}} - \int_0^{R_{max}}R \cdot d CA(R) \\
 &= 0 + \int_0^{R_{max}} R\cdot p(R)\cdot d R\\
 & = ACR,
\end{align}
where ACR is the average certified radius of the testing inputs.

\section{Technical Details of CROWN~\cite{zhang2018crown}} 
\label{crownexample}
Suppose $F$ is a base classifier that maps an input $\mathbf{x}$ to one of $c$ classes $\{1,2,\cdots, c\}$. For instance, the composition of the
encoder and downstream classifier $g\circ f$ is the base classifier $F$ in SEaaS, and a client can treat the downstream classifier $g$ as the base classifier $F$ in REaaS.
We use $H(\mathbf{x})$ to denote the base classifier's last-layer output vector for $\mathbf{x}$, where $H_l(\mathbf{x})$ represents the $l$th entry of $H(\mathbf{x})$ and $l=1,2,\cdots,c$. 
$F(\mathbf{x})$ denotes the predicted label for $\mathbf{x}$, i.e.,  $F(\mathbf{x})=\argmax_{l=1,2,\cdots,c}H_{l}(\mathbf{x})$. Suppose the base classifier predicts label $y$ for $\mathbf{x}$ when there is no adversarial perturbation, i.e., $F(\mathbf{x})=y$. 
A base classifier based certification method derives a certified radius $R$ such that $F(\mathbf{x}+\delta)=y$ for all $\lnorm{\delta}_2 < R$, where $\delta$ is adversarial perturbation. Next, we discuss how to derive the certified radius $R$ for CROWN~\cite{zhang2018crown}, a state-of-the-art base classifier based certification method.

The key idea of CROWN is to bound each entry of the output vector, i.e., $H_{l}(\mathbf{x})$. In particular, when the output of the activation function (e.g., ReLU) can be bounded by two linear functions, CROWN shows that each entry $H_{l}(\mathbf{x})$ can be bounded by two  linear functions $H^{L}_{l}(\mathbf{x})$ and $H^{U}_{l}(\mathbf{x})$. Formally, we have $H^{L}_l(\mathbf{x})\leq H_l(\mathbf{x})  \leq H^{U}_l(\mathbf{x})$, where $l \in \{1,2,\cdots, c\}$. Suppose an adversarial perturbation $\delta$ that satisfies $\lnorm{\delta}_2 < r$ is added to $\mathbf{x}$. Then, we have the following:
\begin{align}
\label{inequality_crown}
  \min_{\lnorm{\delta}_2 < r} H^{L}_l(\mathbf{x} + \delta) \leq H_l(\mathbf{x}+\delta) \leq \max_{\lnorm{\delta}_2 < r} H^{U}_{l}(\mathbf{x}+\delta),
\end{align}
where $l =1,2,\cdots, c$. The base classifier $F$ still predicts label $y$ for $\mathbf{x} + \delta$ when the lower bound of the $y$th entry (i.e., $\min_{\lnorm{\delta}_2 < r} H^{L}_y(\mathbf{x} + \delta)$) is larger than the upper bounds of all other entries (i.e., $\max_{l\neq y} \max_{\lnorm{\delta}_2 < r} H^{U}_{l}(\mathbf{x}+\delta)$). Therefore, CROWN derives the certified radius $R$ as follows:
\begin{align}
    R = \max_{r} r \textit{ s.t. } \min_{\lnorm{\delta}_2 < r} H^{L}_y(\mathbf{x} + \delta) > \max_{l\neq y}\max_{\lnorm{\delta}_2 < r} H^{U}_{l}(\mathbf{x}+\delta). \nonumber
\end{align}
Note that CROWN shows that $\min_{\lnorm{\delta}_2 < r} H^{L}_l(\mathbf{x} + \delta)$ and $\max_{\lnorm{\delta}_2 < r} H^{U}_{l}(\mathbf{x}+\delta)$ have closed-form solutions for $l=1,2,\cdots,c$.

\section{Technical Details of SC based Certification~\cite{cohen2019certified}}
\label{randomizedsmoothingexample}

Given a base classifier $F$ (e.g., $g\circ f$ in SEaaS and $g$ in REaaS), randomized smoothing builds a smoothed classifier $h$ via adding random Gaussian noise to the testing input of the base classifier. The smoothed classifier $h$ provably predicts the same label for the testing input when the $\ell_2$-norm of the perturbation add to the testing input is less than a threshold (i.e., certified radius). Next, we respectively discuss how to build a smoothed classifier, how to derive the certified radius for a testing input, as well as how to train the base classifier to improve the certified radius. 

\myparatight{Building a smoothed classifier}For simplicity, we use $\mathcal{N}(0, \sigma^2 \mathbf{I})$ to denote a zero-mean isotropic Gaussian noise, where $\sigma$ is the standard deviation and $\mathbf{I}$ is an identity matrix. Given an input $\mathbf{x}$, a base classifier $F$, and a zero-mean Gaussian noise $\mathcal{N}(0, \sigma^2 \mathbf{I})$, we define the \emph{label probability} $p_l, l \in \{1,2,\cdots,c\}$, as follows:
\begin{align}
    p_l = \text{Pr}(F (\mathbf{x} + \mathcal{N}(0, \sigma^2 \mathbf{I}))=l).
\end{align}
Roughly speaking, $p_l$ is the probability that the base classifier $F$ predicts label $l$ when taking  $\mathbf{x} + \mathcal{N}(0, \sigma^2 \mathbf{I})$ as input. The smoothed classifier $h$ predicts the label with the largest label probability for the input $\mathbf{x}$:
\begin{align}
    h(\mathbf{x}) = \argmax_{l \in \{1,2,\cdots,c\}} p_l. 
\end{align}
In practice, it is usually computationally intractable to compute the exact label probabilities due to the complexity of the base classifier. In response, we sample $N$ noise $\mathbf{n}_1, \mathbf{n}_2,\cdots, \mathbf{n}_N$ from $\mathcal{N}(0, \sigma^2 \mathbf{I})$. For each $l \in \{1,2,\cdots, c\}$, we denote $N_l = \sum_{j=1}^{N}\mathbb{I}(F(\mathbf{x}+\mathbf{n}_j)=l)$ (called \emph{label frequency}), where $\mathbb{I}$ is an indicator function. The label (denoted as $y$) with the largest label frequency is predicted as the label of $\mathbf{x}$.

\begin{table}[!t]\renewcommand{\arraystretch}{1.3}
    \centering
    \caption{Architecture of the neural network for the encoder.}
    \begin{tabular}{|c|c|} \hline 
    Layer Type & Layer Parameters  \\ \hline
    \multicolumn{2}{|c|}{Input  $32\times 32$} \\ \hline
    Convolution& $16\times 3 \times 3$, strides=$(1, 1)$, padding=1  \\ 
    Activation& ReLU  \\ \hline
    Convolution& $16\times 4 \times 4$, strides=$(2, 2)$, padding=1  \\ 
    Activation& ReLU  \\ \hline
    Convolution& $32\times 3 \times 3$, strides=$(1, 1)$, padding=1  \\ 
    Activation& ReLU  \\ \hline
    Convolution& $32\times 4\times 4$, strides=$(2, 2)$, padding=1  \\ 
    Activation& ReLU  \\ 
    Fully Connected& 256  \\ \hline
    \multicolumn{2}{|c|}{Output} \\ \hline
    \end{tabular} 
    \vspace{-4mm}
    \label{architecture} 
\end{table}

\myparatight{Computing certified radius for an input} Cohen et al.~\cite{cohen2019certified} proved that the smoothed classifier  predicts the label $y$ for the input $\mathbf{x}$ when the $\ell_2$-norm of the adversarial perturbation is less than $\frac{\sigma}{2}\cdot (\Phi^{-1}(\underline{p_{y}}) - \Phi^{-1}(\overline{p_u}))$, where $\underline{p_y}$ is a lower bound of $p_y$ and $\overline{p_u}$ is an upper bound of $\max_{l\neq y}p_{l}$. In other words, we have $h(\mathbf{x}+\delta)=y$ when $\lnorm{\delta}_2 < R= \frac{\sigma}{2}\cdot (\Phi^{-1}(\underline{p_{y}}) - \Phi^{-1}(\overline{p_u}))$, where $R$ is the certified radius. 

To compute the certified radius, we need to estimate a lower bound of $p_y$ and an upper bound of $\max_{l\neq y}p_{l}$. Cohen et al. proposed to estimate such lower or upper bounds  using the standard one-sided Clopper-Pearson method~\cite{clopper1934use} based on label frequencies. In particular, $N_y$ follows a binomial distribution with parameters $N$ and $p_y$. Based on the Clopper-Pearson method, we have the following lower bound for $p_y$:
\begin{align}
\label{cp_lower_bound}
    \underline{p_y} = Beta(\alpha; N_y, N-N_y+1),
\end{align}
where $1-\alpha$ is the confidence level and $Beta(\alpha; \varsigma, \vartheta)$ is the $\alpha$th quantile of the Beta distribution with shape parameters $\varsigma$ and $\vartheta$. Intuitively, with probability at least $1-\alpha$ over the randomness of the sampling of Gaussian noise, we have $p_y \geq \underline{p_y}$. Given the lower bound $\underline{p_y}$ and the fact that the summation of label probabilities is $1$ (i.e., $\sum_{l=1}^c p_l=1$), we can derive $1 - \underline{p_y}$ as an upper bound for $\max_{l\neq y}p_{l}$, i.e., $\max_{l\neq y}p_{l} \leq 1-\underline{p_y}$. Given the lower bound of $p_y$ and the upper bound of $\max_{l\neq y}p_{l}$, we have the certified radius $R = \sigma \cdot \Phi^{-1}(\underline{p_y})$. In other words, with probability at least $1-\alpha$ over the randomness of the sampling of Gaussian noise, we have $g(\mathbf{x}+\delta)=y$ when $\lnorm{\delta}_2 < \Phi^{-1}(\underline{p_y})$.

\myparatight{Training a base classifier with Gaussian noise} The certified radius for an input depends on whether the base classifier $F$ can correctly classify the input with Gaussian noise. However, when the base classifier $F$ is trained using the normal training examples, it is very likely that the base classifier cannot correctly predict the label for an input with Gaussian noise due to the difference between training and testing data distributions. Therefore, Cohen et al.~\cite{cohen2019certified} proposed to add Gaussian noise to the training inputs when training the base classifier. In particular, for each mini-batch of training examples, we add random Gaussian noise to each training input and then use them to update the base classifier.  Such Gaussian noise based training method can significantly improve the certified radius~\cite{cohen2019certified}.

\section{On the Tightness of Equation~\ref{inequality_crown_method}}
\label{tightness_of_crown_equation}
\CR{
We show that the upper and lower bounds in (5) are tight when the encoder consists of one linear layer. Suppose $f(\mathbf{x})= \mathbf{W}\mathbf{x}+\mathbf{b}$, where $\mathbf{W}$ and $\mathbf{b}$ are the weight matrix and bias vector of the linear encoder $f$. For simplicity, we respectively use $\mathbf{W}[i,:]$ and $\mathbf{b}[i]$ to denote the $i$th row of $\mathbf{W}$ and $i$th element of $\mathbf{b}$. Based on Theorem 3.2 in CROWN~\cite{zhang2018crown}, we have $f_i^{U}(\mathbf{x})=\mathbf{W}[i,:]\mathbf{x}+\mathbf{b}[i]$ and $f_i^{L}(\mathbf{x})=\mathbf{W}[i,:]\mathbf{x}+\mathbf{b}[i]$ when $f(\mathbf{x})= \mathbf{W}\mathbf{x}+\mathbf{b}$ (we omit the details for simplicity). Next, we prove the right-hand side of Equation~\ref{inequality_crown_method} is tight. To reach the goal, our idea is to  show  there exists $\delta'$ satisfying $\lnorm{\delta'}_2=\rho_k$ such that $f_i(\mathbf{x}+\delta')>\max_{\lnorm{\delta}_2 < \rho_k} f^{U}_{i}(\mathbf{x}+\delta)$. We first derive an upper bound of $\max_{\lnorm{\delta}_2 < \rho_k} f^{U}_{i}(\mathbf{x}+\delta)$.
\begin{align}
\label{tightness_0}
   & \max_{\lnorm{\delta}_2 < \rho_k}  f_i^U(\mathbf{x}+\delta) \\
    =&  \max_{\lnorm{\delta}_2 < \rho_k} (\mathbf{W}[i,:]\mathbf{x}+\mathbf{W}[i,:]\delta+\mathbf{b}[i]) \\
    \label{tightness_1}
    =& \mathbf{W}[i,:]\mathbf{x} +\mathbf{b}[i] +\max_{\lnorm{\delta}_2 < \rho_k} \mathbf{W}[i,:]\delta \\
    \label{tightness_2}
    \leq & \mathbf{W}[i,:]\mathbf{x} +\mathbf{b}[i]  + \max_{\lnorm{\delta}_2 < \rho_k} \lnorm{\mathbf{W}[i,:]}_2 \lnorm{\delta}_2 \\
    \label{tightness_3}
    <& \mathbf{W}[i,:]\mathbf{x} +\mathbf{b}[i]  + \lnorm{\mathbf{W}[i,:]}_2 \rho_{k}
\end{align}
We have Equation~\ref{tightness_2} from~\ref{tightness_1} based on Cauchy-Schwartz inequality. Our derived upper bound is the same as the one obtained from the closed-form solution of $f_i^U$ based on Corollary 3.3 in~\cite{zhang2018crown}. We let $\delta' = \frac{\rho_k \mathbf{W}[i,:]^T}{\lnorm{\mathbf{W}[i,:]}_2}$, where $T$ represents the transpose operation. We can verify that $\lnorm{\delta'}=\rho_k$. Then, we have:
\begin{align}
 &f_i(\mathbf{x}+\delta') \\
 =& \mathbf{W}[i,:]\mathbf{x} +\mathbf{b}[i] + \mathbf{W}[i,:]\frac{\rho_k \mathbf{W}[i,:]^T}{\lnorm{\mathbf{W}[i,:]}_2} \\
 =& \mathbf{W}[i,:]\mathbf{x} +\mathbf{b}[i]  + \lnorm{\mathbf{W}[i,:]}_2 \rho_{k}.
\end{align}
Combining the above equations with Equation~\ref{tightness_0} and ~\ref{tightness_3}, we know that there exist $\delta'$ satisfying $\lnorm{\delta'}_2=\rho_k$ such that $f_i(\mathbf{x}+\delta')>\max_{\lnorm{\delta}_2 < \rho_k} f^{U}_{i}(\mathbf{x}+\delta)$. Thus, we prove the tightness of the right-hand side of Equation~\ref{inequality_crown_method}. Similarly, we can prove the tightness of the left-hand side of Equation~\ref{inequality_crown_method}. 
}

\section{Applying REaaS to NLP domain}
\label{reaas_nlp_domain}
\CR{
Our REaaS provides two APIs: Feature-API and F2IPerturb-API. 
The Feature-API can be directly applied to the NLP domain. In particular, given a text as input, the Feature-API returns the feature vector produced by a text encoder for it. Next, we discuss how to adapt our F2IPerturb-API to the NLP domain. 
}

\myparatight{\CR{Adapting our F2IPerturb-API to NLP domain}}\CR{
Given a text input and a feature-space certified radius, we can adapt our F2IPerturb-API to provide the input-space certified radius for the input text. In particular, we consider an attacker can replace a small number of words in a text with their synonyms such that the downstream classifier of a client makes incorrect prediction for it. For simplicity, given a text input $\mathbf{x}$, we use $\mathbf{x}[i]$  to denote the $i$th word in $\mathbf{x}$. Moreover, we use $\delta$ to denote an adversarial text perturbation whose length is the same as $\mathbf{x}$, where $\delta[i]$ is an empty string if we don't perturb $\mathbf{x}[i]$; we use $\mathbf{x} \oplus \delta$ to denote the perturbed text obtained by replacing $\mathbf{x}[i]$ as $\delta[i]$ if $\delta[i]$ is not an empty string; we use $\lnorm{\delta}_0$ to denote the number of words in $\delta$ that are not empty string.  
}

\CR{
The input-space certified radius returned by our adapted F2IPerturb-API is the maximum number of words that can be replaced by their synonyms such that the predicted label by the downstream classifier is unchanged. Formally, given a text input $\mathbf{x}$ and a feature-space certified radius $R_F$, our adapted F2IPerturb-API aims to solve the following optimization problem: 
\begin{align}
\label{nlp_api_3_opt_1}
   & R = \max_{r} r \\
    \textit{ s.t. }& \max_{\lnorm{\delta}_0 < r}
    \label{nlp_api_3_opt_2} \lnorm{f(\mathbf{x}\oplus \delta)-f(\mathbf{x})}_2 < R_F.
\end{align}
Similar to the image domain, we can use binary search to solve the above optimization problem. The key difference with the image domain is how to estimate the lower or upper bounds of $f_i(\mathbf{x}\oplus \delta)$. We note that Xu et al.~\cite{xu2020automatic} extend CROWN~\cite{zhang2018crown} for general perturbations, which can be applied to derive lower and upper bounds (denoted as $f_i^L(\mathbf{x}\oplus \delta)$ and $f_i^U(\mathbf{x}\oplus \delta)$)  of $f_i(\mathbf{x}\oplus \delta)$. Moreover, they also provide a computation-efficient method to compute $\min_{\lnorm{\delta}_0 < r} f_i^L(\mathbf{x}\oplus \delta)$ and $\max_{\lnorm{\delta}_0 < r} f_i^U(\mathbf{x}\oplus \delta)$ (please see Theorem 2 in~\cite{xu2020automatic} for details). Given those bounds, our  adapted F2IPerturb-API can be used to obtain the certified radius. Algorithm~\ref{alg:api_3} shows our complete algorithm. The function \textsc{ExtendedCrown} uses the extended CROWN~\cite{xu2020automatic} to obtain lower or upper bounds. 
}

\begin{algorithm}[!t]
   \caption{\CR{\emph{F2IPerturb-API (NLP)}}}
   \label{alg:api_3}
\begin{algorithmic}
\STATE {\bfseries \CR{Input from client:}} \CR{text $\mathbf{x}$ and feature-space certified \\ radius $R_F$}
  \CR{ \STATE {\bfseries Output for client:} text-space certified radius $R$ \\
   \STATE $\rho^L, \rho^U \gets 0, 10$ \\
    \WHILE{$\rho^U - \rho^L > 1$} 
    \STATE $\rho_k = \lceil \frac{\rho^L+\rho^U}{2} \rceil$ \\
    \FOR{$i=1,2,\cdots, d$} 
    \STATE $f_i^L, f_i^U \gets \textsc{ExtendedCrown}(i,\mathbf{x},f)$ \\
   \STATE \CRR{$L_i= \min_{\lnorm{\delta}_0 \leq \rho_k}f_i^L(\mathbf{x}+\delta)-f_i(\mathbf{x})$} \\
   \STATE \CRR{$U_i= \max_{\lnorm{\delta}_0 \leq \rho_k}f_i^U(\mathbf{x}+\delta)-f_i(\mathbf{x})$}\\
    \ENDFOR
    \STATE $R_F^\prime = \sqrt{\sum_{i=1}^d\max(L_i^2, U_i^2)}$ \\
    \IF{$R_F^\prime < R_F$}
    \STATE $\rho^L = \rho_k$ \\
    \ELSE
    \STATE $\rho^U = \rho_k$ \\
    \ENDIF
    \ENDWHILE
   \STATE \textbf{return} $\rho^L$}
\end{algorithmic}
\end{algorithm}

\myparatight{\CR{Experimental setup}}\CR{We use SST-2 dataset~\cite{socher2013recursive} as the pre-training dataset and train a text encoder using supervised learning. 
Following~\cite{xu2020automatic}, we use LSTM as the encoder architecture. Moreover, we use public implementation from~\cite{xu2020automatic} in our experiments.  We consider the same word substitutions as previous work~\cite{huang2019achieving,jia2019certified} when deriving the input-space certified radius for our F2IPerturb-API. 
We use IMDB dataset~\cite{maas2011learning} as the downstream dataset, which is a benchmark dataset used for sentiment analysis in the NLP domain. We train a logistic regression classifier as the downstream classifier. Moreover, we use CROWN~\cite{zhang2018crown} as BC based certification to derive the certified radius of the downstream classifier. 
}

\section{Extending StolenEncoder~\cite{liu2022stolenencoder} to Steal Encoders with our new F2IPerturb-API}
\label{stolenencoder}

\CR{We first introduce StolenEncoder~\cite{liu2022stolenencoder} and then discuss how to extend it to steal encoders with our F2IPerturb-API.}

\myparatight{\CR{StolenEncoder}}\CR{Suppose an attacker has an unlabeled surrogate dataset $\mathcal{D}$. Given black-box access to an encoder (i.e., an attacker can query the Feature-API in our terminology), the attacker can use StolenEncoder to steal a target encoder $f$. Suppose $f_s$ is the stolen encoder. StolenEncoder aims to solve the following optimization problem:
\begin{align}
\label{original_stolen_encoder}
  \min_{f_s} \mathcal{L}(\mathcal{D}) = \frac{1}{|\mathcal{D}|}\sum_{\mathbf{x}\in \mathcal{D}}[d(f(\mathbf{x}), f_s(\mathbf{x})) + \kappa d(f(\mathbf{x}), f_s(\mathcal{A}(\mathbf{x})))], 
\end{align}
where $\kappa$ is a hyperparameter, $d$ is a distance metric (e.g., $\ell_2$-distance), and $\mathcal{A}$ is the data augmentation component (e.g., RandomHorizontalFlip, ColorJitter, and RandomGrayScale). The total number of queries to the Feature-API is $|\mathcal{D}|$ as a single query is made  for each input in $\mathcal{D}$. StolenEncoder uses standard stochastic gradient descent (SGD) to solve the above optimization problem.}

\vspace{-2mm}
\myparatight{\CR{Extending StolenEncoder with our new F2IPerturb-API}} \CR{An attacker can also query F2IPerturb-API in our REaaS. We extend StolenEncoder to leverage our F2IPerturb-API to steal the target encoder $f$. In particular, given an input image $\mathbf{x}$ and a feature-space certified radius $R_{F}$ for $\mathbf{x}$, our F2IPerturb-API returns an input-space certified radius $R$. Therefore, we know the following based on the output of F2IPerturb-API:
\begin{align}
\label{target_encoder_equation}
 \lnorm{f(\mathbf{x})-f(\mathbf{x}+\delta)}_2 < R_F \text{ for } \forall \delta, \lnorm{\delta}_2 < R.   
\end{align}
 We can train the stolen encoder $f_s$ such that it also satisfies Equation~\ref{target_encoder_equation}. Suppose we have a surrogate dataset $\mathcal{D}$. Given an input $\mathbf{x}\in \mathcal{D}$, we first randomly sample a radius from $[R_{F}^{min}, R_{F}^{max}]$ and view it as the feature-space certified radius $R_{F}^{\mathbf{x}}$. Then, we query F2IPerturb-API to obtain the input-space certified radius $R^{\mathbf{x}}$. To make the stolen encoder $f_s$ satisfy Equation~\ref{target_encoder_equation} for $\mathbf{x}$, we randomly sample an input-space perturbation $\delta^{\mathbf{x}}$ that satisfies $\lnorm{\delta^{\mathbf{x}}}_2 = R^{\mathbf{x}}$. Then, we define the following loss:
\begin{align}
    \mathcal{L}'(\mathcal{D}) = \frac{1}{|\mathcal{D}|}\sum_{\mathbf{x}\in \mathcal{D}} &d(f_s(\mathbf{x}), f_s(\mathbf{x}+\delta^{\mathbf{x}}))\cdot \\ \nonumber
    &\mathbb{I}(d(f_s(\mathbf{x}), f_s(\mathbf{x}+\delta^{\mathbf{x}}))> R_F^{\mathbf{x}}),
\end{align}
where $\mathbb{I}$ is an indicator function whose output is 1 if the condition is satisfied and 0 otherwise. 
Intuitively, the loss term in $ \mathcal{L}'(\mathcal{D})$ for an input $\mathbf{x}$ is 0 if $d(f_s(\mathbf{x}), f_s(\mathbf{x}+\delta^{\mathbf{x}})) \leq R_F^{\mathbf{x}}$ and is non-zero otherwise. Combining with the loss of StolenEncoder, our final optimization problem is as follows:
\begin{align}
\label{extend_stolenencoder_loss}
 \min_{f_s}   \mathcal{L}_1(\mathcal{D}) = \mathcal{L}(\mathcal{D}) + \gamma \mathcal{L}'(\mathcal{D}),
\end{align}
where $\gamma$ is a hyperparameter and $\mathcal{L}$ is defined in Equation~\ref{original_stolen_encoder}. The total number of queries is $2|\mathcal{D}|$ as we respectively make one query to the Feature-API and F2IPerturb-API for each input in $\mathcal{D}$. Similarly, we use stochastic gradient descent to solve the optimization problem. Note that our extended StolenEncoder reduces to StolenEncoder when $\gamma=0$.
}

\myparatight{\CR{Experimental setup}}\CR{We use Tiny-ImageNet as the pre-training dataset and use CIFAR10 as the downstream task. Moreover, we randomly sample 10,000 images from STL10 dataset as the surrogate dataset $\mathcal{D}$. We adopt the same $\kappa$, $d$, and $\mathcal{A}$ as StolenEncoder.
We set $R_F^{min}$ and $R_F^{max}$ to be 0.01 and 0.5, respectively. As we use the same surrogate dataset, we give advantages to the extended StolenEncoder since its total number of queries is twice of StolenEncoder. Following~\cite{liu2022stolenencoder}, we use \emph{Stolen Accuracy} as evaluation metrics. In particular, given a stolen encoder and a downstream task, Stolen Accuracy is the classification accuracy of the downstream classifier built upon the stolen encoder for the downstream task.}

\begin{figure*}[tb]
\centering
\subfloat[CIFAR10]{\includegraphics[width=0.33\textwidth]{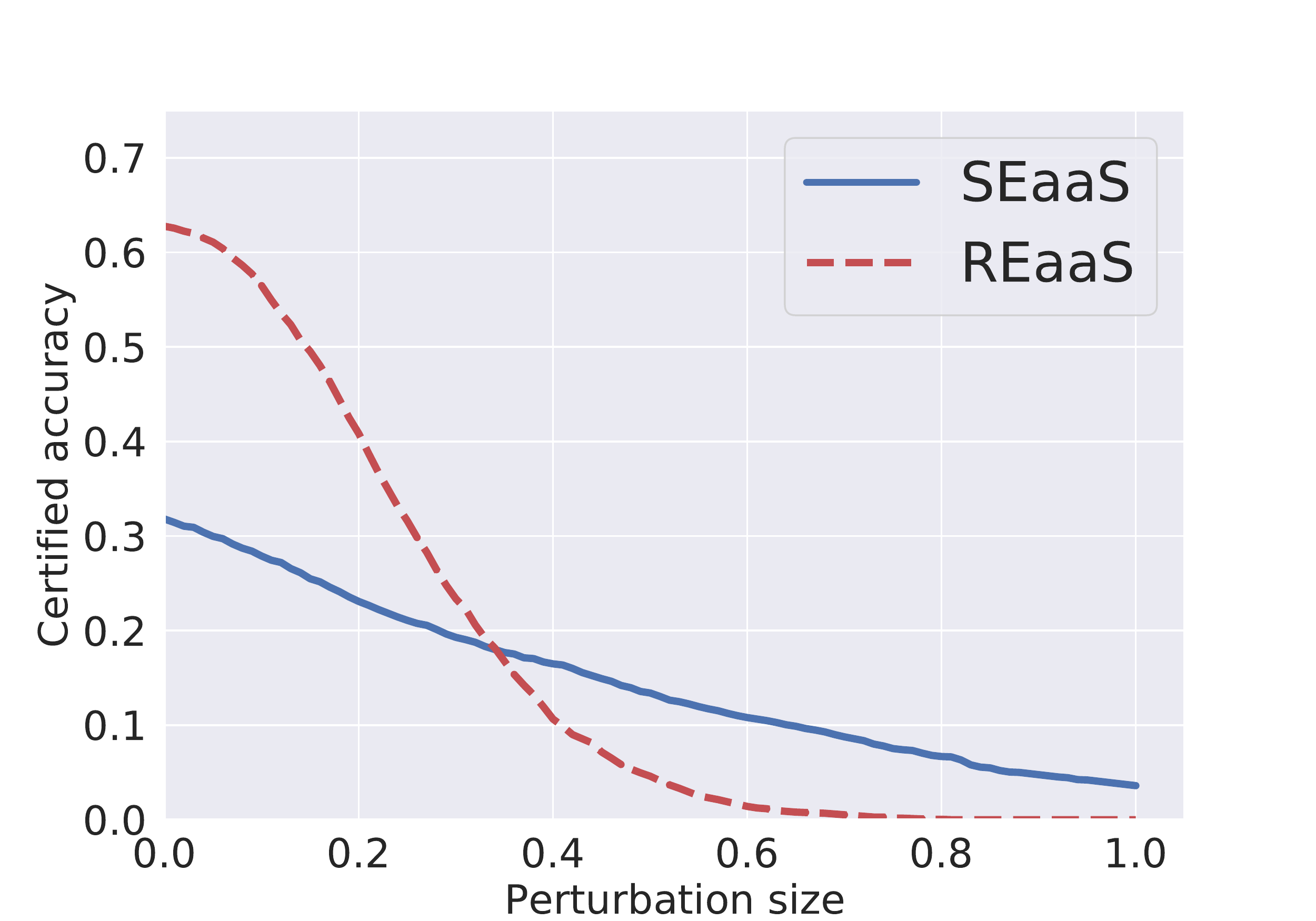}}
\subfloat[SVHN]{\includegraphics[width=0.33\textwidth]{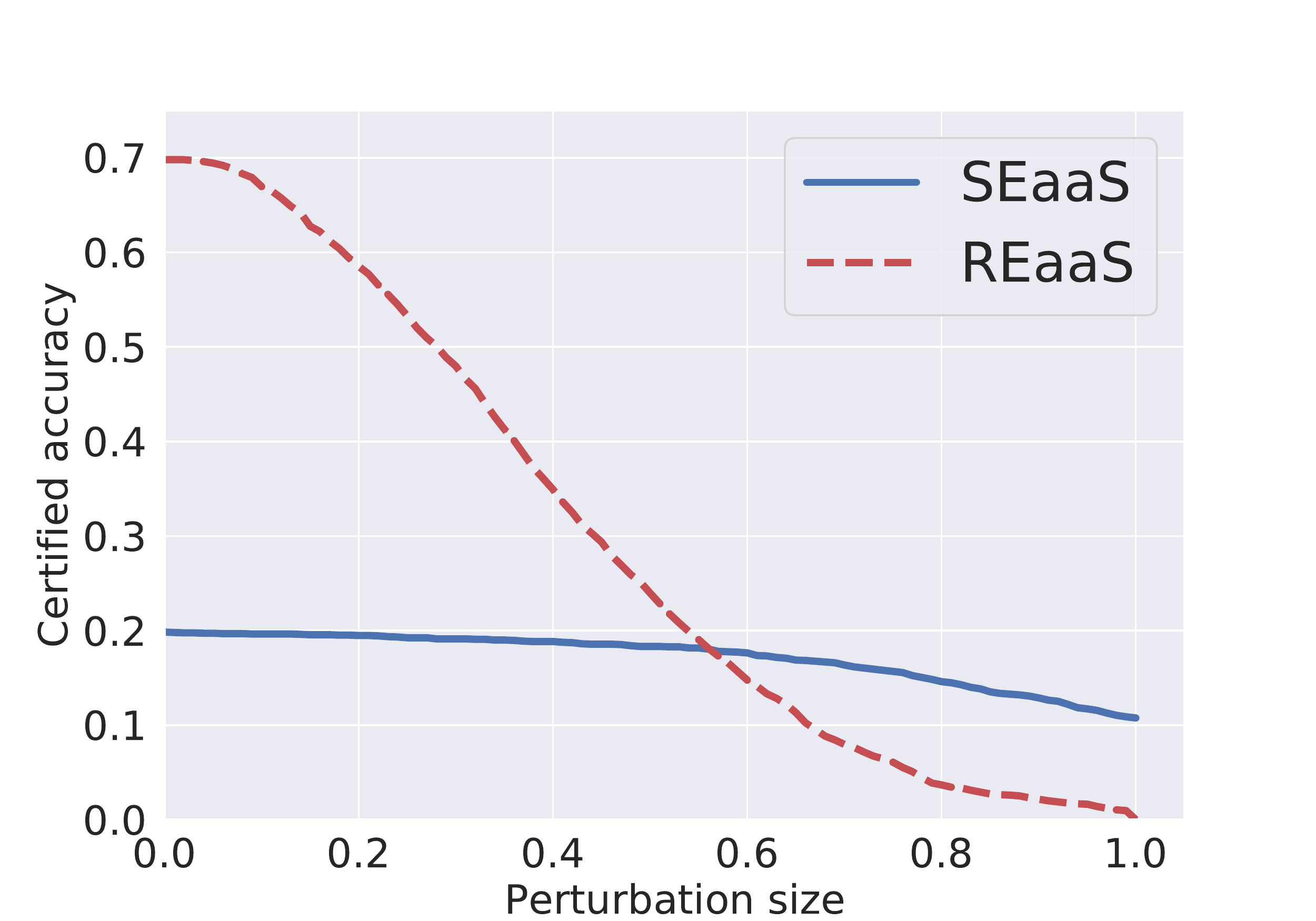}}
\subfloat[STL10]{\includegraphics[width=0.33\textwidth]{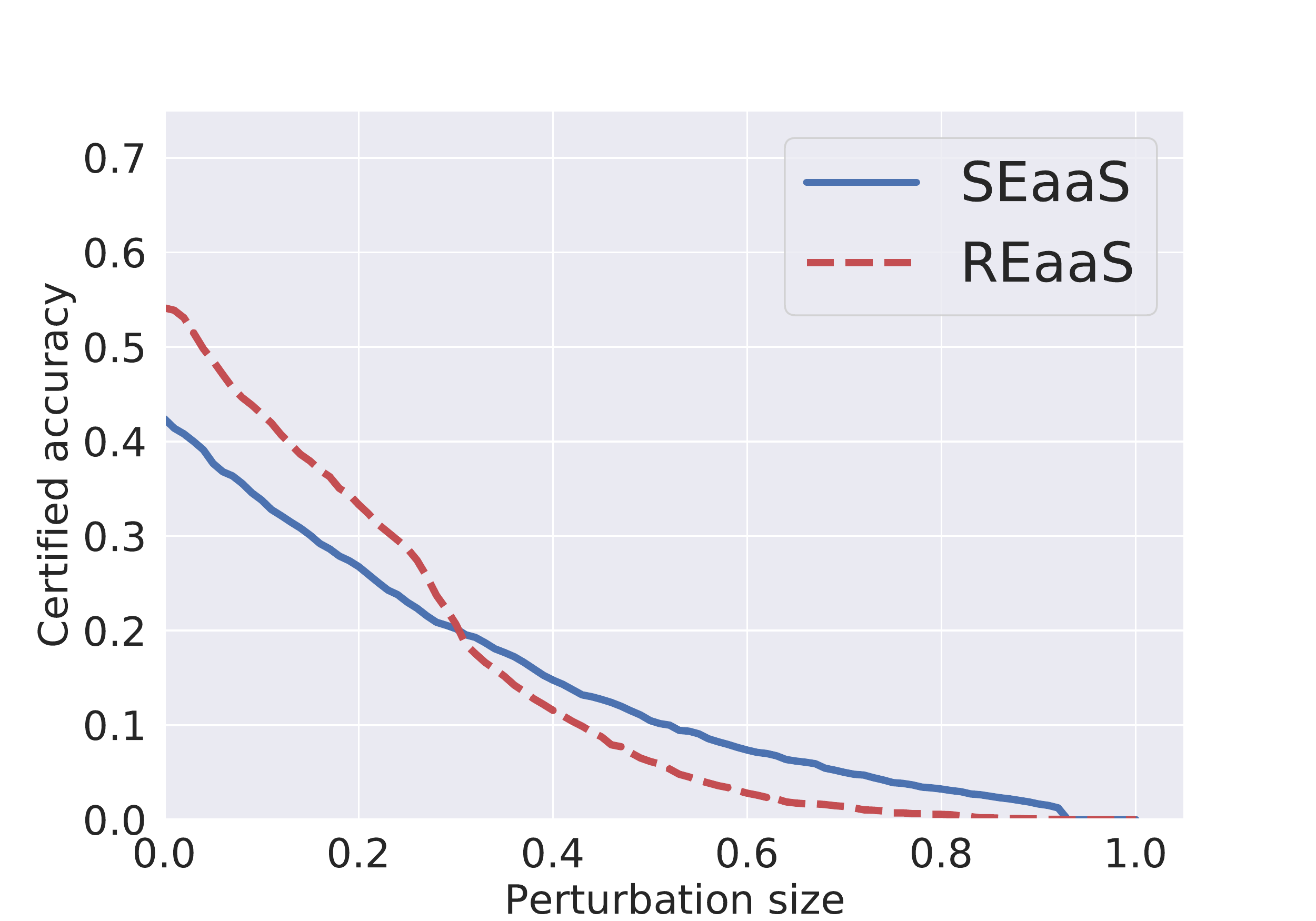}}

\vspace{-3mm}
\subfloat[CIFAR10]{\includegraphics[width=0.33\textwidth]{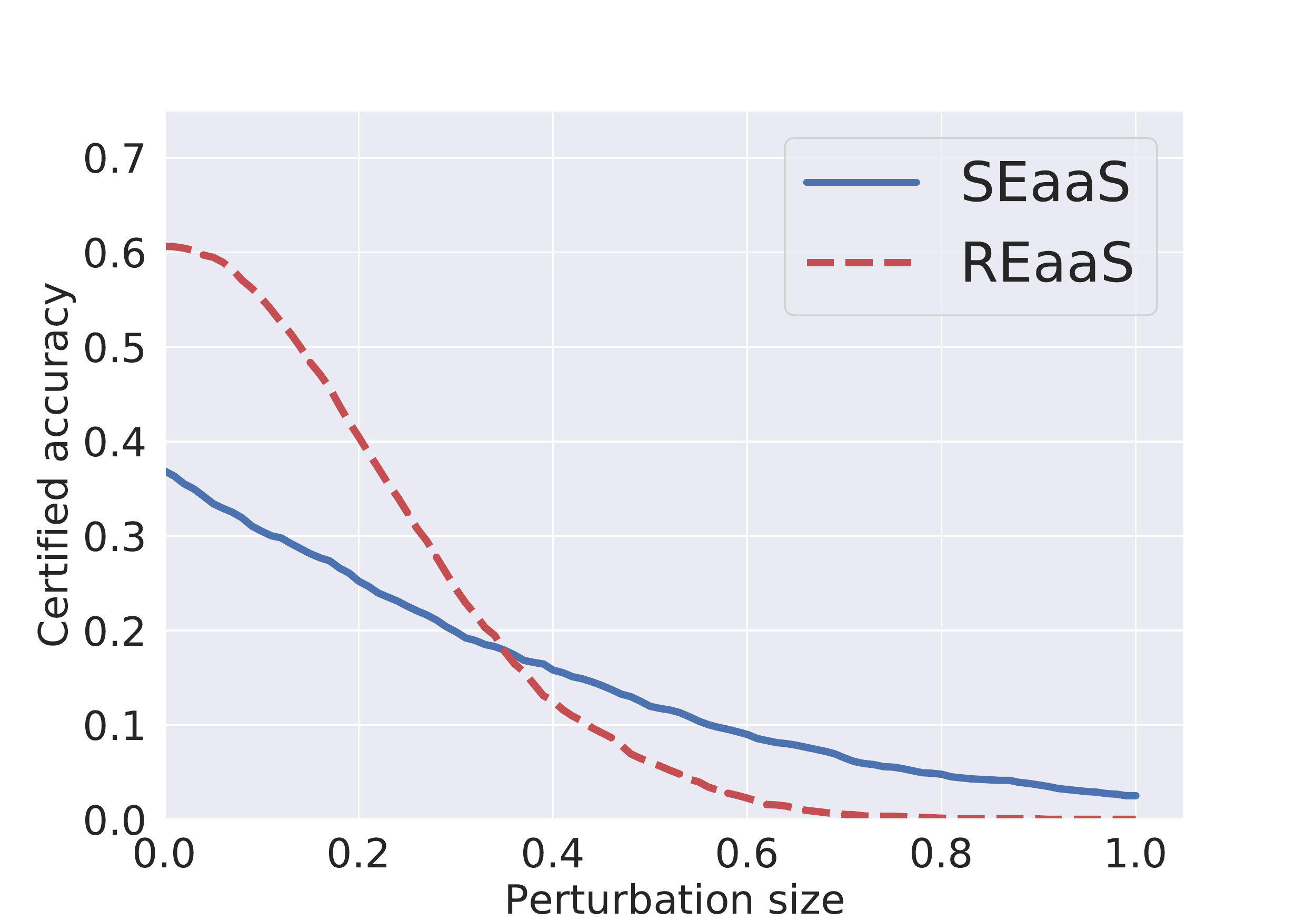}}
\subfloat[SVHN]{\includegraphics[width=0.33\textwidth]{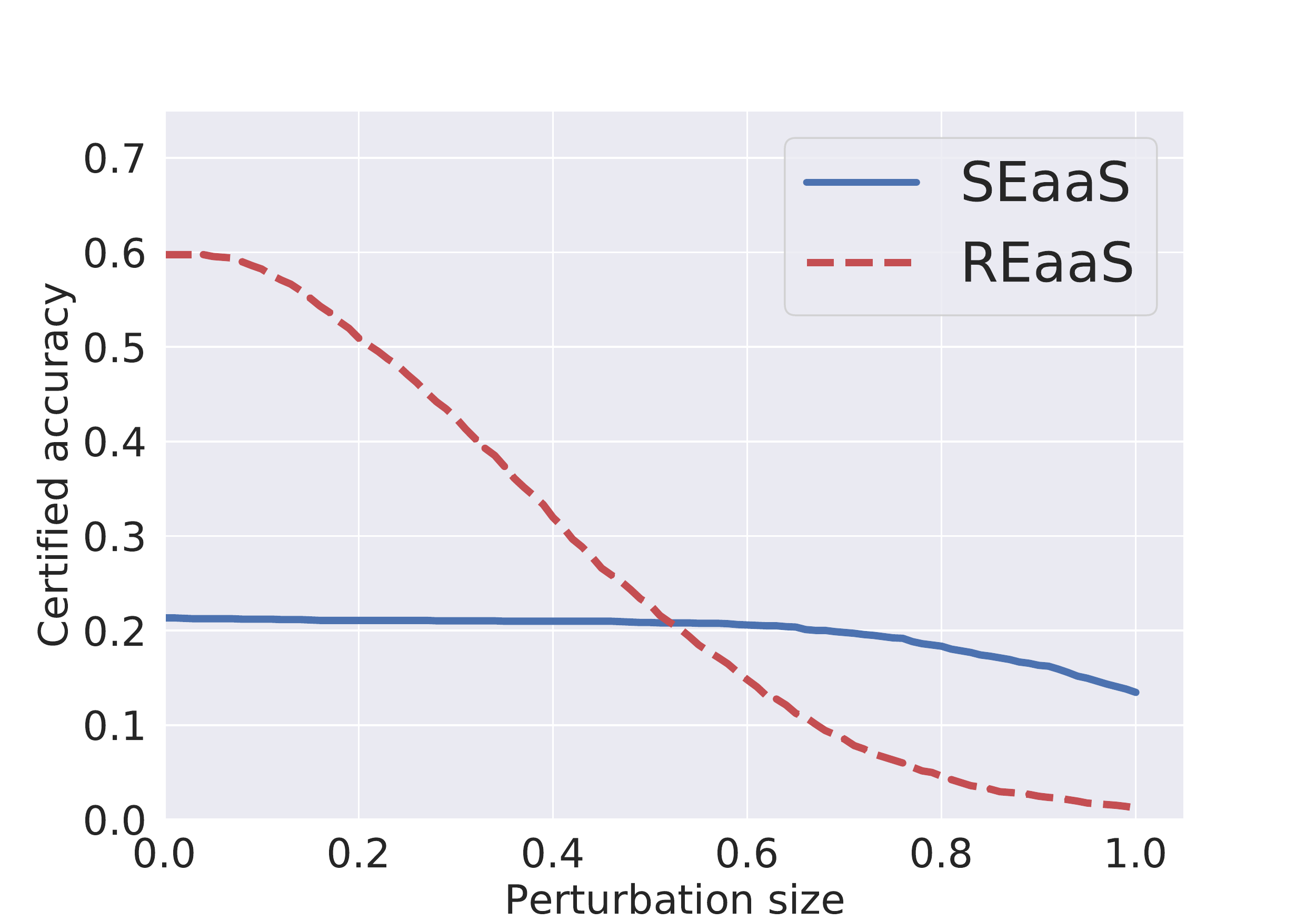}}
\subfloat[Tiny-ImageNet]{\includegraphics[width=0.33\textwidth]{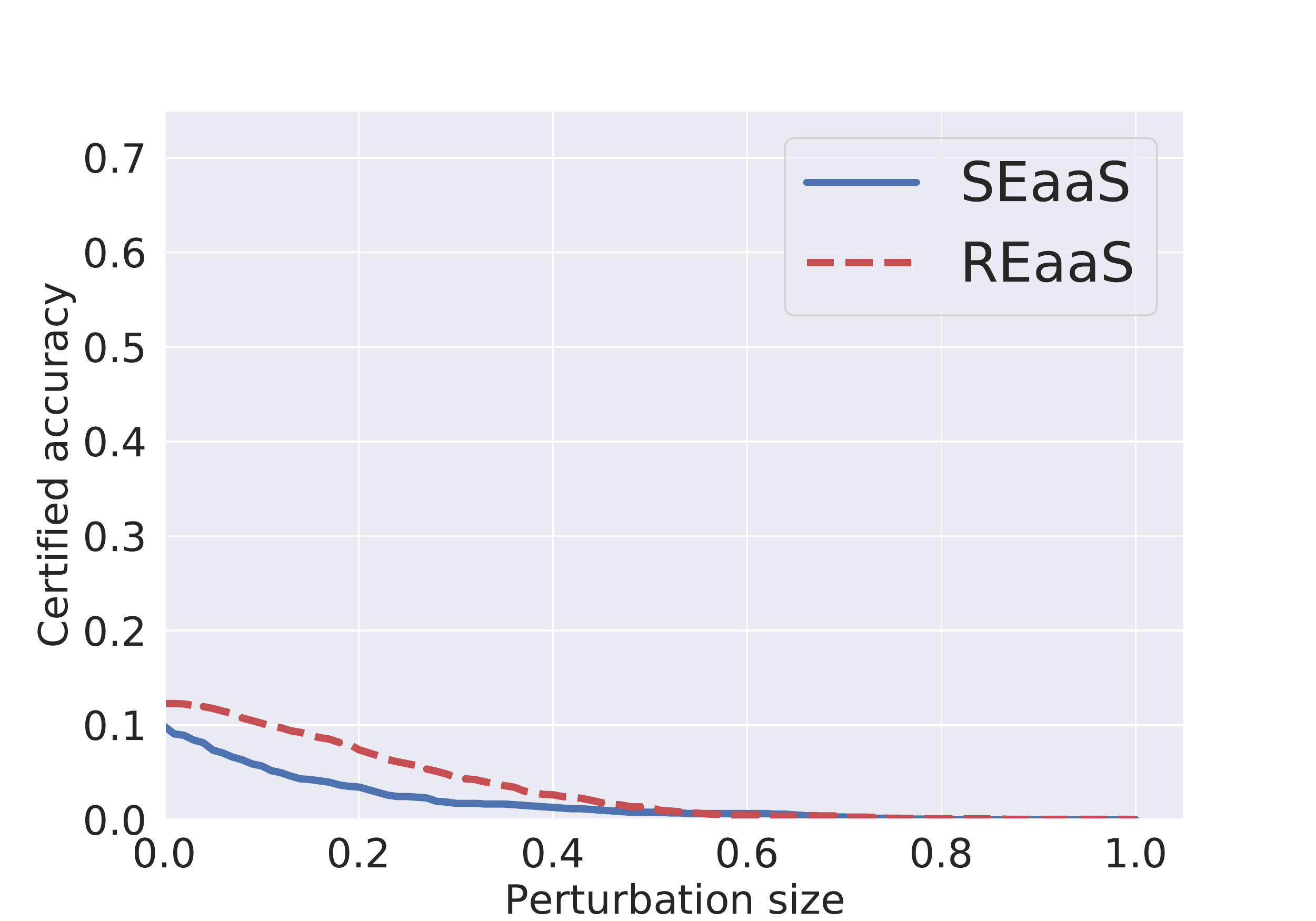}}
\caption{Comparing certified accuracy vs. perturbation size of SC in SEaaS and REaaS for different downstream datasets, where the pre-training dataset is Tiny-ImageNet (first row) and STL10 (second row).} 
\label{fig:overallcomparison1}
\vspace{-4mm}
\end{figure*}

\begin{figure*}[!t]
\centering

\subfloat[$N$]{\includegraphics[width=0.33\textwidth]{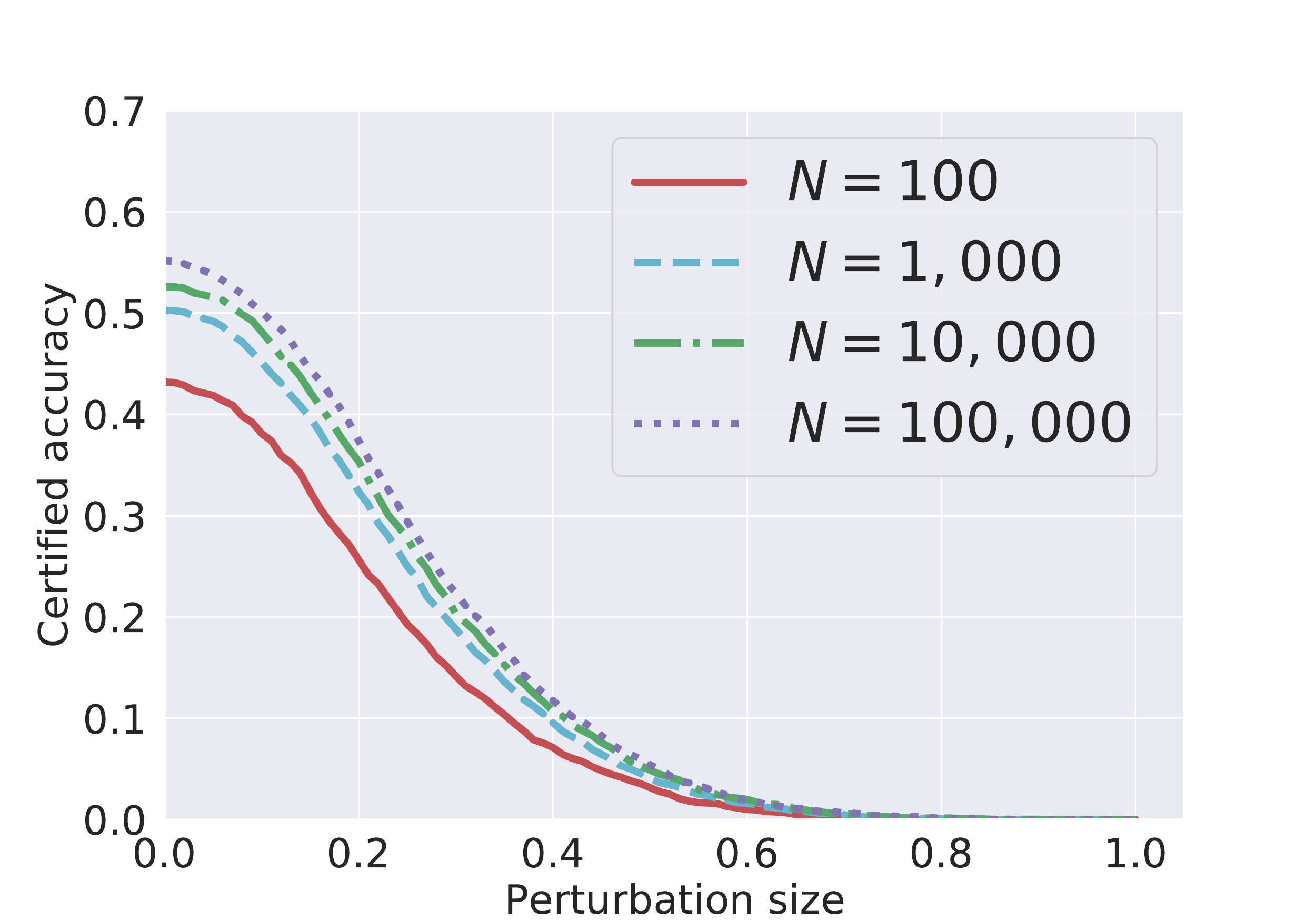}}
\subfloat[$\sigma$]{\includegraphics[width=0.33\textwidth]{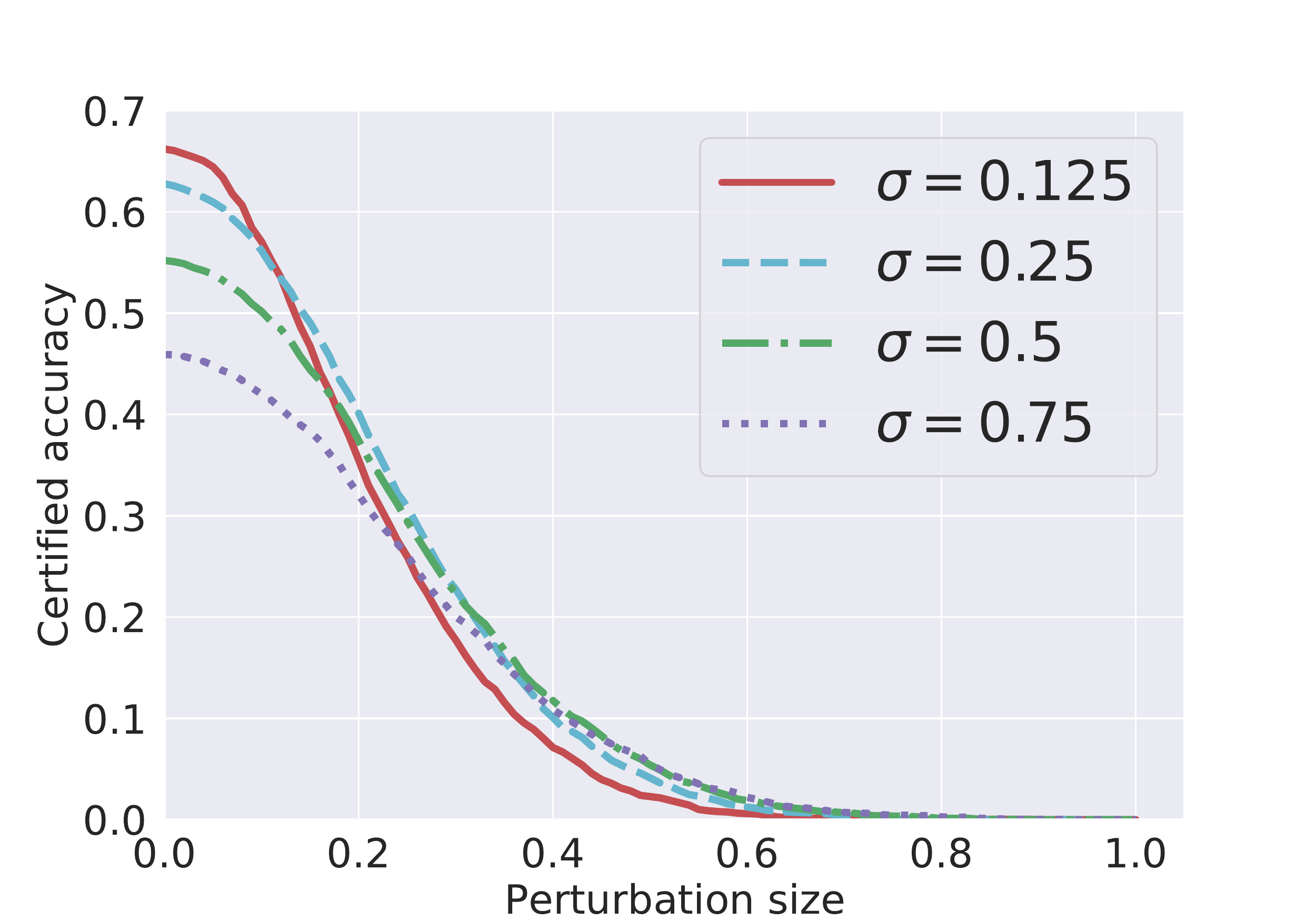}}
\subfloat[$\alpha$]{\includegraphics[width=0.33\textwidth]{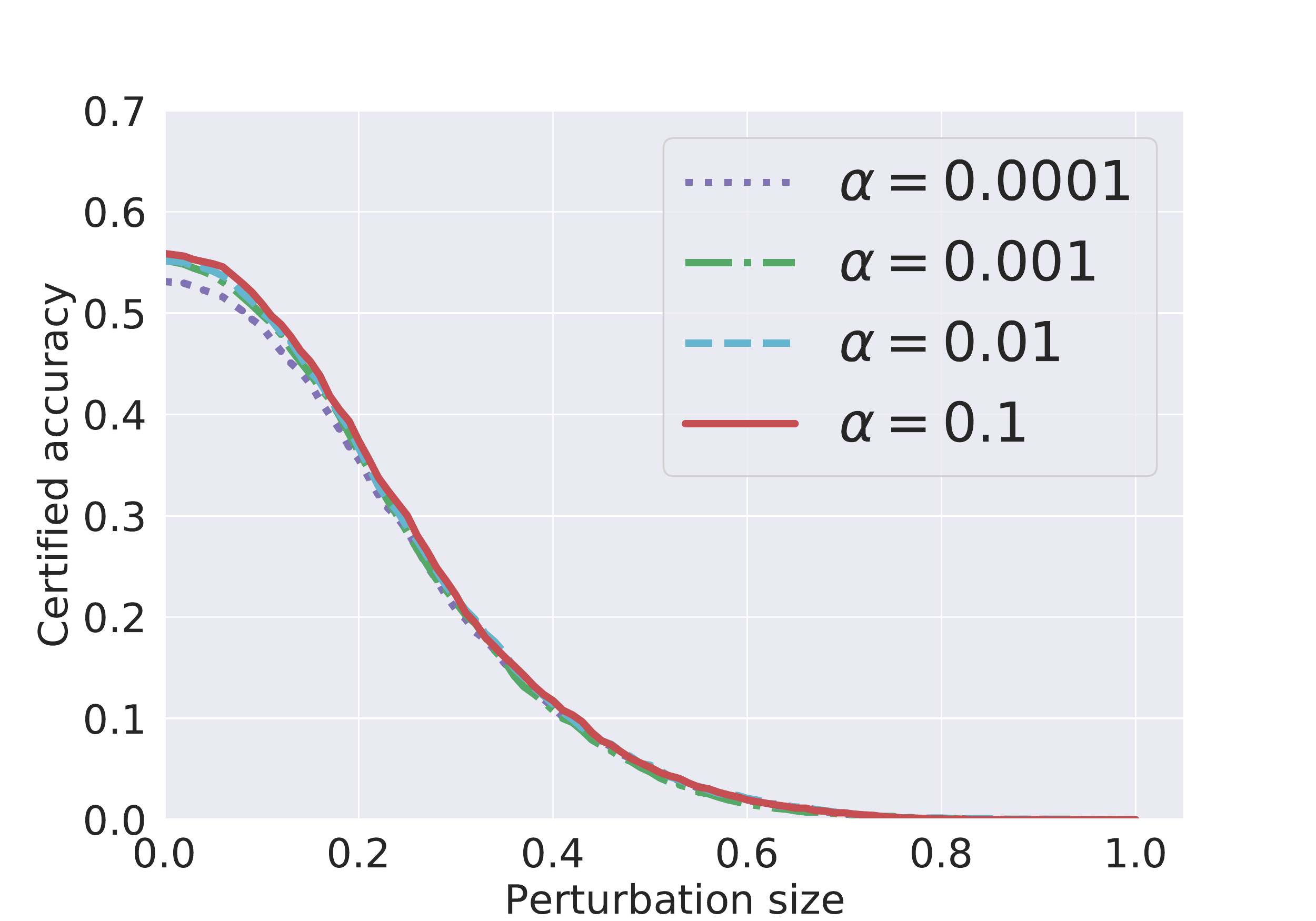}}
\caption{Impact of $N$, $\sigma$, and $\alpha$ on  certified accuracy  of SC in REaaS. The pre-training dataset is Tiny-ImageNet and the downstream dataset is CIFAR10.}
\label{fig:SCparameterCA}
\end{figure*}

%% file: paper.bbl
\begin{thebibliography}{10}
\providecommand{\url}[1]{#1}
\csname url@samestyle\endcsname
\providecommand{\newblock}{\relax}
\providecommand{\bibinfo}[2]{#2}
\providecommand{\BIBentrySTDinterwordspacing}{\spaceskip=0pt\relax}
\providecommand{\BIBentryALTinterwordstretchfactor}{4}
\providecommand{\BIBentryALTinterwordspacing}{\spaceskip=\fontdimen2\font plus
\BIBentryALTinterwordstretchfactor\fontdimen3\font minus
  \fontdimen4\font\relax}
\providecommand{\BIBforeignlanguage}[2]{{%
\expandafter\ifx\csname l@#1\endcsname\relax
\typeout{** WARNING: IEEEtran.bst: No hyphenation pattern has been}%
\typeout{** loaded for the language `#1'. Using the pattern for}%
\typeout{** the default language instead.}%
\else
\language=\csname l@#1\endcsname
\fi
#2}}
\providecommand{\BIBdecl}{\relax}
\BIBdecl

\bibitem{he2019moco}
K.~He, H.~Fan, Y.~Wu, S.~Xie, and R.~Girshick, ``Momentum contrast for
  unsupervised visual representation learning,'' in \emph{CVPR}, 2020.

\bibitem{chen2020simple}
T.~Chen, S.~Kornblith, M.~Norouzi, and G.~Hinton, ``A simple framework for
  contrastive learning of visual representations,'' in \emph{ICML}, 2020.

\bibitem{GPT3-api}
{GPT-3 API}, \url{https://openai.com/blog/customized-gpt-3/}.

\bibitem{clarifai_embedding}
``{General Image Embedding Model},''
  \url{https://www.clarifai.com/models/general-image-embedding}.

\bibitem{Szegedy14}
C.~Szegedy, W.~Zaremba, I.~Sutskever, J.~Bruna, D.~Erhan, I.~Goodfellow, and
  R.~Fergus, ``Intriguing properties of neural networks,'' in \emph{ICLR},
  2014.

\bibitem{carlini2017towards}
N.~Carlini and D.~Wagner, ``Towards evaluating the robustness of neural
  networks,'' in \emph{IEEE S \& P}, 2017.

\bibitem{wong2017provable}
E.~Wong and J.~Z. Kolter, ``Provable defenses against adversarial examples via
  the convex outer adversarial polytope,'' in \emph{ICML}, 2018.

\bibitem{raghunathan2018certified}
A.~Raghunathan, J.~Steinhardt, and P.~Liang, ``Certified defenses against
  adversarial examples,'' in \emph{ICLR}, 2018.

\bibitem{zhang2018crown}
H.~Zhang, T.-W. Weng, P.-Y. Chen, C.-J. Hsieh, and L.~Daniel, ``Efficient
  neural network robustness certification with general activation functions,''
  in \emph{NeurIPS}, 2018.

\bibitem{wang2018formal}
S.~Wang, K.~Pei, J.~Whitehouse, J.~Yang, and S.~Jana, ``Formal security
  analysis of neural networks using symbolic intervals,'' in \emph{USENIX
  Security}, 2018.

\bibitem{cao2017mitigating}
X.~Cao and N.~Z. Gong, ``Mitigating evasion attacks to deep neural networks via
  region-based classification,'' in \emph{ACSAC}, 2017.

\bibitem{lecuyer2019certified}
M.~Lecuyer, V.~Atlidakis, R.~Geambasu, D.~Hsu, and S.~Jana, ``Certified
  robustness to adversarial examples with differential privacy,'' in \emph{IEEE
  S \& P}, 2019.

\bibitem{cohen2019certified}
J.~M. Cohen, E.~Rosenfeld, and J.~Z. Kolter, ``Certified adversarial robustness
  via randomized smoothing,'' \emph{ICML}, 2019.

\bibitem{kim2020adversarial}
M.~Kim, J.~Tack, and S.~J. Hwang, ``Adversarial self-supervised contrastive
  learning,'' in \emph{NeurIPS}, 2020.

\bibitem{goodfellow2014explaining}
I.~J. Goodfellow, J.~Shlens, and C.~Szegedy, ``Explaining and harnessing
  adversarial examples,'' in \emph{ICLR}, 2015.

\bibitem{madry2017towards}
A.~Madry, A.~Makelov, L.~Schmidt, D.~Tsipras, and A.~Vladu, ``Towards deep
  learning models resistant to adversarial attacks,'' in \emph{ICLR}, 2018.

\bibitem{salman2020denoised}
H.~Salman, M.~Sun, G.~Yang, A.~Kapoor, and J.~Z. Kolter, ``Denoised smoothing:
  A provable defense for pretrained classifiers,'' \emph{NeurIPS}, 2020.

\bibitem{pan2009survey}
S.~J. Pan and Q.~Yang, ``A survey on transfer learning,'' \emph{TKDE}, 2009.

\bibitem{chen2020adversarial}
T.~Chen, S.~Liu, S.~Chang, Y.~Cheng, L.~Amini, and Z.~Wang, ``Adversarial
  robustness: From self-supervised pre-training to fine-tuning,'' in
  \emph{CVPR}, 2020.

\bibitem{jiang2020robust}
Z.~Jiang, T.~Chen, T.~Chen, and Z.~Wang, ``Robust pre-training by adversarial
  contrastive learning.'' in \emph{NeurIPS}, 2020.

\bibitem{mises1929praktische}
R.~Mises and H.~Pollaczek-Geiringer, ``Praktische verfahren der
  gleichungsaufl{\"o}sung.'' \emph{ZAMM-Journal of Applied Mathematics and
  Mechanics}, 1929.

\bibitem{krizhevsky2009learning}
A.~Krizhevsky, G.~Hinton \emph{et~al.}, ``Learning multiple layers of features
  from tiny images,'' \emph{Tech Report}, 2009.

\bibitem{netzer2011reading}
Y.~Netzer, T.~Wang, A.~Coates, A.~Bissacco, B.~Wu, and A.~Y. Ng, ``Reading
  digits in natural images with unsupervised feature learning,'' in
  \emph{NeurIPS Workshop on Deep Learning and Unsupervised Feature Learning},
  2011.

\bibitem{coates2011analysis}
A.~Coates, A.~Ng, and H.~Lee, ``An analysis of single-layer networks in
  unsupervised feature learning,'' in \emph{AISTATS}, 2011.

\bibitem{tinyimagenet}
``{Tiny-ImageNet},'' \url{https://www.kaggle.com/c/tiny-imagenet/overview}.

\bibitem{moco_implementation}
``{MoCo},'' \url{https://github.com/facebookresearch/moco}.

\bibitem{balunovic2019adversarial}
M.~Balunovic and M.~Vechev, ``Adversarial training and provable defenses:
  Bridging the gap,'' in \emph{ICLR}, 2019.

\bibitem{zhang2019towards}
H.~Zhang, H.~Chen, C.~Xiao, S.~Gowal, R.~Stanforth, B.~Li, D.~Boning, and C.-J.
  Hsieh, ``Towards stable and efficient training of verifiably robust neural
  networks,'' \emph{ICLR}, 2020.

\bibitem{crown_implementation}
``{CROWN},'' \url{https://github.com/huanzhang12/CROWN-IBP}.

\bibitem{randomsmooth_implementation}
``{Randomized Smoothing},'' \url{https://github.com/locuslab/smoothing}.

\bibitem{alzantot-etal-2018-generating}
M.~Alzantot, Y.~Sharma, A.~Elgohary, B.-J. Ho, M.~Srivastava, and K.-W. Chang,
  ``Generating natural language adversarial examples,'' in \emph{EMNLP}, 2018.

\bibitem{ren-etal-2019-generating}
S.~Ren, Y.~Deng, K.~He, and W.~Che, ``Generating natural language adversarial
  examples through probability weighted word saliency,'' in \emph{ACL}, 2019.

\bibitem{huang2019achieving}
P.-S. Huang, R.~Stanforth, J.~Welbl, C.~Dyer, D.~Yogatama, S.~Gowal,
  K.~Dvijotham, and P.~Kohli, ``Achieving verified robustness to symbol
  substitutions via interval bound propagation,'' \emph{EMNLP-IJCNLP}, 2019.

\bibitem{liu2022stolenencoder}
Y.~Liu, J.~Jia, H.~Liu, and N.~Z. Gong, ``Stolenencoder: Stealing pre-trained
  encoders in self-supervised learning,'' in \emph{CCS}, 2022.

\bibitem{socher-etal-2013-recursive}
R.~Socher, A.~Perelygin, J.~Wu, J.~Chuang, C.~D. Manning, A.~Ng, and C.~Potts,
  ``Recursive deep models for semantic compositionality over a sentiment
  treebank,'' in \emph{EMNLP}, 2013.

\bibitem{maas2011learning}
A.~Maas, R.~E. Daly, P.~T. Pham, D.~Huang, A.~Y. Ng, and C.~Potts, ``Learning
  word vectors for sentiment analysis,'' in \emph{ACL}, 2011.

\bibitem{duchi2013local}
J.~C. Duchi, M.~I. Jordan, and M.~J. Wainwright, ``Local privacy and
  statistical minimax rates,'' in \emph{FOCS}, 2013.

\bibitem{erlingsson2014rappor}
{\'U}.~Erlingsson, V.~Pihur, and A.~Korolova, ``Rappor: Randomized aggregatable
  privacy-preserving ordinal response,'' in \emph{CCS}, 2014.

\bibitem{wang2017locally}
T.~Wang, J.~Blocki, N.~Li, and S.~Jha, ``Locally differentially private
  protocols for frequency estimation,'' in \emph{USENIX Security}, 2017.

\bibitem{costan2016intel}
V.~Costan and S.~Devadas, ``Intel sgx explained.'' \emph{IACR Cryptol. ePrint
  Arch.}, 2016.

\bibitem{bost2015machine}
R.~Bost, R.~A. Popa, S.~Tu, and S.~Goldwasser, ``Machine learning
  classification over encrypted data.'' in \emph{NDSS}, 2015.

\bibitem{gilad2016cryptonets}
R.~Gilad-Bachrach, N.~Dowlin, K.~Laine, K.~Lauter, M.~Naehrig, and J.~Wernsing,
  ``Cryptonets: Applying neural networks to encrypted data with high throughput
  and accuracy,'' in \emph{ICML}, 2016.

\bibitem{jia2022badencoder}
J.~Jia, Y.~Liu, and N.~Z. Gong, ``Badencoder: Backdoor attacks to pre-trained
  encoders in self-supervised learning,'' in \emph{IEEE S\&P}, 2022.

\bibitem{carlini2021poisoning}
N.~Carlini and A.~Terzis, ``Poisoning and backdooring contrastive learning,''
  in \emph{ICLR}, 2021.

\bibitem{liu2022poisonedencoder}
H.~Liu, J.~Jia, and N.~Z. Gong, ``Poisonedencoder: Poisoning the unlabeled
  pre-training data in contrastive learning,'' in \emph{USENIX Security}, 2022.

\bibitem{clopper1934use}
C.~J. Clopper and E.~S. Pearson, ``The use of confidence or fiducial limits
  illustrated in the case of the binomial,'' \emph{Biometrika}, 1934.

\bibitem{xu2020automatic}
K.~Xu, Z.~Shi, H.~Zhang, Y.~Wang, K.-W. Chang, M.~Huang, B.~Kailkhura, X.~Lin,
  and C.-J. Hsieh, ``Automatic perturbation analysis for scalable certified
  robustness and beyond,'' in \emph{NeurIPS}, 2020.

\bibitem{socher2013recursive}
R.~Socher, A.~Perelygin, J.~Wu, J.~Chuang, C.~D. Manning, A.~Y. Ng, and
  C.~Potts, ``Recursive deep models for semantic compositionality over a
  sentiment treebank,'' in \emph{EMMLP}, 2013.

\bibitem{jia2019certified}
R.~Jia, A.~Raghunathan, K.~G{\"o}ksel, and P.~Liang, ``Certified robustness to
  adversarial word substitutions,'' \emph{arXiv preprint arXiv:1909.00986},
  2019.

\end{thebibliography}
